\documentclass[11pt,a4paper,table ]{article}

\usepackage[margin=1in]{geometry}

\usepackage{mathtools}
\usepackage{setspace}
\usepackage[table]{xcolor}
\usepackage{tabularx}

\usepackage{subcaption}
\usepackage{tikz}
\usepackage{url}
\usetikzlibrary{spy}
\usetikzlibrary{decorations.pathreplacing}
\usetikzlibrary{decorations.pathmorphing}
\usetikzlibrary{circuits.logic.US,circuits.logic.IEC,fit}
\usetikzlibrary{hobby}
\usepackage{tabularx}
\usepackage{quiver}
\usepackage{todonotes}

\usepackage{cleveref} 
\usepackage[normalem]{ulem}

\usepackage[utf8]{inputenc}
\usepackage{amsmath}

\usepackage{graphicx}

\usepackage{qcircuit}

\usepackage{eurosym}
\usepackage{tikz}
\tikzset{snake it/.style={decorate, decoration=snake}}
\usetikzlibrary{calc}

\definecolor{qubitcolor}{rgb}{1,1,1}
\definecolor{xcheckcolor}{rgb}{0.89,0,0.13}
\definecolor{zcheckcolor}{rgb}{0,0.5,1}

\newcommand{\op}[1]{\operatorname{#1}}

\newcommand{\im}{\operatorname{im}}

\newcommand{\ftwo}{\mathbb{F}_2}
\DeclareMathOperator{\id}{id}

\usepackage{amsmath}
\usepackage{amsfonts}
\usepackage{amssymb}
\usepackage{amsthm}

\usepackage{physics}

\usepackage{enumitem}
\usepackage{mathrsfs}%
\usepackage{dsfont}

\usepackage{tikz}

\usetikzlibrary{spy}
\usetikzlibrary{decorations.pathreplacing}
\usetikzlibrary{decorations.pathmorphing}
\usetikzlibrary{circuits.logic.US,circuits.logic.IEC,fit}

\usepackage{array}

\usepackage{todonotes}
\usepackage[normalem]{ulem}

\usepackage{cleveref}

\definecolor{good_red}{RGB}{212, 17, 89}
\definecolor{good_blue}{RGB}{26, 133, 255}
\definecolor{good_yellow}{RGB}{255, 193, 7}
\definecolor{good_green}{RGB}{0,77,64}

\definecolor{cell_aut_good_green}{RGB}{163,226,131}

\newcommand{\latticecolor}{gray}
\newcommand{\latticethickness}{thick}
\newcommand{\supportthickness}{2pt}

\let\op\undefined
\newcommand{\op}[1]{\operatorname{#1}}

\DeclareMathOperator{\supp}{supp}

\title{Logical Operators and Derived Automorphisms of Tile Codes }

\newtheorem{theorem}{Theorem}
\newtheorem{proposition}{Proposition}
\newtheorem{corollary}{Corollary}
\newtheorem{lemma}{Lemma}

\newtheorem{example}{Example}
\newtheorem{remark}{Remark}

\author{
Nikolas P. Breuckmann$^{1}$ \hspace{1cm}
Shin Ho Choe$^{2}$\thanks{shinho.choe@meetiqm.com} \hspace{1cm}
Jens Niklas Eberhardt$^{3}$ \\[0.3em]
Francisco Revson Fernandes Pereira$^{2}$ \hspace{1cm}
Vincent Steffan$^{2}\thanks{vincent.steffan@meetiqm.com}$
}

\date{
$^{1}$Breuqmann Ltd., Redcross Village, BS2 0BB, Bristol, United Kingdom \\%
$^{2}$IQM Quantum Computers, Georg-Brauchle-Ring 23–25, 80992 Munich, Germany \\%
$^{3}$Institute of Mathematics, Johannes Gutenberg-Universität, Mainz, Germany \\[1em]
\today
}

\begin{document}

\maketitle

\abstract{
The recently introduced tile codes are a promising alternative to surface codes, combining two-dimensional locality with higher encoding efficiency. While surface codes are well understood in terms of their logical operators and boundary behavior, much less is known about tile codes. In this work, we establish a natural and precise description of their logical operator space. We prove that, under mild assumptions, any tile code admits a canonical symplectic basis of logical operators supported along lattice boundaries, which can be generated efficiently by a simple cellular automaton with the number of update rules only depending on the non-locality of the tile code. 
Further, we develop algebraic and algebro-geometric frameworks for tile codes, by resolving them by translationally invariant Pauli stabilizer models and showing that they arise as derived sections of a Koszul complex on $\mathbb{P}^1 \times \mathbb{P}^1$. Finally, we introduce the concept of derived automorphisms for quantum codes. These are automorphism-like operations that can exist even for codes that do not have symmetries. We explain how derived automorphisms can be implemented for tile codes in a low-overhead and fault-tolerant manner by extending the lattice on one side and shrinking it on the other. While this operation is trivial for the surface code, it induces a product of logical CNOT gates on the encoded information. Our results provide new structural insights into tile codes and lay the groundwork for tile codes as building blocks for fault-tolerant quantum computation.
}

\section{Introduction}\label{sec:intro}

It is widely accepted that quantum error correction is an indispensable ingredient for large-scale quantum computing.
The currently leading proposal for implementing quantum error correction and fault-tolerance is the surface code~\cite{bravyi1998quantum,freedman2001projective}.
This is owed to its good error correction capabilities and easy-to-implement planar layout.
However, recently, there have been several breakthroughs in the construction of quantum codes that are much more efficient and could lead to huge savings in the resource overheads, in particular, so-called quantum low-density parity-check (qLDPC) codes~\cite{breuckmannQuantumLowDensityParityCheck2021a}.

One recently discovered and promising class of qLDPC codes is \emph{tile codes} \cite{steffan2025tilecodeshighefficiencyquantum,liang2025planarquantumlowdensityparitycheck}.
Tile codes were found by generalizing the surface code in the following way:
Physical qubits are arranged on a square grid as for the surface code, but unlike for the surface code, stabilizer checks are allowed to have support inside a box of some fixed side length rather than being nearest-neighbor only.
Hence, stabilizer checks can be non-planar, but they are local in the sense that they are supported within a constant-sized region.
Such mild forms of non-locality would be accessible to hardware platforms with mobility, such as neutral-atom systems~\cite{PhysRevLett.85.2208,RevModPhys.82.2313,PhysRevLett.107.263001}.
Recent developments in superconducting hardware also demonstrate that strict connectivity constraints can be relaxed~\cite{vigneau2025quantumerrordetectionqubitresonator, renger2025superconductingqubitresonatorquantumprocessor, Wallraff2004, blais2004, Song2017, Song2019}.
By construction, all stabilizer checks of tile codes are translationally invariant, so that the code is completely specified by the local set of checks and the boundaries.
These restrictions allow for finding concrete examples by a brute-force computer search.
In follow-up work it has been demonstrated that tile codes admit efficient protocols to implement logical gates~\cite{yang2025planarfaulttolerantquantumcomputation} without any additional connectivity requirements and with moderate physical qubit overhead. In another follow-up work \cite{mathews2025placingroutingquantumldpc}, concrete ways of implementing tile codes and other qLDPC code proposals like bivariate bicycle (BB) codes~\cite{linQuantumTwoblockGroup2023,bravyi2023highthreshold,eberhardt2024pruningqldpccodesbivariate,eberhardt2024logicaloperatorsfoldtransversalgates,chen2025anyontheorytopologicalfrustration, liang2025generalizedtoriccodestwisted, liang2025operatoralgebraalgorithmicconstruction} on superconducting platforms have been investigated: According to \cite{mathews2025placingroutingquantumldpc}, tile codes outperform all other investigated proposals for qLDPC codes significantly with respect to average coupler length and number of hardware components such as through-silicon vias, bump bonds and routing tiers. 

\begin{figure}[htb]
    \centering
    \includegraphics[width=.8\textwidth]{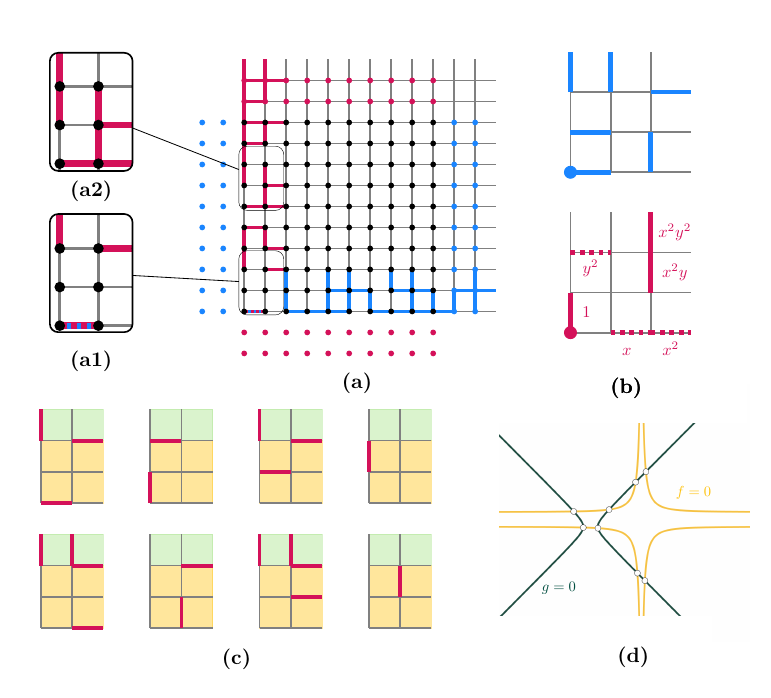}
    \caption{In (a), we show a tile code with a pair of logical operators $\bar{\op X}$ and $\bar{\op Z}$. The stabilizer tiles for this code are shown in (b). The logical $\bar{\op X}$ operators can be constructed using the cellular automaton depicted in (c). In (a1) and (a2), we highlight how the logical operator can be extended using linear combinations of the rules in (c). From an algebraic perspective, the logical dimension of a tile code corresponds to the number of intersection points of the zero sets of the polynomials defining the stabilizer tiles; for an illustration, see (e).}
    \label{fig:main_figure}
\end{figure}

While \cite{steffan2025tilecodeshighefficiencyquantum,liang2025planarquantumlowdensityparitycheck} found several examples of codes with good parameters, there was little control over the structure of the codes.
Their brute-force search is unsatisfactory from a purely theoretical viewpoint alone, but it is also problematic for the development of fault-tolerant logic, as this typically requires control over the logical operators.
In the present work, we remedy this shortcoming twofold:
First, we show that there exists a canonical choice of logical operators in tile codes.
They are sets of pairs of associated $\op X$- and $\op Z$-logicals that only anti-commute with their partner.
They can be localized along the boundary of the tile code, which makes them amenable to lattice surgery techniques.
Interestingly, they arise naturally from the boundary conditions and are constructed via a cellular automaton rule, see \Cref{fig:main_figure}\mbox{(a--c)}.

In the second part of the paper we introduce a framework that explains tile codes in the language of algebraic geometry:
As it turns out, tile codes arise as higher global sections of a shifted Koszul complex
\[\begin{tikzcd}
 {\mathcal{K}^\bullet(\mathbb{P}^1\times \mathbb{P}^1,f,g)\otimes \mathcal{S}:\mathcal{O}(-2D,-2D)\otimes \mathcal{S}} & {\mathcal{O}(-D,-D)^2\otimes \mathcal{S}} & {\mathcal{O}\otimes \mathcal{S}}
 \arrow["{(-g,f)^t}", from=1-1, to=1-2]
 \arrow["{(f,g)}", from=1-2, to=1-3]
\end{tikzcd}\]
of vector bundles on $\mathbb{P}^1\times \mathbb{P}^1$. Here, $f,g \in R = \mathbb{F}_2[x^\pm, y^\pm]$ is a pair of Laurent polynomials corresponding to the stabilizer tiles of the code, see \Cref{fig:main_figure}(b), while $\mathcal{S}$ is a line bundle specifying the size of the bulk and the position of the $\op X$- and $\op Z$-boundary of the tile code.
Using this algebro-geometric framework, we can show that the space of logicals of a tile code is isomorphic to $R/(f,g)$. In this way, the logical dimension of a tile code can be understood as the number of intersection points of the zero sets of $f$ and $g$ over $\overline{\mathbb{F}}_2$, see \Cref{fig:main_figure} (d). This not only provides a different perspective on the structure of logical operators of tile codes, but also gives rise to an understanding of, e.g., tile codes in higher dimensions. For example, in \Cref{sec:algebraoftilecodes} we explain how this lets one understand the structure of 4D versions of tile codes. 

We then go on and introduce \textit{derived automorphisms} for CSS codes inspired by derived categories. The notion of derived automorphisms for quantum codes is a relaxation of the usual automorphisms, which are permutations of qubits and stabilizer checks preserving the code space~\cite{Grassl_2013, breuckmannFoldTransversalCliffordGates2024, eberhardt2024logicaloperatorsfoldtransversalgates,yoder2025tourgrossmodularquantum}. Many codes do not allow for any automorphisms. For example, most tile codes have no automorphisms. We find that for tile codes, there are always derived automorphisms that can be implemented in a fault-tolerant manner with low qubit overhead. The action on the physical qubits consists of extending the tile code to one side and measuring out qubits on the opposite side. We show in \Cref{subsec: pedestrianstylederivedautomorphisms} that this introduces a non-trivial circuit of CNOTs on the logical qubits whose action can be understood in terms of the cellular automaton mentioned before. Moreover, the action on logical operators can also be understood in terms of multiplication by $x$ and $y$ on $R/(f,g).$
We mention that this operation has been studied for the surface code in~\cite{Fowler_2012}: Here, since surface codes only have a single logical qubit, this action merely introduces a change in sign.

We believe that this work, which uses for the first time methods from homological algebra and algebraic geometry to study boundaries of qLDPC codes, opens the door for new research directions:

\begin{enumerate}
    \item How can we optimally combine our low-overhead method of implementing CNOT circuits via derived automorphisms with other proposals for logical operations for tile codes, such as \cite{yang2025planarfaulttolerantquantumcomputation}. 
    \item All results follow under some mild technical assumptions that we explain in \Cref{sec:stabtileson2dplane}. These assumptions are justified since all of the most efficient tile codes fulfill these assumptions~\cite{steffan2025tilecodeshighefficiencyquantum}. Moreover, these assumptions turn out to hold in the `generic' case, see \Cref{sec:algebraoftilecodes}. From a theoretical perspective, it might be interesting to also investigate the remaining corner-cases. 
    \item While first steps to study boundaries in qLDPC codes have been done in \cite{liang2025operatoralgebraalgorithmicconstruction}, our techniques make up the first systematic, formal exploration of boundaries in qLDPC codes using methods from homological algebra and algebraic geometry. We wonder how our techniques help understand boundary condensation phenomena and similar in the context of condensed matter physics. 
\end{enumerate}

The document is structured as follows: In \Cref{sec:stabcsscodes}, we recap the basics on stabilizer CSS codes from the perspective of homological algebra. 
In \Cref{sec:stabtileson2dplane}, we introduce stabilizer tiles on the 2D plane $\mathbb{Z}^2$, recap the construction of tile codes as presented in~\cite{steffan2025tilecodeshighefficiencyquantum}, and present our main findings on the structure of logical operators of tile codes. There, we also present and introduce derived automorphisms for tile codes as an application. 
In \Cref{sec:algebraoftilecodes}, we explain how tile codes can alternatively be understood as resolutions of Koszul complexes. In \Cref{sec:geometryoftilecodes} we give yet another way of explaining the structure of tile codes as Koszul complexes over $\mathbb{P}^1 \times \mathbb{P}^1$. We then discuss the concept of derived automorphisms for general CSS codes in \Cref{sec:derivedautomorphisms}.

\section{Stabilizer CSS codes}\label{sec:stabcsscodes}

A quantum CSS code $\mathcal{Q}\subset (\mathbb{C}^2)^{\otimes n}$ is the joint $+1$ eigenspace of a collection of commuting Pauli operators $S_1, \dots , S_r$ where the operators $S_i$ are elements of $\lbrace \id, X \rbrace^{\otimes n} \cup \lbrace \id, Z \rbrace^{\otimes n}$. Equivalently, we can specify the code by matrices $H_{\op X} \in \mathbb{F}_2^{r_{\op X} \times n}$ and $H_{\op Z} \in \mathbb{F}_2^{r_{\op Z} \times n}$ such that $H_{\op X}H_{\op Z}^{\op{tr}}=0.$ Here, the rows of $H_{\op X}$ and $H_{\op Z}$ correspond to the stabilizer generators $S_i$ that are products of $\op X$ and $\id$ and $\op Z$ and $\id$, respectively.

The matrices $H_{\op X}$ and $H_{\op Z}$ with $H_{\op X}H_{\op Z}^{\op{tr}}=0$ can be viewed from a homological perspective.
Namely, a chain complex $C^\bullet$ of  $\ftwo$-vector spaces equipped with bases
% https://q.uiver.app/#q=WzAsNSxbMSwwLCJDXntpLTF9Il0sWzIsMCwiQ157aX0iXSxbMywwLCJDXntpKzF9Il0sWzAsMCwiXFxjZG90cyJdLFs0LDAsIlxcY2RvdHMiXSxbMCwxLCJkXntpLTF9Il0sWzEsMiwiZF5pIl0sWzMsMF0sWzIsNF1d
\[\begin{tikzcd}
 \cdots & {C^{i-1}} & {C^{i}} & {C^{i+1}} & \cdots
 \arrow[from=1-1, to=1-2]
 \arrow["{d^{i-1}}", from=1-2, to=1-3]
 \arrow["{d^i}", from=1-3, to=1-4]
 \arrow[from=1-4, to=1-5]
\end{tikzcd}\]
gives for each choice of index $i$ a CSS code by taking $H_{\op X}$ and $H_{\op Z}^{\op{tr}}$ as the matrices associated to $d^{i-1}$ and $d^{i}.$

This way, logical $\op X$-operators of the code $\mathcal{Q}$ correspond to the first cohomology group $H^i(C^\bullet)=\ker(d^i)/\im(d^{i-1})$ of the complex. The logical $\op Z$-operators can be identified with the dual space $H^i(C^\bullet)^*.$

Treating CSS codes as chain complexes allows us to import the language of homological algebra. For example, we will compare the spaces of logicals of various CSS codes via chain maps of the associated chain complexes. See for example~\cite{Cowtan_2024}, where several concepts from homological algebra are being used to study CSS codes.

\section{Logical operators and derived automorphisms of tile codes}\label{sec:stabtileson2dplane}

In this section, we will study tile codes and, in particular, the structure of logical operators of tile codes.

\subsection{Stabilizer tiles on the 2D plane}\label{subsec: background: stabilizertileson2Dplane}

Consider the infinite 2D plane $\mathbb{Z}^2$:

\begin{equation}\label{eq: 2dplane}
    \begin{tikzpicture}[scale=0.4, baseline={(current bounding box.center)}]
        % Grid lines
        \foreach \x in {-2,...,2} {
            \draw[\latticecolor, \latticethickness] (-2.5,\x) -- (2.5,\x);
            \draw[\latticecolor, \latticethickness] (\x,-2.5) -- (\x,2.5);
        }
        
        % Dots at intersections36+
        \foreach \x in {-2,...,2}
            \foreach \y in {-2,...,2}
                \fill[black] (\x,\y) circle (0.15);
                
    \end{tikzpicture}
\end{equation}
On each edge of the lattice sits a qubit. Each vertex of the lattice hosts one $\op X$-type and one $\op Z$-type \textit{stabilizer tile} each in a translationally invariant fashion. Here, a $\op X$-type (resp. $\op Z$-type) stabilizer tile is a $\op X$-type (resp. $\op Z$-type) Pauli operator whose support is contained in a box of size $(D+1) \times (D+1)$ of which the south-west corner is the vertex at which the stabilizer tile is placed. We require that all $\op X$-type tiles and all $\op Z$-type tiles commute. An example of a valid pair of tiles with $D=2$ is the following: 

\begin{equation}\label{eq:stabilizertiles}
    \begin{tikzpicture}[scale=0.5, , baseline={(current bounding box.center)}]
        % Define horizontal and vertical qubit coordinates
        \def\horizontalqubits{(1,0),(2,0), (0,2)}
        \def\verticalqubits{(0,0),(2,1),(2,2)}

        \def\boxsize{3}
        
        % Left figure - X-type stabilizer
        \begin{scope}[xshift=0cm]
            % Grid lines
            \draw[\latticecolor, \latticethickness] (3,0) -- (0,0) -- (0,3) ;
            \draw[\latticecolor, \latticethickness] (1,0) -- (1,3);
            \draw[\latticecolor, \latticethickness] (2,0) -- (2,3);
            \draw[\latticecolor, \latticethickness] (0,1) -- (3,1);
            \draw[\latticecolor, \latticethickness] (0,2) -- (3,2);
            \fill[color = good_red, opacity = .0] (0,0) -- (3,0) -- (3,3) -- (0,3) -- cycle;
            % Red stabilizer lines
            \foreach \pos in \horizontalqubits {
                \draw[good_red, line width = \supportthickness] \pos -- ++(1,0);
            }
            \foreach \pos in \verticalqubits {
                \draw[good_red, line width = \supportthickness] \pos -- ++(0,1);
            }
            
            % Black dot in corner
            \draw[fill=good_red, color = good_red] (0,0) circle (0.15);
        \end{scope}

        \def\horizontalqubitsz{(0,0),(0,1), (2,2)}
        \def\verticalqubitsz{(2,0),(0,2),(1,2)}
        % Right figure - Z-type stabilizer
        \begin{scope}[xshift=5cm]
            % Grid lines
            \draw[\latticecolor, \latticethickness] (3,0) -- (0,0) -- (0,3) ;
            \draw[\latticecolor, \latticethickness] (1,0) -- (1,3);
            \draw[\latticecolor, \latticethickness] (2,0) -- (2,3);
            \draw[\latticecolor, \latticethickness] (0,1) -- (3,1);
            \draw[\latticecolor, \latticethickness] (0,2) -- (3,2);
            
            % Blue stabilizer lines
            \foreach \pos in \horizontalqubitsz {
                \draw[good_blue, line width = \supportthickness] \pos -- ++(1,0);
            }
            \foreach \pos in \verticalqubitsz {
                \draw[good_blue, line width = \supportthickness] \pos -- ++(0,1);
            }
            
            % Black dot in corner
            \draw[fill=good_blue, color = good_blue] (0,0) circle (0.15);
        \end{scope}
    \end{tikzpicture}
\end{equation}

We will always assume that $(D+1)$ is the minimal possible box size in which the stabilizer tile fits. More precisely, we will assume that the stabilizer tile always has support on at least one of the dotted edges in each of the following: 
\begin{equation}\label{eq:stabilizertilecorners}
    \begin{tikzpicture}[scale=0.5, baseline={(current bounding box.center)}]
        % Define horizontal and vertical qubit coordinates
        \def\horizontalqubits{(1,0),(2,0), (0,2)}
        \def\verticalqubits{(0,0),(2,1),(2,2)}

        \def\boxsize{3}
        
        % Repeat the figure four times next to each other with even more gap
        \foreach \i in {0,1,2,3} {
            \begin{scope}[xshift=\i*6cm]
                % First plot (bottom-right corner dotted)
                \ifnum\i=0
                    % Draw all edges except the dotted ones
                    \draw[\latticecolor, \latticethickness] (0,0) -- (0,3);
                    \draw[\latticecolor, \latticethickness] (0,0) -- (2,0);
                    \draw[\latticecolor, \latticethickness] (1,0) -- (1,3);
                    \draw[\latticecolor, \latticethickness] (0,1) -- (3,1);
                    \draw[\latticecolor, \latticethickness] (0,2) -- (3,2);
                    \draw[\latticecolor, \latticethickness] (2,1) -- (2,3);
                    
                    % Draw the dotted edges
                    \draw[\latticecolor, \latticethickness, dotted] (2,0) -- (3,0);
                    \draw[\latticecolor, \latticethickness, dotted] (2,0) -- (2,1);
                \fi
                
                % Second plot (bottom-left corner dotted)
                \ifnum\i=1
                    % Draw all edges except the dotted ones
                    \draw[\latticecolor, \latticethickness] (1,0) -- (3,0);
                    \draw[\latticecolor, \latticethickness] (1,0) -- (1,3);
                    \draw[\latticecolor, \latticethickness] (2,0) -- (2,3);
                    \draw[\latticecolor, \latticethickness] (0,1) -- (3,1);
                    \draw[\latticecolor, \latticethickness] (0,2) -- (3,2);
                    \draw[\latticecolor, \latticethickness] (0,1) -- (0,3);
                    
                    % Draw the dotted edges
                    \draw[\latticecolor, \latticethickness, dotted] (0,0) -- (1,0);
                    \draw[\latticecolor, \latticethickness, dotted] (0,0) -- (0,1);
                \fi
                
                % Third plot (top-left corner dotted)
                \ifnum\i=2
                    % Draw all edges except the dotted ones
                    \draw[\latticecolor, \latticethickness] (0,0) -- (3,0);
                    \draw[\latticecolor, \latticethickness] (0,0) -- (0,2);
                    \draw[\latticecolor, \latticethickness] (1,0) -- (1,3);
                    \draw[\latticecolor, \latticethickness] (2,0) -- (2,3);
                    \draw[\latticecolor, \latticethickness] (0,1) -- (3,1);
                    \draw[\latticecolor, \latticethickness] (1,2) -- (3,2);
                    
                    % Draw the dotted edges
                    \draw[\latticecolor, \latticethickness, dotted] (0,2) -- (0,3);
                    \draw[\latticecolor, \latticethickness, dotted] (0,2) -- (1,2);
                \fi
                
                % Fourth plot (top-right corner dotted)
                \ifnum\i=3
                    % Draw all edges except the dotted ones
                    \draw[\latticecolor, \latticethickness] (0,0) -- (3,0);
                    \draw[\latticecolor, \latticethickness] (0,0) -- (0,2);
                    \draw[\latticecolor, \latticethickness] (2,0) -- (2,2);
                    \draw[\latticecolor, \latticethickness] (0,1) -- (3,1);
                    % \draw[\latticecolor, \latticethickness] (1,2) -- (3,2);
                    \draw[\latticecolor, \latticethickness] (0,0) -- (3,0);
                    \draw[\latticecolor, \latticethickness] (0,0) -- (0,3);
                    \draw[\latticecolor, \latticethickness] (1,0) -- (1,3);
                    \draw[\latticecolor, \latticethickness] (0,2) -- (2,2);

                    % Draw the dotted edges
                    \draw[\latticecolor, \latticethickness, dotted] (2,2) -- (2,3);
                    \draw[\latticecolor, \latticethickness, dotted] (2,2) -- (3,2);
                \fi
                
                % Black dot in corner
                \draw[fill=black] (0,0) circle (0.15);
            \end{scope}
        }
    \end{tikzpicture}
\end{equation}

It will be handy to label nodes, horizontal edges, and vertical edges of the lattice $\mathbb{Z}^2$ by monomials in the Laurent polynomial ring $R_{\pm,\pm} = \mathbb{F}_2 [x^\pm , y^\pm ]$. In particular, we label the node $(a,b)$, the horizontal edge connecting $(a,b)$ with $(a+1,b)$ and the vertical edge connecting $(a,b)$ with $(a,b+1)$ by $x^ay^b$ for any $a,b \in \mathbb{Z}$. In that way, any collection of $\op X$-tiles (resp. $\op Z$-tiles) can be specified by an element in $R_{\pm, \pm}$. Any finitely supported $\op X$-type (resp. $\op Z$-type) Pauli operator can be specified by a tuple $P = (P_h,P_v) \in R_{\pm,\pm}^2$.

So, any choice of $\op X$-stabilizer tile as in \Cref{eq:stabilizertiles} corresponds to a pair of Laurent polynomials $f,g$. In this context, \Cref{eq:stabilizertilecorners} means that at least one of the polynomials has $x$-degree $D$ and at least one of the two polynomials has $y$-degree $D$.

We will say that a Pauli $\op Z$-operator $P = (P_h, P_v)$ is \textit{supported on a strip of length $l$} (resp. \textit{on a strip of height $m$}) if there are no $(a,b), (\tilde{a}, \tilde{b}) \in P_h \cup P_v$ such that $|\tilde{a} - a|> l$ (resp. $|\tilde{b} - b|>m$). For example, the condition in \Cref{eq:stabilizertilecorners} ensures that the stabilizer tiles are not contained in strips of length or height $D$. More generally, the following simple observation will be helpful:

\begin{lemma}\label{lem: nostabsinsmallbox}
    A non-empty product of stabilizers satisfying \Cref{eq:stabilizertilecorners} is never supported within a strip of height or width at most  $D$. 
\end{lemma}

We say that the choice of stabilizer tiles gives rise to \textit{topological order} (TO) if any finitely supported Pauli operator that commutes with all stabilizer tiles must itself be a product of stabilizer tiles\footnote{The choice of tiles gives rise to a translationally invariant Pauli Hamiltonian with commuting terms on the 2D plane
\begin{equation}
    H = -\sum_{x \in \mathbb{Z}^2} S_{X,x} - \sum_{z \in \mathbb{Z}^2} S_{Z,z}.
\end{equation}
Here, $S_{X,x}$ and $S_{Z,z}$ are the $\op X$-type and $\op Z$-type stabilizer tiles at the vertex $x$ and $z$ respectively.
Our notion of TO means that the ground space of this Hamiltonian exhibits TO. }. From now on, we will always assume TO.

Later, it will be helpful to define a stronger version of topological order, which we call \textit{total topological order}. Assuming that a pair of stabilizer tiles is confined in boxes of size $(D+1) \times (D+1)$, the stabilizer tiles in the four quadrants all commute:

\begin{equation*}
    \begin{tikzpicture}[scale = .25]
    \begin{scope}
      \foreach \y in {-5,...,0} {
            \draw[\latticecolor, \latticethickness] (-5.5,\y) -- (2,\y);
        }
         \foreach \x in {-5,...,1} {
            \draw[\latticecolor, \latticethickness] (\x,-5) -- (\x,0.5);
        }
        \foreach \x in {-5,...,-1} {
        \foreach \y in {-5,...,0} {
        \draw[fill = black] (\x, \y) circle (.15);
        }
        }
        \foreach \x in {-5,...,-1} {
        \foreach \y in {-7,-6} {
        \draw[color = good_red,fill = good_red] (\x, \y) circle (.15);
        }
        }
        \foreach \x in {0,1} {
        \foreach \y in {-5,...,0} {
        \draw[color = good_blue, fill = good_blue] (\x, \y) circle (.15);
        }
        }
        \node at (-6.5, -2.5) {$\dots$};
        \node at (-2.5, 3) {$\vdots$};
    \end{scope}

    \begin{scope}[xshift = 15cm]
      \foreach \y in {-5,...,0} {
            \draw[\latticecolor, \latticethickness] (-5,\y) -- (-0.5,\y);
        }
         \foreach \x in {-5,...,-1} {
            \draw[\latticecolor, \latticethickness] (\x,-5) -- (\x,0.5);
        }
        \foreach \x in {-5,...,-1} {
        \foreach \y in {-5,...,0} {
        \draw[fill = black] (\x, \y) circle (.15);
        }
        }
        \foreach \x in {-5,...,-1} {
        \foreach \y in {-7,-6} {
        \draw[color = good_red,fill = good_red] (\x, \y) circle (.15);
        }
        }
        \foreach \x in {-6,-7} {
        \foreach \y in {-5,...,0} {
        \draw[color = good_blue, fill = good_blue] (\x, \y) circle (.15);
        }
        }
        \node at (1.5, -2.5) {$\dots$};
        \node at (-2.5, 3) {$\vdots$};
    \end{scope}

    \begin{scope}[yshift = -12cm]
      \foreach \y in {-5,...,0} {
            \draw[\latticecolor, \latticethickness] (-5.5,\y) -- (2,\y);
        }
         \foreach \x in {-5,...,1} {
            \draw[\latticecolor, \latticethickness] (\x,-5.5) -- (\x,1);
        }
        \foreach \x in {-5,...,-1} {
        \foreach \y in {-5,...,-2} {
        \draw[fill = black] (\x, \y) circle (.15);
        }
        }
        \foreach \x in {-5,...,-1} {
        \foreach \y in {-1,0} {
        \draw[color = good_red,fill = good_red] (\x, \y) circle (.15);
        }
        }
        \foreach \x in {0,1} {
        \foreach \y in {-5,...,-2} {
        \draw[color = good_blue, fill = good_blue] (\x, \y) circle (.15);
        }
        }
        \node at (-6.5, -2.5) {$\dots$};
        \node at (-2.5, -7) {$\vdots$};
    \end{scope}

    \begin{scope}[xshift = 15cm, yshift = -12cm]
      \foreach \y in {-5,...,0} {
            \draw[\latticecolor, \latticethickness] (-5,\y) -- (-0.5,\y);
        }
         \foreach \x in {-5,...,-1} {
            \draw[\latticecolor, \latticethickness] (\x,-5.5) -- (\x,1);
        }
        \foreach \x in {-5,...,-1} {
        \foreach \y in {-5,...,-2} {
        \draw[fill = black] (\x, \y) circle (.15);
        }
        }
        \foreach \x in {-5,...,-1} {
        \foreach \y in {-1,0} {
        \draw[color = good_red,fill = good_red] (\x, \y) circle (.15);
        }
        }
        \foreach \x in {-7,-6} {
        \foreach \y in {-5,...,-2} {
        \draw[color = good_blue, fill = good_blue] (\x, \y) circle (.15);
        }
        }
        \node at (1.5, -2.5) {$\dots$};
        \node at (-2.5, -7) {$\vdots$};
    \end{scope}
    \end{tikzpicture}  
\end{equation*}

Here, we draw an edge for each qubit we include in the quadrant. A black dot indicates that the quadrant contains both an $\op X$- and a $\op Z$-type stabilizer tile at this position. Red dots (resp. blue dots) indicate that we only include an $\op X$-type (resp. $\op Z$-type) stabilizer at this position. Note that the supports of the red and blue stabilizer tiles are not entirely contained in the sublattice of qubits. In these cases, we will restrict the support of the tiles accordingly.

We say that the stabilizer tiles give rise to \textit{total topological order} if, for each of the four quadrants depicted above, whenever a finitely supported Pauli operator commutes with all included stabilizers, then it must itself be a product of included stabilizers.

It will be useful to study excitations of the stabilizer tiles. We will only focus on excitations of $\op Z$-type stabilizer tiles; treating excitations of $\op X$-type stabilizer tiles works analogously. For a finitely supported $\op X$-type Pauli operator $P$, we call the element specifying the $\op Z$-type stabilizer tiles anticommuting with it the \textit{syndrome} associated with $P$. We will write  
\begin{equation}\label{eq: excitation map}
  \partial : R_{\pm,\pm} \oplus R_{\pm,\pm} \rightarrow R_{\pm,\pm}  
\end{equation}
for the map associating a syndrome $\partial(P)$ to a Pauli operator $P$. This map is often referred to as the \textit{excitation map}. Then, TO means that if $\partial(P) = 0$, then $P$ is a product of finitely many $\op X$-type stabilizer tiles.

\subsection{Tile codes}\label{subsec: tile codes}

Tile codes \cite{eberhardt2024pruningqldpccodesbivariate, steffan2025tilecodeshighefficiencyquantum,liang2025planarquantumlowdensityparitycheck} are a systematic way of turning stabilizer tiles on the 2D plane $\mathbb{Z}^2$ into finite-size CSS codes. A tile code consists of a finite-size subset $\mathcal{T} \subset \mathbb{Z}^2$ of qubits and finite subsets of $\op X$-type and $\op Z$-type tiles such that all the stabilizer tiles we include, restricted to~$\mathcal{T}$, commute. In this way, a tile code is a well-defined CSS code. 

If the support of a tile is entirely contained in $\mathcal{T}$, we call it a \textit{bulk stabilizer tile}. Else, we refer to it as a \textit{boundary stabilizer tile}. We will refer to the stabilizer tiles whose support overlaps with $\mathcal{T}$ but which are not included in the stabilizer set as \textit{omitted boundary stabilizer tiles}.

We will now recap the construction consisting of four steps from \cite{steffan2025tilecodeshighefficiencyquantum}. For that, choose a valid pair of stabilizer tiles confined in boxes of size $(D+1)\times (D+1)$.

In \textbf{Step 1}, we choose a grid of bulk stabilizers of size $(L-D) \times (M-D)$, for example:

\begin{equation*}\label{eq: bulkstablattices}
  \begin{tikzpicture}[scale = .5]
          \foreach \x in {0,...,3} {
                \foreach \y in {0,...,3} {
                    \draw[fill=black] (\x/2,\y/2) circle (0.1);
                }
            }
  \end{tikzpicture}
\end{equation*}

In \textbf{Step 2}, we draw the layout of qubits, which we choose to be the union of all $(D+1)\times (D+1)$ boxes that are centered at the vertices chosen in Step 1. This will result in a grid of $2LM$ many physical qubits.

\begin{equation*}
    \begin{tikzpicture}[scale = .5]
                    % Vertical lines
            \foreach \x in {0,...,5} {
                \draw[\latticecolor, \latticethickness] (\x/2,0) -- (\x/2,3);
            }
            
            % Horizontal lines
            \foreach \y in {0,...,5} {
                \draw[\latticecolor, \latticethickness] (0,\y/2) -- (3,\y/2);
            }

            \foreach \x in {0,...,3} {
                \foreach \y in {0,...,3} {
                    \draw[fill=black] (\x/2,\y/2) circle (0.1);
                }
            }
    \end{tikzpicture}
\end{equation*}

In \textbf{Step 3}, we add boundary stabilizers. On the north and south side of the grid, we draw $D$ layers of $\op X$-type gauge stabilizer tiles. On the east and west side of the grid, we draw $D$ layers of $\op Z$-type gauge stabilizer tiles. Note that by the choice of qubit lattice, all included stabilizer tiles commute. 

\begin{equation*}
    \begin{tikzpicture}[scale = .5]
        \draw[decorate,decoration={brace},thick] (-0.2,2) -- (-0.2,3) node[midway,left=0.1cm]{$D$};
            \foreach \x in {0,...,5} {
                \draw[\latticecolor, \latticethickness] (\x/2,0) -- (\x/2,3);
            }
            
            % Horizontal lines
            \foreach \y in {0,...,5} {
                \draw[\latticecolor, \latticethickness] (0,\y/2) -- (3,\y/2);
            }
            
            % 3x3 grid of black dots
            \foreach \x in {0,...,3} {
                \foreach \y in {0,...,3} {
                    \draw[fill=black] (\x/2,\y/2) circle (0.1);
                }
            }
            
            % Red dots on top and bottom
            \foreach \x in {0,...,3} {
                \draw[fill=good_red, color = good_red] (\x/2,-0.5) circle (0.1);
                \draw[fill=good_red, color = good_red] (\x/2,-1) circle (0.1);
                \draw[fill=good_red, color = good_red] (\x/2,2) circle (0.1);
                \draw[fill=good_red, color = good_red] (\x/2,2.5) circle (0.1);
            }
            
            % Blue dots on left and right
            \foreach \y in {0,...,3} {
                \draw[fill=good_blue, color = good_blue] (-0.5,\y/2) circle (0.1);
                \draw[fill=good_blue, color = good_blue] (-1,\y/2) circle (0.1);
                \draw[fill=good_blue, color = good_blue] (2,\y/2) circle (0.1);
                \draw[fill=good_blue, color = good_blue] (2.5,\y/2) circle (0.1);
            }
    \end{tikzpicture}
\end{equation*}

In \textbf{Step 4}, we delete qubits that are not supported by any $\op X$-type stabilizer (resp. $\op Z$-type stabilizer). If the support of a stabilizer is empty, we also delete that one. Finally, we add \textit{corner stabilizers} if they commute with all so-far chosen stabilizers. 

The best tile codes found so far in \cite{steffan2025tilecodeshighefficiencyquantum} were all constructed from tiles that satisfy \Cref{eq:stabilizertilecorners} and give rise to total topological order. They also came from constructions that terminated after the third step which turns out not to be a coincidence: 

\begin{lemma}
    Say our stabilizer tiles satisfy condition \Cref{eq:stabilizertilecorners} and give rise to total topological order. Then, the construction terminates after Step 3.
\end{lemma}

\begin{proof}
    One easily checks that there are no qubits that are not supported by any $\op X$-type stabilizer (resp. $\op Z$-type stabilizer) by condition \Cref{eq:stabilizertilecorners}. It is also clear that \Cref{eq:stabilizertilecorners} implies that all stabilizers have support on the lattice chosen in Step 2. Finally, total topological order implies that no corner term is commuting with all stabilizers. Hence, the tile code stays unchanged in Step 4. 
\end{proof}

Note that by \Cref{eq:stabilizertilecorners}, the stabilizer generators of a tile code are independent. Since there are $2D^2$ more physical qubits than stabilizer generators, the logical dimension is  $2D^2$. One can also say something about the logical dimension in the case where \Cref{eq:stabilizertilecorners} is not satisfied. This, however, is more involved, we touch upon this topic in \Cref{sec:algebraoftilecodes}.

\subsection{Logical operators of tile codes}\label{subsec: logicalopsstruc: operatorsonastrip}

In this section we will show that the tile codes admit a choice of basis of logical operators with many beneficial properties. We will first present our main findings and then give intuitive and visual, yet rigorous, proofs. For a visualization of \Cref{thm: logicalops} we refer to \Cref{fig:main_figure}. 

\begin{theorem}\label{thm: logicalops}
    Consider a tile code constructed from stabilizer tiles confined in $(D+1) \times (D+1)$ boxes that satisfy \Cref{eq:stabilizertilecorners} and give rise to total topological order. Then, the following statements hold:
    \begin{enumerate}
        \item The tile code has a symplectic basis of logical operators labelled by the physical qubits in a $D \times D$ box in the bottom left corner. For each physical qubit in this box, exactly one pair of logical $\bar{\op X}$- and $\bar{\op Z}$-operators is supported there.
        \item The logical operators of this symplectic basis can be chosen to live in strips of width (resp. height) $D$ on the left (resp. bottom) of the tile code. 
        \item Within these strips, the representation of the logical operators is unique and can be constructed by a cellular automaton with a set of $2D^2$ rules. 
        \item All of these logical operators are products of omitted boundary stabilizers. 
    \end{enumerate}
\end{theorem}

We will now, without loss of generality, study $\op X$-type operators. For $\op Z$-type operators, everything works analogously. Draw a box of size $D \times D$ on the lattice $\mathbb{Z}^2$. 

\begin{equation*}
    \begin{tikzpicture}[scale=0.5]
        % Grid lines
        \foreach \x in {-3,...,3} {
            \draw[\latticecolor, \latticethickness] (-3.5,\x) -- (3.5,\x);
            \draw[\latticecolor, \latticethickness] (\x,-3.5) -- (\x,3.5);
        }
        \draw[good_yellow, line width = \supportthickness] (-1,0) -- (1,0); 
        \draw[good_yellow, line width = \supportthickness] (-1,1) -- (1,1); 
        \draw[good_yellow, line width = \supportthickness] (-1,0) -- (-1,2);
        \draw[good_yellow, line width = \supportthickness] (0,0) -- (0,2);
    \end{tikzpicture}
\end{equation*}

Any $\op X$-type Pauli operator whose support is contained in this box of size $D \times D$ can only anticommute with the $\op Z$-stabilizer tiles within the following region of width and height $2D$. 

\begin{equation*}
    \begin{tikzpicture}[scale=0.5]
        % Grid lines
        \foreach \x in {-3,...,3} {
            \draw[\latticecolor, \latticethickness] (-3.5,\x) -- (3.5,\x);
            \draw[\latticecolor, \latticethickness] (\x,-3.5) -- (\x,3.5);
        }
        \draw[good_yellow, line width = \supportthickness] (-1,0) -- (1,0); 
        \draw[good_yellow, line width = \supportthickness] (-1,1) -- (1,1); 
        \draw[good_yellow, line width = \supportthickness] (-1,0) -- (-1,2);
        \draw[good_yellow, line width = \supportthickness] (0,0) -- (0,2);

        \foreach \x in {-3, -2, -1, 0} {
        \foreach \y in {-2, -1, 0, 1} {
        \draw[color = good_blue, fill = good_blue] (\x, \y) circle (.15);
        }
        }
    \end{tikzpicture}
\end{equation*}

We split this set of $\op Z$-stabilizers into two regions $R$ and $R'$, where $R$ is the union of the $D$ rows on the bottom and $R'$ is the union of the $D$ rows on the top.

\begin{equation*}
    \begin{tikzpicture}[scale=0.5]
        % Grid lines
        \foreach \x in {-3,...,3} {
            \draw[\latticecolor, \latticethickness] (-3.5,\x) -- (3.5,\x);
            \draw[\latticecolor, \latticethickness] (\x,-3.5) -- (\x,3.5);
        }
        \draw[good_yellow, line width = \supportthickness] (-1,0) -- (1,0); 
        \draw[good_yellow, line width = \supportthickness] (-1,1) -- (1,1); 
        \draw[good_yellow, line width = \supportthickness] (-1,0) -- (-1,2);
        \draw[good_yellow, line width = \supportthickness] (0,0) -- (0,2);

        \foreach \x in {-3, -2, -1, 0} {
        \foreach \y in {-2, -1, 0, 1} {
        \draw[color = good_blue, fill = good_blue] (\x, \y) circle (.15);
        }
        }
        \draw[rounded corners, thick] (-3.5, -2.5) -- (-3.5, -.6) -- (.5, -.6) -- (.5, -2.5) -- cycle;

        \draw[rounded corners, thick] (-3.5, -.4) -- (-3.5, 1.5) -- (0.5, 1.5) -- (0.5, -.4) -- cycle;

        \node at (-1.5, -1.5) {$R$};
        \node at (-1.5, .5) {$R'$};
    \end{tikzpicture}
\end{equation*}

\begin{lemma}
        Assuming total topological order, any $\op X$-type operator in the yellow box anticommutes with at least one stabilizer in $R$ and at least one stabilizer in $R'$.
\end{lemma}
\begin{proof}
    Place the yellow box on the bottom left corner of the north-east quadrant. By condition \Cref{eq:stabilizertilecorners} and \Cref{lem: nostabsinsmallbox}, no product of stabilizers of this quadrant is supported in this box. Hence, by total topological order, any $\op X$-type Pauli operator supported in the box must anticommute with some $\op Z$-type stabilizers of the quadrant. Since the stabilizers in $R$ are discarded in this quadrant, there must be at least one stabilizer in $R'$ anticommuting with the operator. Doing the same argument, putting the yellow box on the top right corner of the south-west quadrant, shows that there must be at least one $\op Z$-type stabilizer in $R$ with which it anticommutes. 
\end{proof}

\begin{corollary}\label{cor: independent syndromes}
    The $2D^2$ single qubit Pauli $\op X$-operators in the box give rise to linearly independent syndromes in both $R$ and $R'$. 
\end{corollary}
\begin{proof}
    If there would be a linear dependency then some non-trivial Pauli $\op X$-operator in the box does not trigger a syndrome. 
\end{proof}

\begin{corollary}
    Any syndrome in $R$ (and any syndrome in $R'$) can be triggered by a Pauli $\op X$-operator in the $D \times D$-sized box. 
\end{corollary}
\begin{proof}
    The space of syndromes in $R$ (resp. in $R'$) is $2D^2$ dimensional. Since the $2D^2$ different single qubit Pauli-$\op X$ in the box trigger linearly independent syndromes they must span the whole space. 
\end{proof}

In fact, we know a little bit more than that. Consider only the bottom row of horizontal and vertical qubits of the yellow box:

\begin{equation*}
    \begin{tikzpicture}[scale=0.5]
        % Grid lines
        \foreach \x in {-3,...,3} {
            \draw[\latticecolor, \latticethickness] (-3.5,\x) -- (3.5,\x);
            \draw[\latticecolor, \latticethickness] (\x,-3.5) -- (\x,3.5);
        }
        \draw[good_yellow, line width = \supportthickness] (-1,0) -- (1,0); 
        \draw[good_yellow, line width = \supportthickness] (-1,0) -- (-1,1);
        \draw[good_yellow, line width = \supportthickness] (0,0) -- (0,1);

        \foreach \x in {-3, -2, -1, 0} {
        \foreach \y in {-2, -1, 0, 1} {
        \draw[color = good_blue, fill = good_blue] (\x, \y) circle (.15);
        }
        }
        \draw[rounded corners, thick] (-3.5, -2.5) -- (-3.5, -.6) -- (.5, -.6) -- (.5, -2.5) -- cycle;

        \draw[rounded corners, thick] (-3.5, -.4) -- (-3.5, 1.5) -- (0.5, 1.5) -- (0.5, -.4) -- cycle;

\draw[rounded corners, thick, good_blue] (-3.3, -2.3) -- (-3.3, -1.7) -- (.3, -1.7) -- (.3, -2.3) -- cycle;

        \node at (-1.5, -1.5) {$R$};
        \node at (-1.5, .5) {$R'$};
    \end{tikzpicture}
\end{equation*}

Then, the Pauli $\op X$-type operators on these $2D$ qubits yield linearly independent syndromes on the lowest row of $\op Z$-type stabilizers in $R$.

\begin{corollary}\label{cor: uniqueextension}
    For all Pauli $\op X$-operators in the $D \times D $ box on the bottom left of a tile code there is a unique extension to a logical operator in a strip of width $D$. 
\end{corollary}

\begin{proof}
    Pick any Pauli-$\op X$ in the box on the bottom left.

\begin{equation*}
    \begin{tikzpicture}[scale=0.5]
        % Grid lines
        \foreach \x in {0, 1, 2, 3} {
            \draw[\latticecolor, \latticethickness] (-1,\x) -- (3.5,\x);
        }
        \foreach \x in {-1, 0, 1, 2, 3} {
            \draw[\latticecolor, \latticethickness] (\x,0) -- (\x,3.5);
        }
        \draw[good_yellow, line width = \supportthickness] (-1,0) -- (1,0); 
        \draw[good_yellow, line width = \supportthickness] (-1,1) -- (1,1); 
        \draw[good_yellow, line width = \supportthickness] (-1,0) -- (-1,2);
        \draw[good_yellow, line width = \supportthickness] (0,0) -- (0,2);

        \foreach \x in {-3,-2,-1, 0} {
        \foreach \y in { 0, 1} {
        \draw[color = good_blue, fill = good_blue] (\x, \y) circle (.15);
        }
        }

        \draw[rounded corners, thick] (-3.4, -.5) -- (-3.4, 1.5) -- (0.5, 1.5) -- (0.5, -.5) -- cycle;

        \node at (-2.5, .5) {$R$};
    \end{tikzpicture}
\end{equation*}
    We have seen that this Pauli $\op X$-operator may anticommute with some $\op Z$-type tile code stabilizers in the region $R$. But, we have also seen that there is a unique $\op Z$-operator in the following region highlighted in yellow triggering the same syndrome in the bottom row of $R$.

\begin{equation*}
    \begin{tikzpicture}[scale=0.5]
        % Grid lines
        \foreach \x in {0, 1, 2, 3} {
            \draw[\latticecolor, \latticethickness] (-1,\x) -- (3.5,\x);
        }
        \foreach \x in {-1, 0, 1, 2, 3} {
            \draw[\latticecolor, \latticethickness] (\x,0) -- (\x,3.5);
        }
        \draw[good_yellow, line width = \supportthickness] (-1,2) -- (-1,3); 
        \draw[good_yellow, line width = \supportthickness] (-1,2) -- (1,2); 
        \draw[good_yellow, line width = \supportthickness] (0,2) -- (0,3);

        \foreach \x in {-3, -2, -1, 0} {
        \foreach \y in {0,1} {
        \draw[color = good_blue, fill = good_blue] (\x, \y) circle (.15);
        }
        }

        \draw[rounded corners, thick] (-3.5, -.5) -- (-3.5, 1.5) -- (0.4, 1.5) -- (0.4, -.5) -- cycle;

        \draw[good_blue, rounded corners, thick] (-3.3, -.3) -- (-3.3, .3) -- (.3, .3) -- (.3, -.3) -- cycle;

        \node at (-2.5, .7) {$R$};
    \end{tikzpicture}
\end{equation*}

 Extending the $\op X$-operator in the corner with this there is no syndrome in the bottom row of $R$ left. Recursively, we find that there is a unique logical operator extending the Pauli $\op X$-operator on a strip such that there are no $\op Z$-checks violated. 
\end{proof}

We emphasize that this extension is unique not only up to stabilizers - there is only one logical operator representative in the strip continuing the pattern in the bottom left box.  The proof of \Cref{cor: uniqueextension} in particular shows the following. 

\begin{corollary}\label{cor: symplecticbasis}
    A tile code admits a symplectic basis of logical operators $\bar{\op X}_1, \bar{\op Z}_1, \dots , \bar{\op X}_{D}, \bar{\op Z}_{D}$ with the following properties:
    \begin{itemize}
        \item The pairs $\bar{\op X}_i, \bar{\op Z}_i$ can be labelled by the physical qubits in the $D \times D$ box on the bottom left of the tile code in the sense that they are supported on this qubit and all other logicals in the symplectic basis are not.
        \item All $\bar{\op X}_i$ are supported in a strip of width $D$ along the left boundary of the tile code. All $\bar{\op Z}_i$ are supported on a strip of height $D$ on the bottom of the tile code. 
    \end{itemize}
\end{corollary}

We depict an example of such a pair of logical operators in \Cref{fig:main_figure} (a). 

The proof of \Cref{cor: uniqueextension} shows that the elements of this symplectic basis arises from a cellular automaton. For more details on cellular automata we refer to~\cite{cellular_automaton}. We will now describe how to get the $2D^2$ rules that let one construct logical operators. Recall that any single qubit Pauli $\op X_i$ in the green box gives rise to excited $\op Z$-type stabilizer tiles in $R'$. Let $a(\op X_i) = \prod_j \op X_j$ be the unique Pauli operator that excites the same syndrome in $R'$, in formulas, $\partial (\op X_i)| _ {R'} = \partial (a(\op X_i)) | _{R'}$. 

\begin{proposition}
    The logical $\op X$- and $\op Z$-operators of the canonical symplectic basis in \Cref{cor: symplecticbasis} can be constructed using a cellular automaton consisting of $2D^2$ rules each. The rules can be inferred from the paragraph above. 
\end{proposition}

For the stabilizer tiles in \Cref{eq:stabilizertiles}, the 8 rules are as follows: 

\begin{equation}\label{eq: cellular automaton}
\raisebox{-0.5\height}{\includegraphics[width=0.3\textwidth]{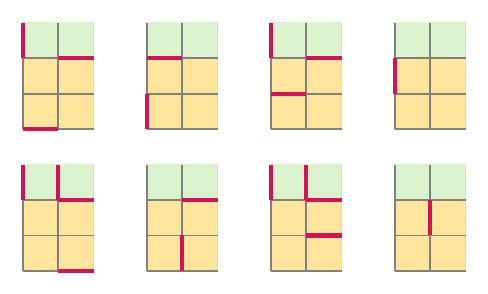}}
\end{equation}

Let us illustrate in an example how to construct a logical operator step-by-step using this cellular automaton:

\begin{equation*}
    \begin{tikzpicture}[scale = .6]
      
            \foreach \x in {0,...,1} {
                \draw[\latticecolor, \latticethickness] (\x/2,0) -- (\x/2,3);
            }
            
            % Horizontal lines
            \foreach \y in {0,...,5} {
                \draw[\latticecolor, \latticethickness] (0,\y/2) -- (1,\y/2);
            }
            \draw[color = good_red, line width = \supportthickness] (0,0) -- (.5,0);

            \foreach \x in {0,...,1} {
                \foreach \y in {0,...,3} {
                    \draw[fill=black] (\x/2,\y/2) circle (0.1);
                }
            }
            
            % Red dots on top and bottom
            \foreach \x in {0,...,1} {
                \draw[fill=good_red, color = good_red] (\x/2,-0.5) circle (0.1);
                \draw[fill=good_red, color = good_red] (\x/2,-1) circle (0.1);
                \draw[fill=good_red, color = good_red] (\x/2,2) circle (0.1);
                \draw[fill=good_red, color = good_red] (\x/2,2.5) circle (0.1);
            }
            
            % Blue dots on left and right
            \foreach \y in {0,...,3} {
                \draw[fill=good_blue, color = good_blue] (-0.5,\y/2) circle (0.1);

            }

            \node[scale = .8] at (1.8, 1) {$\dots$};

            \node at (3,1) {$\mapsto$};

            \begin{scope}[xshift = 4.5cm]
            \draw[good_yellow, fill=good_yellow, opacity=.4] (0,0) -- (1,0) -- (1,1) -- (0,1) -- cycle;
      \draw[cell_aut_good_green,  fill=cell_aut_good_green,  opacity=.4] (0,1)  -- (1,1) -- (1,1.5) -- (0,1.5)  -- cycle;
      
            \foreach \x in {0,...,1} {
                \draw[\latticecolor, \latticethickness] (\x/2,0) -- (\x/2,3);
            }
            
            % Horizontal lines
            \foreach \y in {0,...,5} {
                \draw[\latticecolor, \latticethickness] (0,\y/2) -- (1,\y/2);
            }
            \draw[color = good_red, line width = \supportthickness] (0,0) -- (.5,0);

            \draw[color = good_red, line width = \supportthickness] (0,1) -- (0,1.5);

            \draw[color = good_red, line width = \supportthickness] (0.5,1) -- (1,1);
                \foreach \x in {0,...,1} {
                \foreach \y in {0,...,3} {
                    \draw[fill=black] (\x/2,\y/2) circle (0.1);
                }
            }
            
            % Red dots on top and bottom
            \foreach \x in {0,...,1} {
                \draw[fill=good_red, color = good_red] (\x/2,-0.5) circle (0.1);
                \draw[fill=good_red, color = good_red] (\x/2,-1) circle (0.1);
                \draw[fill=good_red, color = good_red] (\x/2,2) circle (0.1);
                \draw[fill=good_red, color = good_red] (\x/2,2.5) circle (0.1);
            }
            
            % Blue dots on left and right
            \foreach \y in {0,...,3} {
                \draw[fill=good_blue, color = good_blue] (-0.5,\y/2) circle (0.1);

            }

            \node[scale = .8] at (1.8, 1) {$\dots$};

            \node at (3,1) {$\mapsto$};
            \end{scope}

\begin{scope}[xshift = 9cm]
\draw[good_yellow, fill=good_yellow, opacity=.4] (0,0.5) -- (1,0.5) -- (1,1.5) -- (0,1.5) -- cycle;
      \draw[cell_aut_good_green,  fill=cell_aut_good_green,  opacity=.4] (0,1.5)  -- (1,1.5) -- (1,2) -- (0,2)  -- cycle;
    \foreach \x in {0,...,1} {
                \draw[\latticecolor, \latticethickness] (\x/2,0) -- (\x/2,3);
            }
            
            % Horizontal lines
            \foreach \y in {0,...,5} {
                \draw[\latticecolor, \latticethickness] (0,\y/2) -- (1,\y/2);
            }
            \draw[color = good_red, line width = \supportthickness] (0,0) -- (.5,0);

            \draw[color = good_red, line width = \supportthickness] (0,1) -- (0,1.5);

            \draw[color = good_red, line width = \supportthickness] (0,1.5) -- (0,2);

            \draw[color = good_red, line width = \supportthickness] (0.5,1) -- (1,1);
            
            \draw[color = good_red, line width = \supportthickness] (0.5,1.5) -- (1,1.5);

            \draw[color = good_red, line width = \supportthickness] (0.5,1.5) -- (0.5,2);

            % 3x3 grid of black dots
            \foreach \x in {0,...,1} {
                \foreach \y in {0,...,3} {
                    \draw[fill=black] (\x/2,\y/2) circle (0.1);
                }
            }
            
            % Red dots on top and bottom
            \foreach \x in {0,...,1} {
                \draw[fill=good_red, color = good_red] (\x/2,-0.5) circle (0.1);
                \draw[fill=good_red, color = good_red] (\x/2,-1) circle (0.1);
                \draw[fill=good_red, color = good_red] (\x/2,2) circle (0.1);
                \draw[fill=good_red, color = good_red] (\x/2,2.5) circle (0.1);
            }
            
            % Blue dots on left and right
            \foreach \y in {0,...,3} {
                \draw[fill=good_blue, color = good_blue] (-0.5,\y/2) circle (0.1);

            }

            \node[scale = .8] at (1.8, 1) {$\dots$};

            \node at (3,1) {$\mapsto$};
\end{scope}

\begin{scope}[xshift = 13.5cm]
\draw[good_yellow, fill=good_yellow, opacity=.4] (0,1) -- (1,1) -- (1,2) -- (0,2) -- cycle;
      \draw[cell_aut_good_green,  fill=cell_aut_good_green,  opacity=.4] (0,2)  -- (1,2) -- (1,2.5) -- (0,2.5)  -- cycle;
    \foreach \x in {0,...,1} {
                \draw[\latticecolor, \latticethickness] (\x/2,0) -- (\x/2,3);
            }
            
            % Horizontal lines
            \foreach \y in {0,...,5} {
                \draw[\latticecolor, \latticethickness] (0,\y/2) -- (1,\y/2);
            }
            \draw[color = good_red, line width = \supportthickness] (0,0) -- (.5,0);

            \draw[color = good_red, line width = \supportthickness] (0,1) -- (0,1.5);

            \draw[color = good_red, line width = \supportthickness] (0,1.5) -- (0,2);

            \draw[color = good_red, line width = \supportthickness] (0.5,1) -- (1,1);
            
            \draw[color = good_red, line width = \supportthickness] (0.5,1.5) -- (1,1.5);

            \draw[color = good_red, line width = \supportthickness] (0.5,1.5) -- (0.5,2);

            \draw[color = good_red, line width = \supportthickness] (0,2) -- (.5,2);

            % 3x3 grid of black dots
            \foreach \x in {0,...,1} {
                \foreach \y in {0,...,3} {
                    \draw[fill=black] (\x/2,\y/2) circle (0.1);
                }
            }
            
            % Red dots on top and bottom
            \foreach \x in {0,...,1} {
                \draw[fill=good_red, color = good_red] (\x/2,-0.5) circle (0.1);
                \draw[fill=good_red, color = good_red] (\x/2,-1) circle (0.1);
                \draw[fill=good_red, color = good_red] (\x/2,2) circle (0.1);
                \draw[fill=good_red, color = good_red] (\x/2,2.5) circle (0.1);
            }
            
            % Blue dots on left and right
            \foreach \y in {0,...,3} {
                \draw[fill=good_blue, color = good_blue] (-0.5,\y/2) circle (0.1);

            }

            \node[scale = .8] at (1.8, 1) {$\dots$};

            \node at (3,1) {$\mapsto$};
\end{scope}

\begin{scope}[xshift = 18cm]
\draw[good_yellow, fill=good_yellow, opacity=.4] (0,1.5) -- (1,1.5) -- (1,2.5) -- (0,2.5) -- cycle;
      \draw[cell_aut_good_green,  fill=cell_aut_good_green,  opacity=.4] (0,3)  -- (1,3) -- (1,2.5) -- (0,2.5)  -- cycle;
    \foreach \x in {0,...,1} {
                \draw[\latticecolor, \latticethickness] (\x/2,0) -- (\x/2,3);
            }
            
            % Horizontal lines
            \foreach \y in {0,...,5} {
                \draw[\latticecolor, \latticethickness] (0,\y/2) -- (1,\y/2);
            }
            \draw[color = good_red, line width = \supportthickness] (0,0) -- (.5,0);

            \draw[color = good_red, line width = \supportthickness] (0,1) -- (0,1.5);

            \draw[color = good_red, line width = \supportthickness] (0,1.5) -- (0,2);

            \draw[color = good_red, line width = \supportthickness] (0.5,1) -- (1,1);
            
            \draw[color = good_red, line width = \supportthickness] (0.5,1.5) -- (1,1.5);

            \draw[color = good_red, line width = \supportthickness] (0.5,1.5) -- (0.5,2);

            \draw[color = good_red, line width = \supportthickness] (0,2) -- (.5,2);

            \draw[color = good_red, line width = \supportthickness] (0,2.5) -- (.5,2.5);

            \draw[color = good_red, line width = \supportthickness] (0.5,2.5) -- (1,2.5);

            \draw[color = good_red, line width = \supportthickness] (0.5,2.5) -- (0.5,3);
            % 3x3 grid of black dots
            \foreach \x in {0,...,1} {
                \foreach \y in {0,...,3} {
                    \draw[fill=black] (\x/2,\y/2) circle (0.1);
                }
            }
            
            % Red dots on top and bottom
            \foreach \x in {0,...,1} {
                \draw[fill=good_red, color = good_red] (\x/2,-0.5) circle (0.1);
                \draw[fill=good_red, color = good_red] (\x/2,-1) circle (0.1);
                \draw[fill=good_red, color = good_red] (\x/2,2) circle (0.1);
                \draw[fill=good_red, color = good_red] (\x/2,2.5) circle (0.1);
            }
            
            % Blue dots on left and right
            \foreach \y in {0,...,3} {
                \draw[fill=good_blue, color = good_blue] (-0.5,\y/2) circle (0.1);

            }

            \node[scale = .8] at (1.8, 1) {$\dots$};

            % \node at (3,1) {$\mapsto$};
\end{scope}
            
            \end{tikzpicture}
\end{equation*}

Here, in all steps except the first one, one has to apply linear combinations of the rules in \Cref{eq: cellular automaton}. For example, in the second step, one has to apply rules four and five. We also visualize the growing of logical operators via cellular automata in \Cref{fig:main_figure}.

The fact that logical operators can be easily understood in terms of cellular automata will be a crucial component for understanding so-called \textit{derived automorphisms} in the next section.

We finish this section with some remarks regarding connections to other works. 

\begin{remark}
An alternative way of constructing a finite-size CSS code from translationally invariant Pauli stabilizer-tiles on the 2D plane is by introducing periodicity to construct so-called bivariate bicycle (BB) codes~\cite{bravyi2023highthreshold, linQuantumTwoblockGroup2023}. The logical dimension of a BB code of a certain size turns out to be determined by the period of this cellular automaton. 
\end{remark}

\begin{remark}
    For hypergraph product codes~\cite{Tillich_2014} a construction for a canonical basis of logical operators has been found in the past in~\cite{Quintavalle_2023}. In fact, it has been shown that hypergraph-product \textit{La-Cross codes}~\cite{Pecorari_2025,eberhardt2024pruningqldpccodesbivariate} are instances of tile codes where the stabilizers are induced by univariate polynomials. It turns out that for this special case these bases are the same when choosing the right pivot qubits in the method of \cite{Quintavalle_2023}.
\end{remark}

\begin{remark}
The concept of constructing logical operators via cellular automata is related to logical operators seen as fractal operators, see \cite{Haah2013} for more details on fractal operators.
\end{remark}

The logical operators admit one more feature. We mention that this statement holds even when only assuming topological order and \emph{total} topological order is not necessary. A statement similar to this was used in~\cite{yang2025planarfaulttolerantquantumcomputation} to construct fault-tolerant logical operations on tile codes. For convenience of the reader, we state it here again.

\begin{proposition}
    Consider a $\op Z$-type logical operator $\bar{\op Z}$ of a tile code. Then, $\bar{\op Z}$ is the product of omitted boundary $\op Z$-stabilizers restricted to the tile code lattice. 
\end{proposition}

\subsection{Derived automorphisms of tile codes}\label{subsec: pedestrianstylederivedautomorphisms}

In this section, we define the \emph{derived $x$-} and \emph{$y$-automorphisms} $T_x$ and $T_y$, which are logical Clifford operations on the tile codes that map $\op X$-operators to $\op X$-operators and $\op Z$-operators to $\op Z$-operators. 
Then, we explain how these operations can be implemented and how the basis constructed in \Cref{subsec: logicalopsstruc: operatorsonastrip} lets us understand the action on the encoded information. Derived automorphisms are a broader, new concept that we will discuss in more generality in \Cref{sec:derivedautomorphisms}. There, the choice of name will also become clear. We mention that the notion of derived automorphisms for quantum codes is a relaxation to the usual automorphisms, which are permutations of qubits and stabilizer checks preserving the code space~\cite{Grassl_2013, breuckmannFoldTransversalCliffordGates2024, eberhardt2024logicaloperatorsfoldtransversalgates,yoder2025tourgrossmodularquantum}. In this section, we will only discuss how derived automorphisms can be understood for tile codes; for a more general discussion, we refer to \Cref{sec:derivedautomorphisms}.

\begin{figure}
\centering
\begin{tikzpicture}[scale = 0.2, baseline=(current bounding box.center)]
\def\horizontalqubitsx{(0, 0), (0, 4), (0, 5), (0, 8), (0, 9), (0, 10), (0, 11), (1, 2), (1, 3), (1, 5), (1, 6), (1, 9), (1, 11)}
\def\verticalqubitsx{(0, 2), (0, 3), (0, 6), (0, 7), (0, 8), (0, 9), (0, 10), (0, 11), (1, 3), (1, 5), (1, 6), (1, 8), (1, 10), (1, 11)}

                \foreach \y in {0,...,11} {
                \draw[\latticecolor, \latticethickness, opacity = .2] (-1,\y) -- (0,\y);
            }
            \draw[\latticecolor, \latticethickness, opacity = .2] (-1,1) -- (-1, 12);
        \foreach \x in {0,1,...,11} {
            \draw[\latticecolor, \latticethickness] (\x,0) -- (\x, 12);
        }
        \foreach \x in {0,...,11} {
            \draw[\latticecolor, \latticethickness] (\x,1) -- (\x, 12);
        }            
        % Horizontal lines
        \foreach \y in {0,...,11} {
            \draw[\latticecolor, thin] (0,\y) -- (12,\y);
        }
    \foreach \pos in \horizontalqubitsx {
            \draw[good_red, line width = 1pt] \pos -- ++(1,0);
        }
    \foreach \pos in \verticalqubitsx {
            \draw[good_red, line width = 1pt] \pos -- ++(0,1);
        }
        % 3x3 grid of black dots
        \foreach \x in {0,...,9} {
            \foreach \y in {0,...,9} {
                \draw[fill=black] (\x,\y) circle (0.15);
            }
        }
        % Red dots on top and bottom
        \foreach \x in {0,...,9} {
        \foreach \y in {-2,-1,10,11} {
            \draw[fill=good_red, color = good_red] (\x,\y) circle (0.15);
        }
        }
        
        % Blue dots on left and right
        \foreach \y in {0,...,9} {
        \foreach \x in {-2,-1,10,11} {
            \draw[fill=good_blue, color = good_blue] (\x,\y) circle (0.15);
        }
        }

\node at (16,6.8) {$T_x$};
\node at (16,5) {$\mapsto$};

\node at (5.5, -4.5) {\textbf{(a)}};

        \begin{scope}[xshift = 22cm]
            \foreach \x in {0,1,...,11} {
            \draw[\latticecolor, thin] (\x,0) -- (\x, 12);
        }
        \foreach \x in {0,...,11} {
            \draw[\latticecolor, thin] (\x,1) -- (\x, 12);
        }            
                        \foreach \y in {0,...,11} {
                \draw[\latticecolor, \latticethickness, opacity = .2] (12,\y) -- (13,\y);
            }
            \draw[\latticecolor, \latticethickness, opacity = .2] (12,0) -- (12, 12);
        % Horizontal lines
        \foreach \y in {0,...,11} {
            \draw[\latticecolor, \latticethickness] (0,\y) -- (12,\y);
        }
    \foreach \pos in \horizontalqubitsx {
            \draw[good_red, line width = 1pt] \pos++(1,0) -- ++(1,0);
        }
    \foreach \pos in \verticalqubitsx {
            \draw[good_red, line width = 1pt] \pos++(1,0) -- ++(0,1);
        }
        % 3x3 grid of black dots
        \foreach \x in {0,...,9} {
            \foreach \y in {0,...,9} {
                \draw[fill=black] (\x,\y) circle (0.15);
            }
        }
        % Red dots on top and bottom
        \foreach \x in {0,...,9} {
        \foreach \y in {-2,-1,10,11} {
            \draw[fill=good_red, color = good_red] (\x,\y) circle (0.15);
        }
        }
        
        % Blue dots on left and right
        \foreach \y in {0,...,9} {
        \foreach \x in {-2,-1,10,11} {
            \draw[fill=good_blue, color = good_blue] (\x,\y) circle (0.15);
        }
        }
                
\node at (16,5) {$\simeq$};
\node at (5.5, -4.5) {\textbf{(b)}};
        \end{scope}

\def\horizontalqubits{(0, 4), (0, 5), (0, 7), (0, 8), (0, 11), (1, 0), (1, 2), (1, 3), (1, 4), (1, 8), (1, 10), (1, 6)}
        \def\verticalqubits{(0, 9), (0, 2), (0, 3), (0, 5), (0, 11), (0, 6), (1, 2), (1, 3), (1, 6), (1, 7), (1, 8), (1, 9), (1, 10), (1, 11)}

        \begin{scope}[xshift = 44cm]
            \foreach \x in {0,1,...,11} {
            \draw[\latticecolor, thin] (\x             ,0) -- (\x, 12);
        }
        \foreach \x in {0,...,11} {
            \draw[\latticecolor, thin] (\x,1) -- (\x, 12);
        }            
                        \foreach \y in {0,...,11} {
                \draw[\latticecolor, \latticethickness, opacity = .2] (12,\y) -- (13,\y);
            }
            \draw[\latticecolor, \latticethickness, opacity = .2] (12,0) -- (12, 12);
        % Horizontal lines
        \foreach \y in {0,...,11} {
            \draw[\latticecolor, \latticethickness] (0,\y) -- (12,\y);
        }
    \foreach \pos in \horizontalqubits {
            \draw[good_red, line width = 1pt] \pos++(0,0) -- ++(1,0);
        }
    \foreach \pos in \verticalqubits {
            \draw[good_red, line width = 1pt] \pos++(0,0) -- ++(0,1);
        }
        % 3x3 grid of black dots
        \foreach \x in {0,...,9} {
            \foreach \y in {0,...,9} {
                \draw[fill=black] (\x,\y) circle (0.15);
            }
        }
        % Red dots on top and bottom
        \foreach \x in {0,...,9} {
        \foreach \y in {-2,-1,10,11} {
            \draw[fill=good_red, color = good_red] (\x,\y) circle (0.15);
        }
        }
        
        % Blue dots on left and right
        \foreach \y in {0,...,9} {
        \foreach \x in {-2,-1,10,11} {
            \draw[fill=good_blue, color = good_blue] (\x,\y) circle (0.15);
        }
        }
        \node at (5.5, -4.5) {\textbf{(c)}};
        \end{scope}
\end{tikzpicture}
\caption{An example of a derived automorphism for a tile code: We track the logical operator $\bar{\op X}_1$ visualized in (a) where we stick to the labeling convention also used in \Cref{fig:main_figure}. By extending the lattice to the left and measuring out qubits on the right as described in \Cref{thm: derivedautomorphismpedestrian}, the logical operator $\bar{\op X}_1$ gets mapped to the logical operator in (b). Multiplying by stabilizers to bring this operator in canonical form, we easily see that this logical operator is equivalent to  $\bar{\op X}_5$.}\label{fig: derivedautos}
\end{figure}
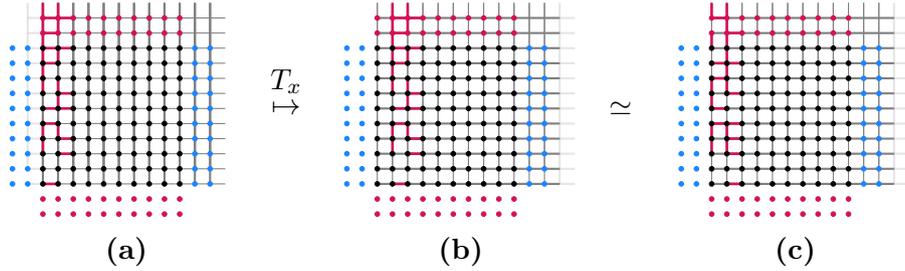

We define the derived $x$-automorphism $T_x$ by the action on the logical operators. 
First, we define its action on the logical $\op X$-operator.
By \Cref{cor: symplecticbasis}, without loss of generality, we may assume that any logical $\op X$-operator is fully supported in the strip of width $D$ along the left boundary.
We define the action of $T_x$ on the logical $\op X$-operators by shifting its support by $+1$ in the $x$-direction, see \Cref{fig: derivedautos}.

Next, we define the action of $T_x$ on the logical $\op Z$-operators.
By \Cref{cor: symplecticbasis}, we may assume without loss of generality that a non-trivial logical $\op Z$-operator $\overline{\op Z}$ is supported in the strip of width $D$ along the bottom boundary.
As an intermediate step, we obtain $\overline{\op Z}_{\mathrm{shifted}}$ by first shifting $\overline{\op Z}$ by $+1$ in the $x$ direction and then restricting to the tile code lattice.
\begin{equation}\label{eq: Zshifted}
    \begin{tikzpicture}[scale=0.3, baseline={(current bounding box.center)}]
        \begin{scope}[xshift=0cm]
        \def\horizontalqubits{(1, 0), (3, 0), (4, 0), (5, 0), (5, 1), (6, 1), (7, 0), (8, 0), (8, 1), (9, 0), (9, 1), (10, 0), (11, 1)}
        \def\verticalqubits{(3, 0), (3, 1), (5, 0), (5, 1), (6, 0), (6, 1), (7, 0), (8, 1), (9, 0), (9, 1), (10, 0), (11, 0)}
        \def\verticalyellow{(0,0),(0,1)}
        \def\horizontalyellow{(0,0),(0,1)}
            % Grid lines
            \foreach \foreach \x in {0,...,11} {
                \draw[\latticecolor, \latticethickness] (\x, 0) -- (\x, 3);
            }
            \draw[\latticecolor, \latticethickness] (0,0) -- (12,0);
            \draw[\latticecolor, \latticethickness] (0,1) -- (12,1);
            \draw[\latticecolor, \latticethickness] (0,2) -- (12,2);

            \fill[color = good_blue, opacity = .0] (0,0) -- (3,0) -- (3,3) -- (0,3) -- cycle;
            % Red stabilizer lines
            \foreach \pos in \horizontalqubits {
                \draw[good_blue, line width = \supportthickness] \pos -- ++(1,0);
            }
            \foreach \pos in \verticalqubits {
                \draw[good_blue, line width = \supportthickness] \pos -- ++(0,1);
            }
            \foreach \pos in \horizontalyellow {
                \draw[good_yellow, line width = \supportthickness] \pos -- ++(1,0);
            }
            \foreach \pos in \verticalyellow {
                \draw[good_yellow, line width = \supportthickness] \pos -- ++(0,1);
            }

            \draw [decorate,decoration={brace,amplitude=5pt,mirror,raise=2ex}]
            (0.7,0) -- (12.3,0) node[midway,yshift=-2.5em]{$\overline{\op Z}_{\mathrm{shifted}}$};

        \end{scope}
        \draw[color = good_red, fill = good_red] (0,0) circle (.2);
        \draw[color = good_red, fill = good_red] (0,1) circle (.2);
        \draw[color = good_red, fill = good_red] (0,-1) circle (.2);
        \draw[color = good_red, fill = good_red] (0,-2) circle (.2);
    \end{tikzpicture}
\end{equation}
By construction, the Pauli $\overline{\op Z}_{\mathrm{shifted}}$ supports none of the $D$ qubits (yellow in \Cref{eq: Zshifted}) at the bottom left vertical strip of the lattice.
We construct the desired logical operator $T_x(\overline{\op Z})$ by extending $\overline{\op Z}_{\mathrm{shifted}}$ to include some Pauli $\op Z$ on the vertical strip.
The product becomes a logical operator if and only if the $D$ syndromes (red in \Cref{eq: Zshifted}) are trivial.
By \Cref{cor: independent syndromes}, one can always find a unique extension of $\overline{\op Z}_{\mathrm{shifted}}$ such that the extended Pauli is a logical operator.

The derived $y$ automorphism $T_y$ is defined analogously to $T_x$ where we swap the roles of $\op X$ and $\op Z$ as well as the $x$- and $y$-direction. It turns out that the derived automorphism preserves symplecticity of the logical basis.

\begin{proposition}
    For our symplectic basis $\bar{\op X}_1, \bar{\op Z}_1, \dots $ constructed in \Cref{thm: logicalops}, the set $T_x(\bar{\op X}_1), T_x(\bar{\op Z}_1), \dots $ is again a symplectic basis of the space of logical operators. 
\end{proposition}

\begin{proof}
    It is clear by construction that since $\langle \bar{\op X}_i, \bar{\op Z}_j \rangle = \delta_{ij}$ also $\langle T_x(\bar{\op X}_i), T_x(\bar{\op Z}_j) \rangle = \delta_{ij}$ holds. Moreover, by construction, it is clear that all $T_x(\bar{\op X}_i)$ and $ T_x(\bar{\op Z}_j)$ commute with all stabilizers of the tile code. This finishes the proof. 
\end{proof}

One can implement $T_x^\epsilon $ for $\epsilon = \pm 1$ on a tile code state by ``sliding'' a tile code in the $x$ direction. The procedure is precisely described in the following theorem. For a visualization, see \Cref{fig: derivedautos}.

\begin{theorem}\label{thm: derivedautomorphismpedestrian}
    Suppose we apply the following operations to a tile code state $\ket{\psi}$ on the lattice~$\mathcal{L}$.
    \begin{enumerate}[label={(P\arabic*)}]
    \item Extend the lattice by adding one column on the left-hand side (resp. right-hand side) if $\epsilon = 1$ or $\epsilon = -1$, respectively, and prepare ancilla qubits on the edges in the new column as $\ket{0}$. Denote $\mathcal{L}^+$ the extended lattice. \label{it: ancilla prep}
    \item Measure all $\op X$-stabilizer generators of the new tile code defined on the extended lattice. \label{it:stabilizer measurement}
    \item Measure the qubits at the right (resp. left) columns if $\epsilon = 1$ (resp. $\epsilon = -1$) in $\op Z$-basis. Denote the subset of the measured qubits whose measurement outcome is $-1$ by $M$. Denote $\mathcal{L}'$ the shifted lattice obtained by removing edges of the measured qubits from the extended lattice $\mathcal{L}^+$.\label{it:single qubit measurement}
    \item Denote $\left(\overline{\op X}'_i, \overline{\op Z}'_i\right)_{i=1}^k$  the logical operators on the shifted lattice $\mathcal{L}'$ obtained by translating the logical operators $(\overline{\op X}_i, \overline{\op Z}_i)$ of the original lattice $\mathcal{L}$.
    Apply the logical operators $\prod_{i = 1}^k {\overline{\op X}'}_i^{j_i}$ of the tile code on the shifted lattice 
    where $j_i = 0$ if the overlap $|M \cap \supp (\overline{\op Z}_i)|$ of $M$ and the support of $\overline{\op Z}_i$ is even, and $j_i = 1$ otherwise. \label{it: logical frame correction}
    \end{enumerate}
    Then, the output state is a logical state $T_x^{\epsilon} \ket{\psi}$ of the tile code defined on the shifted lattice $\mathcal{L}'$   up to a Pauli frame update which is determined by the measurement outcomes of \ref{it:stabilizer measurement} and \ref{it:single qubit measurement}.
    
\end{theorem}
\begin{proof}
    We only give a proof for the case with $\epsilon = 1$. The case with $\epsilon = -1$ works analogously.

    For the proof, we track the stabilizer group of the logical state $\ket{\psi}$.
    We first consider the case where $\ket{\psi}$ is a $\pm 1$ eigenstate of the logical $\bar{\op Z}$-operators.
    Denote $\mathcal{S}_{\op X}$ and $\mathcal{S}_{\op Z}$ the  sets of $\op X$- and $\op Z$-type tile code stabilizer generators on the original lattice $\mathcal{}$.
    The stabilizer group of $\ket{\psi}$ is generated by
    \begin{align}
        \mathcal{S}_{\op X} \cup \mathcal{S}_{\op Z} \cup \{\sigma_1 \overline{\op Z}_1, \dots, \sigma_k\overline{\op Z}_k\} \qquad \text{with} \qquad \sigma_1, \dots, \sigma_k \in \{-1, 1\} \,.
    \end{align}
    Denote $\mathcal{A}$ the set of ancilla qubits and $\op Z_a$ for $a \in \mathcal{A}$ the Pauli $\op Z$-operator at the qubit $a$ in \ref{it: ancilla prep}. The stabilizer group of the entire state after \ref{it: ancilla prep} is generated by 
    \begin{align} \label{eq: Pauli subgroup after initialization}
        \mathcal{S}_{\op X} \cup \mathcal{S}_{\op Z} \cup \{\sigma_1 \overline{\op Z}_1, \dots, \sigma_k \overline{\op Z}_k\}\cup \{Z_a\}_{a \in \mathcal{A}} \,.
    \end{align}
    Denote $\mathcal{S}^+_{\op X}$ and $\mathcal{S}^+_{\op Z}$ the $\op X$-and $\op Z$-type tile code stabilizer generators on the extended lattice $\mathcal{L}^+$. 
    It is clear by construction that $\mathcal{S}_{\op Z} \cup \{\op Z_a\}_{a \in \mathcal{A}}$ and $\tilde{\mathcal{S}}_{\op Z} \cup \{\op Z_a\}_{a \in \mathcal{A}}$ generate the same Pauli subgroup.
    We also replace the logical operators $\overline{\op Z}_i$ with the unique extension $\overline{\op Z}^+_i := P_i \overline{\op Z}_i$ where $\op Z$-type Pauli $P_i$ is the multi-qubit $\op Z$-Pauli supported on the ancilla qubits $\mathcal{A}$ such that $\overline{\op Z}^+_i$ is a logical $\op Z$-operator on the extended lattice $\mathcal{L}^+$.
    Therefore, we can write the generating set in \Cref{eq: Pauli subgroup after initialization} as
    \begin{align}\label{eq: Pauli subgroup after initialization 2}
        \mathcal{S}_{\op X} \cup \mathcal{S}^+_{\op Z} \cup \{\sigma_1 \overline{\op Z}^+_1, \dots, \sigma_k \overline{\op Z}^+_k\}\cup \{\op Z_a\}_{a \in \mathcal{A}} \,.
    \end{align}

    In \ref{it:stabilizer measurement}, we measure all stabilizers in the set $\mathcal{S}^+_{\op X} \setminus \mathcal{S}_{\op X}$, which is the set of $\op X$-stabilizers at the left boundary of the extended lattice,
    as well as the tile code stabilizers $\mathcal{S}_{\op X}$ of the original lattice $\mathcal{L}$.
    We observe how the generating subset $\{\op Z_a\}_{\mathcal{A}}$ evolves during the measurement procedure.
    Indeed, the generating subset $\{\op Z_a\}_{\mathcal{A}}$ is transformed into a set of multi-qubit Pauli $\op Z$-operators which are supported on the left boundary and commute with all $\op X$-stabilizers of the extended lattice $\mathcal{L}^+$. 
    By the total topological order condition, those Pauli $\op Z$-operators must be $\op Z$-stabilizers of the extended lattice.
    Therefore, the stabilizer group of the state on the entire system after \ref{it:stabilizer measurement}, ignoring the sign in front of the stabilizers in $\mathcal{S}^+ \setminus \mathcal{S}$ caused by the measurement result, is generated by
    \begin{align}\label{eq: stabilizer group after stabilizer measurement}
        \mathcal{S}^+_{\op X} \cup \mathcal{S}^+_{\op Z} \cup \{\sigma_1 \overline{\op Z}^+_1, \dots, \sigma_k \overline{\op Z}^+_k\} \,.
    \end{align}

    Now, we observe how the stabilizer group in \Cref{eq: stabilizer group after stabilizer measurement} evolves in \ref{it:single qubit measurement}.
    Using the argument analogous to that showing how the generating subset $\{\op Z_a\}$ evolves to a subset of $\mathcal{S}_{\op Z}^+$, one can show that the generating subset $\mathcal{S}_{\op X}^+$ evolves to $\mathcal{S}'_{\op X}$, the set of $\op X$-stabilizers on the shifted lattice $\mathcal{L}'$.
    Note also that $\mathcal{S}^+_{\op Z}$ evolves into $\mathcal{S}'_{\op Z}$ when we ignore the sign of the stabilizers caused by the single-qubit measurement result.
    One can also see from the definition of the derived automorphism $T_x$ that the logical operator $\overline{\op Z}_i^+$ and the single-qubit Paulis on the measured qubits generate $(-1)^{|M \cap \supp(\overline{\op Z}_i)|} T_x(\overline{\op Z}_i)$.
    Therefore, the stabilizer group of the state on the remaining qubit after \ref{it:single qubit measurement} is generated by
    \begin{align}
        \mathcal{S}'_{\op X} \cup \mathcal{S}'_{\op Z} \cup \{ (-1)^{|M \cap \supp(\overline{\op Z}_i)|} \sigma_i T_x(\overline{\op Z}_i) \}_{i = 1}^k \,.
    \end{align}

    After applying the logical operator $\prod_{i = 1}^k {\overline{\op X}'}^{j_i}_i$ in \ref{it: logical frame correction}, the stabilizer group has the generating set
    \begin{align}
        \mathcal{S}_{\op X}' \cup \mathcal{S}_{\op Z}' \cup \{ \sigma_1 T_x(\overline{\op Z}_1), \dots, \sigma_k T_x(\overline{\op Z}_k) \} \,.
    \end{align}
    In an analogous way, one can track stabilizers assuming that the initial tile code state is an eigenstate of logical $\op X$ operators. This finishes the proof.
\end{proof}

This protocol is exactly the tile code version of moving a patch of surface code used in lattice surgery protocols~\cite{Fowler_2012, Litinski_2019}. 
We note that another method of deforming a tile code, involving both extending and shrinking, also appears in the logical Pauli measurement protocol of Ref.~\cite{yang2025planarfaulttolerantquantumcomputation}. Unlike \Cref{thm: derivedautomorphismpedestrian}, they include only a selection of $\op X$-stabilizer tiles (resp. $\op Z$-stabilizer tiles): In this way, they gauge products of logical operators of the code and are able to measure these products in that way.

Similarly as for the surface code, the sliding protocol can also be implemented in a fault-tolerant way with the following modification. In \ref{it:stabilizer measurement}, we measure $\op Z$ stabilizers of the extended lattice as well, and we repeat the measurement $d$ times, where $d$ is the minimal distance of the tile code.
After \ref{it:single qubit measurement}, we measure $\op X$ and $\op Z$ stabilizers of the shifted code $d$ times again.
Since stabilizers of the original code are products of Pauli $\op Z$-operators in $M$ and boundary stabilizers in the slided version,  the parity of $| M \cap \supp (\overline{\op Z})|$ can be reliably inferred even in the presence of measurement errors.

\begin{remark}
    The action of the protocol described in \Cref{thm: derivedautomorphismpedestrian} on the logical operators can also be understood as a cellular automaton: Using topological order, one can see that for any $\op X$-type operator supported on a $D\times D$ box, there is a unique way of extending the support to the left to make it commute with one additional row of stabilizers.

    \[\begin{tikzcd}[row sep=large]
	& {\begin{tikzpicture}[scale = .3]
	    \foreach \x in {0,1,2} {
        \draw[color = \latticecolor, line width = 1pt] (\x,0) -- (\x,2);
        }
        \foreach \y in {0,1} {
        \draw[color = \latticecolor, line width = 1pt] (0,\y) -- (3,\y);
        }
	\end{tikzpicture}} \\
	{
    \begin{tikzpicture}[scale = .3]
	    \foreach \x in {1,2} {
        \draw[color = \latticecolor, line width = 1pt] (\x,0) -- (\x,2);
        }
        \draw[color = \latticecolor, line width = 1pt, opacity = .3] (0,0) -- (0,2);
        \foreach \y in {0,1} {
        \draw[color = \latticecolor, line width = 1pt, opacity = .3] (0,\y) -- (1,\y);
        \draw[color = \latticecolor, line width = 1pt] (1,\y) -- (3,\y);
        }
	\end{tikzpicture}
    } && {
    \begin{tikzpicture}[scale = .3]
	    \foreach \x in {0,1} {
        \draw[color = \latticecolor, line width = 1pt] (\x,0) -- (\x,2);
        }
        \foreach \y in {0,1} {
        \draw[color = \latticecolor, line width = 1pt] (0,\y) -- (2,\y);
        \draw[color = \latticecolor, line width = 1pt, opacity = .5] (2,\y) -- (3,\y);
        }
        \draw[color = \latticecolor, line width = 1pt, opacity = .5] (2,0) -- (2,2);
	\end{tikzpicture}
    }
	\arrow[from=1-2, to=2-1]
	\arrow[from=1-2, to=2-3]
\end{tikzcd}\]

    With that, for the two tiles in \Cref{eq:stabilizertiles}, the following eight patterns completely describe the action of $T_x$. This can again be interpreted as a cellular automaton. 
    \begin{equation*}
        \begin{tikzpicture}[scale=0.4]
  \foreach \i in {0,...,7} {
    \pgfmathtruncatemacro{\col}{mod(\i,4)}
    \pgfmathtruncatemacro{\row}{floor(\i/4)}
    \begin{scope}[xshift=\col*4cm, yshift=-\row*3cm, local bounding box=cell-\i]

      % grid lines
      \foreach \x in {0,1}            {\draw[black, thin] (-1,\x) -- (2,\x);}
      \foreach \x in {-1,0,1}       {\draw[black, thin] (\x,0)  -- (\x,2);}

      % labels

      \ifnum\i=0\relax
        \draw[good_red, line width = \supportthickness] (1,0) -- (0,0);
      \fi
      \ifnum\i=1\relax
        \draw[good_red, line width = \supportthickness] (0,0) -- (0,1);
      \fi
      \ifnum\i=2\relax
        \draw[good_red, line width = \supportthickness] (1,1) -- (0,1);
      \fi
      \ifnum\i=3\relax
        \draw[good_red, line width = \supportthickness] (-0,2) -- (-0,1);

      \fi
      \ifnum\i=4\relax
        \draw[good_red, line width = \supportthickness] (1,0) -- (2,0);
        \draw[good_red, line width = \supportthickness] (1,0) -- (0,0) -- (-1,0) -- (-1,1) -- (0,1);
      \fi
      \ifnum\i=5\relax
        \draw[good_red, line width = \supportthickness] (1,0) -- (1,1);
        \draw[good_red, line width = \supportthickness] (-1,0) -- (0,0);
        % \draw[good_red, line width = \supportthickness] (2,0) -- (3,0);
      \fi
      \ifnum\i=6\relax
        \draw[good_red, line width = \supportthickness] (1,1) -- (2,1);
        \draw[good_red, line width = \supportthickness] (-1,1) -- (-1,2);
        % \draw[good_red, line width = \supportthickness] (2,1) -- (2,2);
        % \draw[good_red, line width = \supportthickness] (2,0) -- (3,0);
      \fi
      \ifnum\i=7\relax
        \draw[good_red, line width = \supportthickness] (1,1) -- (1,2);
        \draw[good_red, line width = \supportthickness] (-1,0) -- (0,0);
        \draw[good_red, line width = \supportthickness] (-1,1) -- (0,1);
      \fi
    \end{scope}
  }
\end{tikzpicture}
    \end{equation*}

An example of a whole logical operator being moved one step to the left can be seen in \Cref{fig: derivedautos}. There, the first rule of this cellular automaton is being applied. This idea will be made more formal in \Cref{sec:algebraoftilecodes}.
\end{remark}

\begin{example}\label{example: pedestrianderivedauto}
We can track the action of $T_x$ on all logical $\op X$- and $\op Z$-operators of the tile code whose stabilizer tiles are those in \Cref{eq:stabilizertiles}. The action on logical operators can be described by two matrices
\begin{equation*}
T_x = \begin{pmatrix} A & 0 \\ 0 & B \end{pmatrix} \in \mathbb{F}_2^{16 \times 16} ,
        A = {\tiny\begin{pmatrix}
        0 & 0 & 0 & 0 & 1 & 1 & 0 & 1 \\
        0 & 0 & 0 & 0 & 1 & 0 & 0 & 0 \\ 
        0 & 0 & 0 & 0 & 1 & 0 & 0 & 1 \\
        0 & 0 & 0 & 0 & 0 & 0 & 1 & 0 \\
        1 & 0 & 0 & 0 & 1 & 0 & 0 & 0 \\
        0 & 1 & 0 & 0 & 0 & 0 & 0 & 0 \\
        0 & 0 & 1 & 0 & 0 & 0 & 1 & 0 \\
        0 & 0 & 0 & 1 & 0 & 0 & 0 & 0 
        \end{pmatrix}},  
        B = {\tiny\begin{pmatrix}
        0 & 0 & 0 & 0 & 0 & 1 & 0 & 0 \\
        1 & 0 & 0 & 0 & 1 & 0 & 0 & 1 \\
        0 & 0 & 0 & 0 & 0 & 1 & 0 & 1 \\
        0 & 0 & 1 & 0 & 0 & 0 & 1 & 0 \\
        1 & 0 & 0 & 0 & 0 & 0 & 0 & 0 \\
        0 & 1 & 0 & 0 & 0 & 0 & 0 & 0 \\
        0 & 0 & 1 & 0 & 0 & 0 & 0 & 0 \\
        0 & 0 & 0 & 1 & 0 & 0 & 0 & 0
        \end{pmatrix}},
\end{equation*}
where the columns of $A$ and $B$ represent $T_x(\overline{\op X}_i)$ and $T_x(\overline{\op Z}_i)$.

Analogously, one can calculate the action of $T_y$ on the logical operators.
The matrix form of the automorphism~$T_y$ is
\begin{equation*}
    T_y = \begin{pmatrix} C & 0 \\ 0 & D \end{pmatrix} \in \mathbb{F}_2^{16 \times 16} \,, C = {\tiny\begin{pmatrix}
            1 & 0 & 0 & 1 & 0 & 0 & 0 & 1 \\
            0 & 0 & 1 & 0 & 0 & 0 & 0 & 0 \\
            1 & 0 & 0 & 0 & 0 & 0 & 0 & 0 \\
            0 & 1 & 0 & 0 & 0 & 0 & 0 & 0 \\
            0 & 0 & 0 & 0 & 1 & 0 & 0 & 1 \\
            0 & 0 & 0 & 1 & 0 & 0 & 1 & 0 \\
            0 & 0 & 0 & 0 & 1 & 0 & 0 & 0 \\
            0 & 0 & 0 & 0 & 0 & 1 & 0 & 0
        \end{pmatrix}}, 
        D = {\tiny\begin{pmatrix}
            0 & 0 & 0 & 1 & 0 & 0 & 1 & 0 \\
            0 & 0 & 1 & 0 & 0 & 0 & 0 & 0 \\
            1 & 0 & 0 & 1 & 0 & 0 & 1 & 0 \\
            0 & 1 & 0 & 0 & 0 & 0 & 0 & 0 \\
            0 & 0 & 0 & 1 & 0 & 0 & 1 & 1 \\
            0 & 0 & 0 & 0 & 0 & 0 & 1 & 0 \\
            0 & 0 & 0 & 1 & 1 & 0 & 1 & 1 \\
            0 & 0 & 0 & 0 & 0 & 1 & 0 & 0
        \end{pmatrix}},
\end{equation*}
where the columns of $C$ and $D$ represent $T_y(\overline{\op X}_i)$ and $T_y(\overline{\op Z}_i)$.

One can see that $T_y = T_x^{150}$ and they generate the same cyclic subgroup of $\operatorname{Sp}(16, \mathbb{F}_2)$ of order $217$. This can be understood more rigorously from an algebraic perspective, see \Cref{example: algebraofderivedauto}.

From the matrices $A$ and $B$, one can easily construct a circuit composed of logical CNOT and SWAP gates that is actually implemented by $T_x$:

\[
\Qcircuit @C=1em @R=1em {
\lstick{1}& \qswap & \qw & \qw & \qw & \qw  & \ctrl{4} & \qw & \qswap &  \targ & \targ & \qw  & \qw\\
\lstick{2}& \qw & \qswap & \qw & \qw & \qw   & \qw & \qw & \qswap \qwx[-1]  & \ctrl{-1} & \qw & \ctrl{1} & \qw\\
\lstick{3}& \qw & \qw & \qswap & \qw & \qw  & \qw  & \ctrl{4} & \qswap  & \qw & \ctrl{-2} & \targ & \qw\\
\lstick{4}& \qw & \qw & \qw & \qswap & \qw  & \qw  & \qw & \qswap \qwx[-1]  & \qw & \qw & \qw  & \qw\\
\lstick{5}& \qswap \qwx[-4] & \qw  & \qw & \qw   & \qw  & \targ & \qw & \qw & \qw & \qw & \qw & \qw   \\
\lstick{6}& \qw & \qswap \qwx[-4] & \qw & \qw & \qw & \qw & \qw  & \qw & \qw & \qw & \qw & \qw  \\
\lstick{7}& \qw & \qw & \qswap \qwx[-4] & \qw  & \qw & \qw  & \targ & \qw & \qw & \qw & \qw & \qw  \\
\lstick{8}& \qw & \qw & \qw & \qswap \qwx[-4]  & \qw & \qw  & \qw & \qw & \qw & \qw & \qw &\qw
}
\]
\end{example}

\section{Homological algebra of tile codes}\label{sec:algebraoftilecodes}
In this section, we describe how tile codes can be `resolved' by infinite dimensional codes, that arise as Koszul complexes in (Laurent)-polynomial rings. 
\subsection{Background on Koszul complexes}
For a commutative ring $R$ and elements $f,g\in R$, the \emph{Koszul complex} is defined as 
% https://q.uiver.app/#q=WzAsMyxbMCwwLCJLKFxcZnR3b1t4XlxccG0seV5cXHBtXSxmLGcpOiBcXGZ0d29beF5cXHBtLHleXFxwbV0iXSxbMSwwLCJcXGZ0d29beF5cXHBtLHleXFxwbV1eMiJdLFsyLDAsIlxcZnR3b1t4XlxccG0seV5cXHBtXSJdLFswLDEsIihnLGYpXnQiXSxbMSwyLCIoZixnKSJdXQ==
\begin{equation}
\label{eq:koszulcomplex}
\begin{tikzcd}
	{K(R,f,g)^\bullet: R}& {R^2} & {R}
	\arrow["{(-g,f)^t}", from=1-1, to=1-2]
	\arrow["{(f,g)}", from=1-2, to=1-3]
\end{tikzcd}
\end{equation}
By convention, the terms in the complex are in cohomological degrees $-2,$ $-1$ and $0$. One can show that the cohomology groups of the Koszul have the following form \cite{eberhardt2024logicaloperatorsfoldtransversalgates}
\begin{align*}
   H^i(K(R,f,g)^\bullet)\cong \begin{cases}
   R/(f,g) & \text{ if }i=0,\\
   \frac{(f)\cap (g)}{(fg)} & \text{ if }i=-1,\\
   \operatorname{ann}_R(f)\cap \operatorname{ann}_R(g)  & \text{ if }i=-2 \text{ and}\\
    0 & \text{ otherwise}.
\end{cases}
\end{align*}
Here, we denote by $(r_1,\dots,r_n)\subset R$ the ideal generated by a elements $r_1,\dots, r_n\in R$. Moreover, we denote by $\operatorname{ann}_R(r)=\{s\in R\,|\, rs=0\}$ the annihilator of $r\in R$.

Recall that elements $f,g\in R$ form a \emph{Koszul-regular sequence}, if $$H^i(K(R,f,g)^\bullet)= 0\text{ for }i=-2,-1.$$ So $f,g\in R$ are a Koszul-regular sequence if and only if the Koszul complex $K(R,f,g)^\bullet$ is a resolution of $R/(f,g)$. 

We are particularly interested in the case that $f,g\neq 0$ and $R$ is a unique factorization domain, so for example a (Laurent)-polynomial ring over a field. In this case, we always have that $H^{-2}(K(R,f,g)^\bullet)=0$. Hence, $f,g\in R$ form a Koszul-regular sequence if and only if $(f)\cap (g)=(fg)$ which is equivalent to the statement that $f$ and $g$ have no (non-unit) common divisors in $R$.

We will also need Koszul complexes in the following slightly more general form. Assume that $f:R_1\to R$ and $g:R_2\to R$ are maps of a free rank one $R$-modules to $R$, then we obtain the Koszul complex
\begin{equation}
\label{eq:shiftedkoszulcomplex}
\begin{tikzcd}[column sep=40pt]
    	{K(R,f,g)^\bullet: R_1\otimes_R R_2}& {R_1\oplus R_2} & {R}
	\arrow["{ (-1\otimes g,f\otimes 1)^t}"{above = 3pt}, from=1-1, to=1-2, yshift = 0pt]
	\arrow["{(f,g)}"{above = 3pt}, from=1-2, to=1-3, yshift = 0pt]
    \end{tikzcd}
\end{equation}

By choosing identifications of $R_1,R_2$ with $R$, this is isomorphic to the above definition. However, it will be useful for us in order to take care of certain shifts.

\subsection{Codes on infinite lattices as Koszul complexes}\label{sec:infinitelatticekoszul}
We will now explain in detail how Koszul complexes are related to the CSS codes on a $2D$-plane, as introduced in \Cref{sec:stabtileson2dplane}. For this, we consider the (Laurent)-polynomial rings 
\begin{equation}\label{eq:defoflaurentpolynomials}
R_{\epsilon_1\epsilon_2}=\ftwo[x^{\epsilon_1},y^{\epsilon_2}]\text{ for }\epsilon_1,\epsilon_2\in \{+,-,\pm\}. 
\end{equation}
For example, $R_{\pm-}=\ftwo[x^{\pm1},y^{-1}]$. These rings naturally have a basis by monomials $x^ay^b$ which we can identify with coordinates $(a,b)\in \mathbb{Z}^2.$ Moreover, we abbreviate $R=R_{\pm\pm}.$

To obtain the CSS code on the unbounded infinite $2D$-plane, \Cref{eq: 2dplane}, we consider the Koszul complex
% https://q.uiver.app/#q=WzAsMyxbMCwwLCJLKFxcZnR3b1t4XlxccG0seV5cXHBtXSxmLGcpOiBcXGZ0d29beF5cXHBtLHleXFxwbV0iXSxbMSwwLCJcXGZ0d29beF5cXHBtLHleXFxwbV1eMiJdLFsyLDAsIlxcZnR3b1t4XlxccG0seV5cXHBtXSJdLFswLDEsIihnLGYpXnQiXSxbMSwyLCIoZixnKSJdXQ==
\[\begin{tikzcd}
	{K^\bullet=K^\bullet(R,f,g): R} & {R^2} & {R}
	\arrow["{(g,f)^t}", from=1-1, to=1-2]
	\arrow["{(f,g)}", from=1-2, to=1-3].
\end{tikzcd}\]

The basis vectors of the terms $K^i(R,f,g)$ in the correspond to the coordinates of $\op X$-checks, vertical/horizontal qubits and $\op Z$-checks, for $i=-2,-1$ and $0$, respectively.

The polynomials $f,g$ specify the stabilizer tiles in the following way.
The vertical and horizontal qubits in the $\op X$-tile correspond to $f$ and $g$, respectively, while the vertical and horizontal qubits in the $\op Z$-tile correspond to $(xy)^Dg(x^{-1},y^{-1})$ and $(xy)^Df(x^{-1},y^{-1})$, respectively.
The fact that the stabilizer tiles are confined within a box of size $B$ is equivalent to the assumption that $f,g\in \ftwo[x,y]$ are polynomials of maximal degree $D=B-1$ in both $x$ and $y$.
For example, the polynomials $f=1+x^2y+x^2y^2$ and $g=x+x^2+y^2$, correspond to the stabilizer tiles
\begin{equation*}
    \begin{tikzpicture}[scale=0.6]
        % Define horizontal and vertical qubit coordinates
        \def\horizontalqubits{(1,0),(2,0), (0,2)}
        \def\verticalqubits{(0,0),(2,1),(2,2)}

        \def\boxsize{3}
        
        % Left figure - X-type stabilizer
        \begin{scope}[xshift=0cm]
            % Grid lines
            \draw[\latticecolor, \latticethickness] (3,0) -- (0,0) -- (0,3) ;
            \draw[\latticecolor, \latticethickness] (1,0) -- (1,3);
            \draw[\latticecolor, \latticethickness] (2,0) -- (2,3);
            \draw[\latticecolor, \latticethickness] (0,1) -- (3,1);
            \draw[\latticecolor, \latticethickness] (0,2) -- (3,2);
            \fill[color = good_red, opacity = .0] (0,0) -- (3,0) -- (3,3) -- (0,3) -- cycle;
            % Red stabilizer lines
            \foreach \pos in \horizontalqubits {
                \draw[good_red, line width = \supportthickness, dotted] \pos -- ++(1,0);
            }
            \foreach \pos in \verticalqubits {
                \draw[good_red, line width = \supportthickness] \pos -- ++(0,1);
            }
            
            % Black dot in corner
            \draw[fill=good_red, color = good_red] (0,0) circle (0.15);
            % label horizontal qubits
            \node[scale = .7, color = good_red] at (1.5,-.4) {$x^{\phantom{1}}$};
            \node[scale = .7, color = good_red] at (2.5,-.4) {$x^2$};
            \node[scale = .7, color = good_red] at (0.5,1.6) {$y^2$};
% label vertical qubits
            \node[scale = .7, color = good_red] at (0.4,.5) {$1$};
            \node[scale = .7, color = good_red] at (2.6,1.5) {$x^2y$};
            \node[scale = .7, color = good_red] at (2.6,2.5) {$x^2y^2$};
        \end{scope}

        \def\horizontalqubitsz{(0,0),(0,1), (2,2)}
        \def\verticalqubitsz{(2,0),(0,2),(1,2)}
        % Right figure - Z-type stabilizer
        \begin{scope}[xshift=5cm]
            % Grid lines
            \draw[\latticecolor, \latticethickness] (3,0) -- (0,0) -- (0,3) ;
            \draw[\latticecolor, \latticethickness] (1,0) -- (1,3);
            \draw[\latticecolor, \latticethickness] (2,0) -- (2,3);
            \draw[\latticecolor, \latticethickness] (0,1) -- (3,1);
            \draw[\latticecolor, \latticethickness] (0,2) -- (3,2);
            
            % Blue stabilizer lines
            \foreach \pos in \horizontalqubitsz {
                \draw[good_blue, line width = \supportthickness] \pos -- ++(1,0);
            }
            \foreach \pos in \verticalqubitsz {
                \draw[good_blue, line width = \supportthickness, dotted] \pos -- ++(0,1);
            }
            
            % Black dot in corner
            \draw[fill=good_blue, color = good_blue] (0,0) circle (0.15);
% label horizontal qubits
            \node[scale = .7, color = good_blue] at (0.5,-.4) {$1$};
            \node[scale = .7, color = good_blue] at (0.5,.6) {$y$};
            \node[scale = .7, color = good_blue] at (2.6,1.6) {$x^2y^2$};
% label vertical qubits
            \node[scale = .7, color = good_blue] at (2.4,.5) {$x^2$};
            \node[scale = .7, color = good_blue] at (0.4,2.5) {$y^2$};
            \node[scale = .7, color = good_blue] at (1.5,2.5) {$xy^2$};
        \end{scope}
    \end{tikzpicture}
\end{equation*}
Moreover, we mention that the excitation map $\partial$ defined in \Cref{eq: excitation map} can be identified with the differential $(f,g)$ in the Koszul complex.

Similarly to the Koszul complex corresponding to the the unbounded infinite plane discussed above, we also consider Koszul complexes $K^\bullet_{\epsilon_1\epsilon_2}$ associated to coordinate quadrants and half planes by using the rings $R_{\epsilon_1\epsilon_2}$ for $\epsilon_i\in \{\pm,+,-\}.$

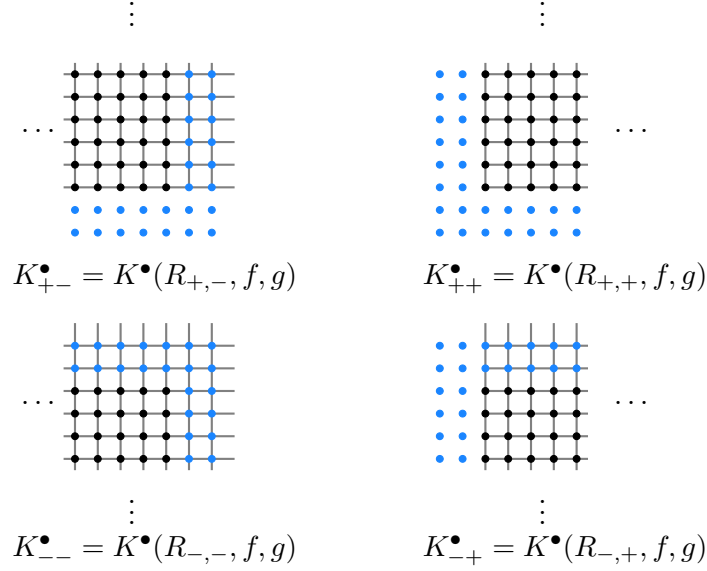
\begin{figure}
    \begin{center}
        \begin{tikzpicture}[scale = .3]
    \begin{scope}
      \foreach \y in {-5,...,0} {
            \draw[\latticecolor, \latticethickness] (-5.5,\y) -- (2,\y);
        }
         \foreach \x in {-5,...,1} {
            \draw[\latticecolor, \latticethickness] (\x,-5) -- (\x,0.5);
        }
        \foreach \x in {-5,...,-1} {
        \foreach \y in {-5,...,0} {
        \draw[fill = black] (\x, \y) circle (.15);
        }
        }
        \foreach \x in {-5,...,-1, 0, 1} {
        \foreach \y in {-7,-6} {
        \draw[color = good_blue,fill = good_blue] (\x, \y) circle (.15);
        }
        }
        \foreach \x in {0,1} {
        \foreach \y in {-5,...,0} {
        \draw[color = good_blue, fill = good_blue] (\x, \y) circle (.15);
        }
        }
        \node at (-6.5, -2.5) {$\dots$};
        \node at (-2.5, 3) {$\vdots$};

        \node at (-1.5, -9) {$K_{+-}^\bullet=K^\bullet (R_{+,-},f, g)$};
    \end{scope}

    \begin{scope}[xshift = 18cm]
      \foreach \y in {-5,...,0} {
            \draw[\latticecolor, \latticethickness] (-5,\y) -- (-0.5,\y);
        }
         \foreach \x in {-5,...,-1} {
            \draw[\latticecolor, \latticethickness] (\x,-5) -- (\x,0.5);
        }
        \foreach \x in {-5,...,-1} {
        \foreach \y in {-5,...,0} {
        \draw[fill = black] (\x, \y) circle (.15);
        }
        }
        \foreach \x in {-7, -6, -5,...,-1} {
        \foreach \y in {-7,-6} {
        \draw[color = good_blue,fill = good_blue] (\x, \y) circle (.15);
        }
        }
        \foreach \x in {-6,-7} {
        \foreach \y in {-5,...,0} {
        \draw[color = good_blue, fill = good_blue] (\x, \y) circle (.15);
        }
        }
        \node at (1.5, -2.5) {$\dots$};
        \node at (-2.5, 3) {$\vdots$};

        \node at (-1.5, -9) {$K_{++}^\bullet=K^\bullet (R_{+,+},f, g)$};
    \end{scope}

    \begin{scope}[yshift = -12cm]
      \foreach \y in {-5,...,0} {
            \draw[\latticecolor, \latticethickness] (-5.5,\y) -- (2,\y);
        }
         \foreach \x in {-5,...,1} {
            \draw[\latticecolor, \latticethickness] (\x,-5.5) -- (\x,1);
        }
        \foreach \x in {-5,...,-1} {
        \foreach \y in {-5,...,-2} {
        \draw[fill = black] (\x, \y) circle (.15);
        }
        }
        \foreach \x in {-5,...,-1,0,1} {
        \foreach \y in {-1,0} {
        \draw[color = good_blue,fill = good_blue] (\x, \y) circle (.15);
        }
        }
        \foreach \x in {0,1} {
        \foreach \y in {-5,...,-2} {
        \draw[color = good_blue, fill = good_blue] (\x, \y) circle (.15);
        }
        }
        \node at (-6.5, -2.5) {$\dots$};
        \node at (-2.5, -7) {$\vdots$};

        \node at (-1.5, -9) {$K_{--}^\bullet=K^\bullet (R_{-,-},f,g)$};
    \end{scope}

    \begin{scope}[xshift = 18cm, yshift = -12cm]
      \foreach \y in {-5,...,0} {
            \draw[\latticecolor, \latticethickness] (-5,\y) -- (-0.5,\y);
        }
         \foreach \x in {-5,...,-1} {
            \draw[\latticecolor, \latticethickness] (\x,-5.5) -- (\x,1);
        }
        \foreach \x in {-5,...,-1} {
        \foreach \y in {-5,...,-2} {
        \draw[fill = black] (\x, \y) circle (.15);
        }
        }
        \foreach \x in {-7, -6,-5,...,-1} {
        \foreach \y in {-1,0} {
        \draw[color = good_blue,fill = good_blue] (\x, \y) circle (.15);
        }
        }
        \foreach \x in {-7,-6} {
        \foreach \y in {-5,...,-2} {
        \draw[color = good_blue, fill = good_blue] (\x, \y) circle (.15);
        }
        }
        \node at (1.5, -2.5) {$\dots$};
        \node at (-2.5, -7) {$\vdots$};

        \node at (-1.5, -9) {$K_{-+}^\bullet=K^\bullet(R_{-,+},f, g)$};
    \end{scope}
    \end{tikzpicture}  
    
    \end{center}
    \caption{Koszul complexes and associated CSS codes supported on the four different quadrants.}\label{fig:quadrants}
\end{figure}

For example, the Koszul complex 
\[\begin{tikzcd}
	{K^\bullet_{++}=K^\bullet(R,f,g): R_{++}} & {R_{++}^2} & {R_{++}}
	\arrow["{(g,f)^t}", from=1-1, to=1-2]
	\arrow["{(f,g)}", from=1-2, to=1-3]
\end{tikzcd}\]
corresponds to a CSS code supported on the right-upper quadrant, see \Cref{fig:quadrants}.

Whenever $\epsilon_1=-$ and or $\epsilon_2=-,$ the polynomials $f,g$ are no longer elements in $R_{\epsilon_1\epsilon_2}$ and we need to incorporate a shift. For this, define
$$s_{\epsilon_1\epsilon_2}=x^{-D(\epsilon_1)}y^{-D(\epsilon_1)}$$
where $D(\epsilon)=D$ if $\epsilon=-$ and $D(\epsilon)=0$, else. Then, multiplication with $f,g$ yields a map $s_{\epsilon_1\epsilon_2}R_{\epsilon_1\epsilon_2}\to R_{\epsilon_1\epsilon_2}.$ 
Whe then obtain the Koszul complex, see \Cref{eq:shiftedkoszulcomplex},
\begin{equation}\label{eq:allthekoszulcomplexes}
    \begin{tikzcd}
	{K^\bullet_{\epsilon_1\epsilon_2}=K(R,f,g): s_{\epsilon_1\epsilon_2}^{2}R_{\epsilon_1\epsilon_2}} & {s_{\epsilon_1\epsilon_2}R_{\epsilon_1\epsilon_2}^2} & {R_{\epsilon_1\epsilon_2}.}
	\arrow["{(g,f)^t}", from=1-1, to=1-2]
	\arrow["{(f,g)}", from=1-2, to=1-3]
\end{tikzcd}
\end{equation}
See \Cref{fig:quadrants} for the corresponding CSS codes on the four quadrants.

With this definition, we will now define an algebraic analog of the topological order conditions considered in \Cref{sec:stabtileson2dplane}.  Namely, we say that the polynomials $f,g$ satisfy \emph{algebraic topological order} if for all $\epsilon_1,\epsilon_2\in \{+,-,\pm\}$ we have
\begin{enumerate}
    \item
    $H^{i}(K_{\epsilon_1\epsilon_2})=0$ for all $i\neq 0$ and
    \item the map $R_{\epsilon_1\epsilon_2}/(s_{\epsilon_1\epsilon_2}f,s_{\epsilon_1\epsilon_2}g)\to R/(f,g)$ is an isomorphism.
\end{enumerate}

\begin{remark}
\begin{enumerate}
    \item  The first condition holds if and only if $s_{\epsilon_1\epsilon_2}f,s_{\epsilon_1\epsilon_2}g$ form a Koszul-regular sequence in $R_{\epsilon_1\epsilon_2}$ or, equivalently, if $s_{\epsilon_1\epsilon_2}f,s_{\epsilon_1\epsilon_2}g$ have no common non-unit factors in $R_{\epsilon_1\epsilon_2}$.
    \item The second condition is equivalent to $x^{-\epsilon_1},y^{-\epsilon_2}\in \ftwo[x^{\epsilon_1},y^{\epsilon_2}]/(s_{\epsilon_1\epsilon_2}f,s_{\epsilon_1\epsilon_2}g).$
\end{enumerate}
\end{remark}

We will assume this algebraic topological order from now. In fact, one may show that this implies total topological order defined in \Cref{subsec: background: stabilizertileson2Dplane}.
One can verify that all tile codes found in \cite{steffan2025tilecodeshighefficiencyquantum} obey algebraic topological order.
\subsection{Resolving tile codes by Koszul complexes}\label{subsec: Resolving tile codes by Koszul complexes}
We now explain how the tile codes introduced in \Cref{subsec: tile codes} can be constructed via an `inclusion-exclusion principle' using the Koszul complexes on the unbounded infinite plane, half planes and quadrants, as well as shifts thereof.

For this, we will the various natural inclusions of the different Koszul complexes. For example, there is a natural map $y^kK_{++}\to K_{+,\pm}$ which maps the upper-right quadrant, shifted up by $k$ steps, into the right half plane.

In a first approximation, we consider codes supported on a horizontal strips, which are either bounded to the left, bounded to the right or unbounded in the horizontal direction. For example, the unbounded strip has the following form.

\begin{equation*}
\begin{tikzpicture}[scale = .5]
    \foreach \x in {0,...,10} {
                \draw[\latticecolor, \latticethickness] (\x/2,0) -- (\x/2,3.5);
            }
            
            % Horizontal lines
            \foreach \y in {0,...,6} {
                \draw[\latticecolor, \latticethickness] (-.5,\y/2) -- (5.5,\y/2);
            }

            %   \foreach \y in {0,...,4}{
            % \foreach \x in {-2,-1}{
            % \draw[fill = good_blue, draw = good_blue] (\x/2, \y/2) circle (0.1);
            % }
            % }
              \foreach \y in {0,...,4}{
            \foreach \x in {0,...,10}{
            \draw[fill = black] (\x/2, \y/2) circle (0.1);
            }
            }
              \foreach \y in {5,6}{
            \foreach \x in {0,...,10}{
            \draw[fill = good_red, draw = good_red] (\x/2, \y/2) circle (0.1);
            }
            }
                 \foreach \y in {-2,-1}{
            \foreach \x in {0,...,10}{
            \draw[fill = good_red, draw = good_red] (\x/2, \y/2) circle (0.1);
            }
            }
            \node[scale = .8] at (6.5, 1) {$\dots$};
            \node[scale = .8] at (-1.5, 1) {$\dots$};
\end{tikzpicture}                
\end{equation*}

Strip codes of height $M$ arise from the following short exact sequence of chain complexes
% https://q.uiver.app/#q=WzAsNSxbMSwwLCJLX3tzdHJpcH0iXSxbMiwwLCJLXlxcYnVsbGV0KFJfe1xccG0sK30sZixnKVxcb3BsdXMgeV57LTF9S15cXGJ1bGxldChSX3tcXHBtLC19LHleey1EfWYseV57LUR9ZykiXSxbMywwLCJLXlxcYnVsbGV0KFJfe1xccG0sXFxwbX0sZixnKSJdLFs0LDAsIjAiXSxbMCwwLCIwIl0sWzAsMV0sWzEsMl0sWzIsM10sWzQsMF1d
\begin{equation}\label{eq:shortexactsequenceofstripcodes}
    \begin{tikzcd}
	0 & {y^{M+D}K_{\epsilon,+}^\bullet\oplus y^{D-1}K_{\epsilon,-}^\bullet} & {K_{\epsilon,\pm}^\bullet} & {K_{\epsilon,\op{strip}}^\bullet} & 0.
	\arrow[from=1-1, to=1-2]
	\arrow[from=1-2, to=1-3]
	\arrow[from=1-3, to=1-4]
	\arrow[from=1-4, to=1-5]
\end{tikzcd}
\end{equation}

For example, the strip code bounded to the left, arises by removing two quadrants from a half-plane, as the following picture shows.
\begin{equation*}
    \begin{tikzpicture}[scale = .5]
    %first term of the chain complex. 
    \begin{scope}[shift = {(-7,-5)}]
            \foreach \x in {0,...,5} {
                \draw[\latticecolor, \latticethickness] (\x/2,-.25) -- (\x/2,1.5);
            }
            
            % Horizontal lines
            \foreach \y in {0,...,2} {
                \draw[\latticecolor, \latticethickness] (0,\y/2) -- (3,\y/2);
            }
            \foreach \y in {0,1,2}{
            \foreach \x in {-2,-1}{
            \draw[fill = good_blue, draw = good_blue] (\x/2, \y/2) circle (0.1);
            }
            }  
            \foreach \y in {1,2}{
            \foreach \x in {0,...,5}{
            \draw[fill = good_blue, draw = good_blue] (\x/2, \y/2) circle (0.1);
            }
            } 
            \foreach \y in {0}{
            \foreach \x in {0,...,5}{
            \draw[fill = black,] (\x/2, \y/2) circle (0.1);
            }
            } 
        \end{scope}

    \begin{scope}[shift = {(-7,0)}]
            \foreach \x in {0,...,5} {
                \draw[\latticecolor, \latticethickness] (\x/2,0) -- (\x/2,1.5);
            }
            
            % Horizontal lines
            \foreach \y in {0,...,2} {
                \draw[\latticecolor, \latticethickness] (0,\y/2) -- (3,\y/2);
            }
            \foreach \y in {0,1,2}{
            \foreach \x in {-2,-1}{
            \draw[fill = good_blue, draw = good_blue] (\x/2, \y/2) circle (0.1);
            }
            }  
             \foreach \y in {-2,-1}{
            \foreach \x in {-2,...,5}{
            \draw[fill = good_blue, draw = good_blue] (\x/2, \y/2) circle (0.1);
            }
            } 
            \foreach \y in {0,1,2}{
            \foreach \x in {0,...,5}{
            \draw[fill = black,] (\x/2, \y/2) circle (0.1);
            }
            } 
        \end{scope}
        \node[scale = .8] at (-6, 2.5) {$\vdots$};
        \node[scale = .8] at (-6, -5.75) {$\vdots$};
        \node[scale = .8] at (-3, 0.5) {$\dots$};
        \node[scale = .8] at (-3, -4.5) {$\dots$};
        \node at (-2,-2) {$\rightarrow$};

    %second term of the chain complex
        \begin{scope}[shift = {(0,-5)}]
            \foreach \x in {0,...,5} {
                \draw[\latticecolor, \latticethickness] (\x/2,-.25) -- (\x/2,6.5);
            }
            
            % Horizontal lines
            \foreach \y in {0,...,12} {
                \draw[\latticecolor, \latticethickness] (0,\y/2) -- (3,\y/2);
            }
             \foreach \y in {0,...,12}{
            \foreach \x in {-2,-1}{
            \draw[fill = good_blue, draw = good_blue] (\x/2, \y/2) circle (0.1);
            }
            }
              \foreach \y in {0,...,12}{
            \foreach \x in {0,...,5}{
            \draw[fill = black] (\x/2, \y/2) circle (0.1);
            }
            }
        \end{scope}

            \node at (5.5,-2) {$\rightarrow$};
        \node[scale = .8] at (2, -5.75) {$\vdots$};
        \node[scale = .8] at (2, 2.5) {$\vdots$};
        \node[scale = .8] at (4, -2) {$\dots$};
%third term of the chain complex
            \begin{scope}[shift = {(8,-3.5)}]
            \foreach \x in {0,...,5} {
                \draw[\latticecolor, \latticethickness] (\x/2,0) -- (\x/2,3.5);
            }
            
            % Horizontal lines
            \foreach \y in {0,...,6} {
                \draw[\latticecolor, \latticethickness] (0,\y/2) -- (3,\y/2);
            }

              \foreach \y in {0,...,4}{
            \foreach \x in {-2,-1}{
            \draw[fill = good_blue, draw = good_blue] (\x/2, \y/2) circle (0.1);
            }
            }
              \foreach \y in {0,...,4}{
            \foreach \x in {0,...,5}{
            \draw[fill = black] (\x/2, \y/2) circle (0.1);
            }
            }
              \foreach \y in {5,6}{
            \foreach \x in {0,...,5}{
            \draw[fill = good_red, draw = good_red] (\x/2, \y/2) circle (0.1);
            }
            }
                 \foreach \y in {-2,-1}{
            \foreach \x in {0,...,5}{
            \draw[fill = good_red, draw = good_red] (\x/2, \y/2) circle (0.1);
            }
            }
            \node[scale = .8] at (4, 1) {$\dots$};
        \end{scope}
        
    \end{tikzpicture}
\end{equation*}

We will now show how this resolution on strip codes allows to compute their logical operators.
\begin{lemma}\label{lem:logicalsofstripcodes}
    %Let $\epsilon\in \{+,-,\pm\}$. 
    Let $\epsilon\in \{+,-,\pm\}.$ Then $H^i(K_{\epsilon\op{strip}}^\bullet)=0$ for all $i\neq -1$ and there is an isomorphism
    $$\partial_{\op{bottom}}:H^{-1}(K_{\epsilon\op{strip}}^\bullet)\to R/(f,g)$$
    defined by
    \[\label{def:deltabottom}
    \partial_{\op{bottom}}([a,b])=\operatorname{tr_{y<0}}(af+bg)\in R/(f,g)
    \]
    for a representative $(a,b)$ of a cohomology class. Here, for a polynomial $h$, we denote by $\operatorname{tr_{y< 0}}(h)$ the polynomial that arises by setting all non-negative powers of $y$ in $h$ to zero.
\end{lemma}
\begin{proof}
    For ease of notation, assume that $\epsilon=\pm.$ The other cases follow by the same argument.
     The algebraic total order condition implies that $H^i(K_{\pm\epsilon'}^\bullet)=0$ for $i\neq 0$ and that the maps $H^{0}(K_{\pm\epsilon'}^\bullet)\to H^{0}(K_{\pm\pm}^\bullet)=R/(f,g)$ are isomorphism for all $\epsilon'\in \{+,-\}$. This implies that the long exact sequence of cohomology groups associated to the short exact sequence in \Cref{eq:shortexactsequenceofstripcodes} has the following form
% https://q.uiver.app/#q=WzAsNSxbMywwLCJIXnswfShLX3tcXHBtXFxwbX0pIl0sWzIsMCwiSF57MH0oS197XFxwbS19KVxcb3BsdXMgSF57MH0oS197XFxwbSt9KSJdLFsxLDAsIkheey0xfShLX3tcXHBtXFxvcHtzdHJpcH19KSJdLFswLDAsIjAiXSxbNCwwLCIwIl0sWzIsMSwiXFxkZWx0YSJdLFszLDJdLFswLDRdLFsxLDAsIlxccGkiXV0=
\[
\begin{tikzcd}
	0 & {H^{-1}(K_{\pm\op{strip}}^\bullet)} & {H^{0}(K_{\pm-}^\bullet)\oplus H^{0}(K_{\pm+}^\bullet)} & {H^{0}(K_{\pm\pm}^\bullet)} & 0
	\arrow[from=1-1, to=1-2]
	\arrow["\delta", from=1-2, to=1-3]
	\arrow["\pi", from=1-3, to=1-4]
	\arrow[from=1-4, to=1-5]
\end{tikzcd}
\label{eq:lesfromsesstripcode}
\]
and all other terms vanish. In particular, $H^i(K_{\epsilon\op{strip}}^\bullet)=0$ for all $i\neq -1$.

Since both components of the map $\pi$, namely $H^{0}(K^\bullet_{\pm-})\to H^{0}(K^\bullet_{\pm\pm})$ and $H^{0}(K_{\pm+}^\bullet)\to H^{0}(K^\bullet_{\pm\pm})$, are isomorphism by algebraic topological order, it follows that both components of the boundary map $\delta$ are isomorphism as well. 
In particular, the boundary map $\delta$ composed with the projection to $H^{0}(K_{\pm-})$ yields the desired isomorphism $$\partial_{\op{bottom}}: H^{-1}(K^\bullet_{\epsilon\op{strip}})\to H^{0}(K_{\pm-})\cong R/(f,g).\qedhere$$

\end{proof}
Finally, the tile code arises from the following short exact sequence
% https://q.uiver.app/#q=WzAsNSxbMiwwLCJLXlxcYnVsbGV0X3srXFxvcHtzdHJpcH19XFxvcGx1cyB4XntMK0R9S15cXGJ1bGxldF97LVxcb3B7c3RyaXB9fSJdLFszLDAsIktfe1xccG1cXG9we3N0cmlwfX0iXSxbMSwwLCJLXlxcYnVsbGV0X3tcXG9we3RpbGV9fSJdLFswLDAsIjAiXSxbNCwwLCIwIl0sWzAsMV0sWzIsMF0sWzEsNF0sWzMsMl1d
\begin{equation}
   \label{eq:sesdefiningtilecodes}
\begin{tikzcd}
	0 & {K^\bullet_{\op{tile}}} & {K^\bullet_{+\op{strip}}\oplus x^{L+D}K^\bullet_{-\op{strip}}} & {K_{\pm\op{strip}}^\bullet} & 0.
	\arrow[from=1-1, to=1-2]
	\arrow[from=1-2, to=1-3]
	\arrow[from=1-3, to=1-4]
	\arrow[from=1-4, to=1-5]
\end{tikzcd} 
\end{equation}
Visually, this means that the tile code is the intersection of the two (shifted) strip codes bounded to the left and right, respectively, inside the unbounded strip code.

\begin{equation*}
    \begin{tikzpicture}[scale = .4]
            \foreach \x in {0,...,5} {
                \draw[\latticecolor, \latticethickness] (\x/2,0) -- (\x/2,3);
            }
            
            % Horizontal lines
            \foreach \y in {0,...,5} {
                \draw[\latticecolor, \latticethickness] (0,\y/2) -- (3,\y/2);
            }
            
            % 3x3 grid of black dots
            \foreach \x in {0,...,3} {
                \foreach \y in {0,...,3} {
                    \draw[fill=black] (\x/2,\y/2) circle (0.1);
                }
            }
            
            % Red dots on top and bottom
            \foreach \x in {0,...,3} {
                \draw[fill=good_red, color = good_red] (\x/2,-0.5) circle (0.1);
                \draw[fill=good_red, color = good_red] (\x/2,-1) circle (0.1);
                \draw[fill=good_red, color = good_red] (\x/2,2) circle (0.1);
                \draw[fill=good_red, color = good_red] (\x/2,2.5) circle (0.1);
            }
            
            % Blue dots on left and right
            \foreach \y in {0,...,3} {
                \draw[fill=good_blue, color = good_blue] (-0.5,\y/2) circle (0.1);
                \draw[fill=good_blue, color = good_blue] (-1,\y/2) circle (0.1);
                \draw[fill=good_blue, color = good_blue] (2,\y/2) circle (0.1);
                \draw[fill=good_blue, color = good_blue] (2.5,\y/2) circle (0.1);
            }
        \node at (4.5,1) {$\rightarrow$};

        \begin{scope}[shift = {(7,2)}]
                      \foreach \x in {0,...,3} {
                \draw[\latticecolor, \latticethickness] (\x/2,0) -- (\x/2,3);
            }
            
            % Horizontal lines
            \foreach \y in {0,...,5} {
                \draw[\latticecolor, \latticethickness] (0,\y/2) -- (1.8,\y/2);
            }
            
            % 3x3 grid of black dots
            \foreach \x in {0,...,3} {
                \foreach \y in {0,...,3} {
                    \draw[fill=black] (\x/2,\y/2) circle (0.1);
                }
            }
            
            % Red dots on top and bottom
            \foreach \x in {0,...,3} {
                \draw[fill=good_red, color = good_red] (\x/2,-0.5) circle (0.1);
                \draw[fill=good_red, color = good_red] (\x/2,-1) circle (0.1);
                \draw[fill=good_red, color = good_red] (\x/2,2) circle (0.1);
                \draw[fill=good_red, color = good_red] (\x/2,2.5) circle (0.1);
            }
            
            % Blue dots on left and right
            \foreach \y in {0,...,3} {
                \draw[fill=good_blue, color = good_blue] (-0.5,\y/2) circle (0.1);
                \draw[fill=good_blue, color = good_blue] (-1,\y/2) circle (0.1);
                % \draw[fill=good_blue, color = good_blue] (2,\y/2) circle (0.1);
                % \draw[fill=good_blue, color = good_blue] (2.5,\y/2) circle (0.1);
            }
            \node[scale = .6] at (2.5,1) {$\dots$};
            \node at (3.5,-1.5) {$\oplus$};
        \end{scope}

        \begin{scope}[shift = {(13,-1)}]
                      \foreach \x in {0,...,5} {
                \draw[\latticecolor, \latticethickness] (\x/2,0) -- (\x/2,3);
            }
            
            % Horizontal lines
            \foreach \y in {0,...,5} {
                \draw[\latticecolor, \latticethickness] (-.3,\y/2) -- (3,\y/2);
            }
            
            % 3x3 grid of black dots
            \foreach \x in {0,...,3} {
                \foreach \y in {0,...,3} {
                    \draw[fill=black] (\x/2,\y/2) circle (0.1);
                }
            }
            
            % Red dots on top and bottom
            \foreach \x in {0,...,3} {
                \draw[fill=good_red, color = good_red] (\x/2,-0.5) circle (0.1);
                \draw[fill=good_red, color = good_red] (\x/2,-1) circle (0.1);
                \draw[fill=good_red, color = good_red] (\x/2,2) circle (0.1);
                \draw[fill=good_red, color = good_red] (\x/2,2.5) circle (0.1);
            }
            
            % Blue dots on left and right
            \foreach \y in {0,...,3} {
                % \draw[fill=good_blue, color = good_blue] (-0.5,\y/2) circle (0.1);
                % \draw[fill=good_blue, color = good_blue] (-1,\y/2) circle (0.1);
                \draw[fill=good_blue, color = good_blue] (2,\y/2) circle (0.1);
                \draw[fill=good_blue, color = good_blue] (2.5,\y/2) circle (0.1);
            }
            \node[scale = .6] at (-1,1) {$\dots$};

        \end{scope}

        \node at (17.5,1) {$\rightarrow$};

        \begin{scope}[shift = {(20,0)}]
    \foreach \x in {0,...,5} {
                \draw[\latticecolor, \latticethickness] (\x/2,0) -- (\x/2,3);
            }
            
            % Horizontal lines
            \foreach \y in {0,...,5} {
                \draw[\latticecolor, \latticethickness] (-.4,\y/2) -- (2.9,\y/2);
            }
            
            % 3x3 grid of black dots
            \foreach \x in {0,...,5} {
                \foreach \y in {0,...,3} {
                    \draw[fill=black] (\x/2,\y/2) circle (0.1);
                }
            }
            
            % Red dots on top and bottom
            \foreach \x in {0,...,5} {
                \draw[fill=good_red, color = good_red] (\x/2,-0.5) circle (0.1);
                \draw[fill=good_red, color = good_red] (\x/2,-1) circle (0.1);
                \draw[fill=good_red, color = good_red] (\x/2,2) circle (0.1);
                \draw[fill=good_red, color = good_red] (\x/2,2.5) circle (0.1);
            }
            \node[scale = .6] at (4,1) {$\dots$};
            \node[scale = .6] at (-1,1) {$\dots$};
            % Blue dots on left and right
            % \foreach \y in {0,...,3} {
            %     \draw[fill=good_blue, color = good_blue] (-0.5,\y/2) circle (0.1);
            %     \draw[fill=good_blue, color = good_blue] (-1,\y/2) circle (0.1);
            %     \draw[fill=good_blue, color = good_blue] (2,\y/2) circle (0.1);
            %     \draw[fill=good_blue, color = good_blue] (2.5,\y/2) circle (0.1);
            % }
        \end{scope}
    \end{tikzpicture}
\end{equation*}
This resolutions of tile codes via strip codes allows to identify the space of logicals of tile codes and strip codes.
\begin{lemma}\label{lem:logicalsoftilecodes}
The inclusion $K^\bullet_{\op{tile}}\to K^\bullet_{\pm\op{strip}}$ is a quasi-isomorphism, that is, it induces an isomorphism $H^i(K^\bullet_{\op{tile}})\to K^\bullet_{\pm\op{strip}}$ for all $i\in \mathbb{Z}.$
\end{lemma}
\begin{proof}
Using \Cref{lem:logicalsofstripcodes}, the inclusion $K^\bullet_{+\op{strip}}\to K^\bullet_{\pm\op{strip}}$ and $x^{L+D}K^\bullet_{+\op{strip}}\to K^\bullet_{\pm\op{strip}}$ are quasi-isomorphisms and the long exact sequence associated to the short exact sequence of chain complexes in \Cref{eq:sesdefiningtilecodes} is given by
    % https://q.uiver.app/#q=WzAsMTAsWzAsMCwiMCJdLFsxLDAsIkheey0xfShLXlxcYnVsbGV0X3tcXG9we3RpbGV9fSkiXSxbMiwwLCJLXlxcYnVsbGV0X3srXFxvcHtzdHJpcH19XFxvcGx1cyB4XntMK0R9S15cXGJ1bGxldF97LVxcb3B7c3RyaXB9fSJdLFszLDAsIktfe1xccG1cXG9we3N0cmlwfX1eXFxidWxsZXQiXSxbNCwwLCIwIl0sWzEsMSwiUi8oZixnKSJdLFsyLDEsIlIvKGYsZylcXG9wbHVzIFIvKGYsZykiXSxbMywxLCJSLyhmLGcpIl0sWzQsMSwiMCJdLFswLDEsIjAiXSxbMCwxXSxbMSwyXSxbMiwzXSxbMyw0XSxbMSw1LCJcXHdyIl0sWzUsNiwiKFxcaWQsXFxpZCledCJdLFsyLDYsIlxcd3IiXSxbMyw3LCJcXHdyIl0sWzYsNywiKFxcaWQsXFxpZCkiXSxbOSw1XSxbNyw4XV0=
\[\begin{tikzcd}[column sep = small]
	0 & {H^{-1}(K^\bullet_{\op{tile}})} & {H^{-1}(K^\bullet_{+\op{strip}})\oplus H^{-1}(x^{L+D}K^\bullet_{-\op{strip}}}) & {H^{-1}(K_{\pm\op{strip}}^\bullet)} & 0 \\
	0 & {R/(f,g)} & {R/(f,g)\oplus R/(f,g)} & {R/(f,g)} & 0
	\arrow[from=1-1, to=1-2]
	\arrow[from=1-2, to=1-3]
	\arrow["\wr", from=1-2, to=2-2]
	\arrow[from=1-3, to=1-4]
	\arrow["\wr", from=1-3, to=2-3]
	\arrow[from=1-4, to=1-5]
	\arrow["\wr", from=1-4, to=2-4]
	\arrow[from=2-1, to=2-2]
	\arrow["{(\id,\id)^t}", from=2-2, to=2-3]
	\arrow["{(\id,\id)}", from=2-3, to=2-4]
	\arrow[from=2-4, to=2-5]
\end{tikzcd}\]
where the right vertical maps are given by the map $\partial_{\op{bottom}}$ as defined in \Cref{lem:logicalsofstripcodes}.
\end{proof}
Putting together \Cref{lem:logicalsofstripcodes} and \Cref{lem:logicalsoftilecodes}, we obtain the following explicit description of the space of logicals of the tile code.
\begin{theorem}\label{cor:logicalsoftilecode}
    We have $H^i(K^\bullet_{\op{tile}})=0$ for all $i\neq -1$ and the map 
    $\partial_{\op{bottom}}: H^{-1}(K^\bullet_{\op{tile}})\to R/(f,g)$
    defined as in \Cref{def:deltabottom} is an isomorphism.
\end{theorem}
With this result, we can immediately compute the logical dimension.
\begin{theorem}\label{thm:logicaldimensiontilecode}
The logical dimension of a tile code is given by
$$\dim_{\mathbb{F}_2}H^{-1}(K^\bullet_{\op{tile}})=\dim_{\mathbb{F}_2}R/(f,g)=2D^2.$$
\end{theorem}
\begin{proof}
    Since $H^{-2}(K^\bullet_{\op{tile}})=H^{0}(K^\bullet_{\op{tile}})=0,$ the logical dimension is given by
    \begin{align*}
        \dim_{\mathbb{F}_2}H^{-1}(K^\bullet_{\op{tile}})&=\dim_{\mathbb{F}_2}K^{-1}_{\op{tile}}-\dim_{\mathbb{F}_2}K^{0}_{\op{tile}}-\dim_{\mathbb{F}_2}K^{-2}_{\op{tile}}\\
        &=2LM-(L+D)(M-D)-(L-D)(M+D)
        =2D^2
    \end{align*}
    as claimed.
\end{proof}

The effect of the map $\partial_{\text{bottom}}$ on an $\op X$-logical $P$ in a tile code can be visualised as follows. First, one embeds the logical operator into a the infinite grid. Second, one computes the boundary exitation $\partial(P)$, that is, the set of coordinates of $\op Z$-stabilizers in the infinite grid that anti-commute with $P$. These exitation are localized along the top and bottom boundary of the tile code, as the following picture visualizes.
\begin{equation*}
    \begin{tikzpicture}[scale = .7]
            \foreach \x in {0,...,5} {
                \draw[\latticecolor, \latticethickness] (\x/2,0) -- (\x/2,3);
            }
            
            % Horizontal lines
            \foreach \y in {0,...,5} {
                \draw[\latticecolor, \latticethickness] (0,\y/2) -- (3,\y/2);
            }
            \draw[color = good_red, line width = \supportthickness] (0,0) -- (.5,0);

            \draw[color = good_red, line width = \supportthickness] (0,1) -- (0,1.5);

            \draw[color = good_red, line width = \supportthickness] (0,1.5) -- (0,2);

            \draw[color = good_red, line width = \supportthickness] (0.5,1) -- (1,1);
            
            \draw[color = good_red, line width = \supportthickness] (0.5,1.5) -- (1,1.5);

            \draw[color = good_red, line width = \supportthickness] (0.5,1.5) -- (0.5,2);

            \draw[color = good_red, line width = \supportthickness] (0,2) -- (.5,2);

            \draw[color = good_red, line width = \supportthickness] (0,2.5) -- (.5,2.5);

            \draw[color = good_red, line width = \supportthickness] (0.5,2.5) -- (1,2.5);

            \draw[color = good_red, line width = \supportthickness] (0.5,2.5) -- (0.5,3);
            % 3x3 grid of black dots
            \foreach \x in {0,...,3} {
                \foreach \y in {0,...,3} {
                    \draw[fill=black] (\x/2,\y/2) circle (0.1);
                }
            }
            
            % Red dots on top and bottom
            \foreach \x in {0,...,3} {
                \draw[fill=good_red, color = good_red] (\x/2,-0.5) circle (0.1);
                \draw[fill=good_red, color = good_red] (\x/2,-1) circle (0.1);
                \draw[fill=good_red, color = good_red] (\x/2,2) circle (0.1);
                \draw[fill=good_red, color = good_red] (\x/2,2.5) circle (0.1);
            }
            
            % Blue dots on left and right
            \foreach \y in {0,...,3} {
                \draw[fill=good_blue, color = good_blue] (-0.5,\y/2) circle (0.1);
                \draw[fill=good_blue, color = good_blue] (-1,\y/2) circle (0.1);
                \draw[fill=good_blue, color = good_blue] (2,\y/2) circle (0.1);
                \draw[fill=good_blue, color = good_blue] (2.5,\y/2) circle (0.1);
            }

            \node[scale = 1.2] at (5.5,1) {$\mapsto$};

        \begin{scope}[shift = {(9,0)}]
            \foreach \x in {-2,...,7} {
                \draw[\latticecolor, very thin] (\x/2,-1.5) -- (\x/2,4);
            }
            
            % Horizontal lines
            \foreach \y in {-2,...,7} {
                \draw[\latticecolor, very thin] (-1.5,\y/2) -- (4,\y/2);
            }
        
            \foreach \x in {0,...,5} {
                \draw[\latticecolor, \latticethickness] (\x/2,0) -- (\x/2,3);
            }
            
            % Horizontal lines
            \foreach \y in {0,...,5} {
                \draw[\latticecolor, \latticethickness] (0,\y/2) -- (3,\y/2);
            }

            \draw[color = good_red, line width = \supportthickness] (0,0) -- (.5,0);

            \draw[color = good_red, line width = \supportthickness] (0,1) -- (0,1.5);

            \draw[color = good_red, line width = \supportthickness] (0,1.5) -- (0,2);

            \draw[color = good_red, line width = \supportthickness] (0.5,1) -- (1,1);
            
            \draw[color = good_red, line width = \supportthickness] (0.5,1.5) -- (1,1.5);

            \draw[color = good_red, line width = \supportthickness] (0.5,1.5) -- (0.5,2);

            \draw[color = good_red, line width = \supportthickness] (0,2) -- (.5,2);

            \draw[color = good_red, line width = \supportthickness] (0,2.5) -- (.5,2.5);

            \draw[color = good_red, line width = \supportthickness] (0.5,2.5) -- (1,2.5);

            \draw[color = good_red, line width = \supportthickness] (0.5,2.5) -- (0.5,3);

    %top syndrome
            \draw[fill = good_blue, draw = good_blue] (-.5,2.5) circle (0.1);
            \draw[fill = good_blue, draw = good_blue] (0,2.5) circle (0.1);
            \draw[fill = good_blue, draw = good_blue] (0.5,2.5) circle (0.1);

            \draw[fill = good_blue, draw = good_blue] (.5,2) circle (0.1);

%bottom syndrome
            \draw[fill = good_blue, draw = good_blue] (0,-.5) circle (0.1);
            \draw[fill = good_blue, draw = good_blue] (-1,-1) circle (0.1);

            \draw[ draw = good_blue] (0,-.5) circle (0.15);
            \draw[ draw = good_blue] (-1,-1) circle (0.15);

        \end{scope}
        
    \end{tikzpicture}
\end{equation*}
Third, we disregard the top boundary exitations.
The remaining bottom boundary exitations can then the interpreted as a Laurent polynomial $\partial_{\text{bottom}}(P)\in R.$ To make this construction invariant under adding $\op X$-stabilizers to $P$ we mod out the ideal generated by $f,g$ and obtain $\partial_{\text{bottom}}(P)\in R/(f,g).$

Analogously to the relations between strip codes extending in the horizontal direction, one may also consider vertical strip codes, say  $K^\bullet_{\pm\op{strip}}.$ Then the projection map $K^\bullet_{\pm\op{strip}}\to K_{\op{tile}}$ which sends all qubits and checks to zero that are not in the tile code is a quasi-isomorphism as well. This can be shown by a similar argument.

\begin{remark}
Combining the two short exact sequences in \Cref{eq:shortexactsequenceofstripcodes} and in \Cref{eq:sesdefiningtilecodes} yields the following double complex
% https://q.uiver.app/#q=WzAsMyxbMCwwLCJDXjAiXSxbMSwwLCJDXjEiXSxbMiwwLCJDXjIiXSxbMCwxXSxbMSwyXV0=
\[\begin{tikzcd}
	{C^{\bullet,\bullet}: C^{0,\bullet}} & {C^{1,\bullet}} & {C^{2,\bullet}}
	\arrow[from=1-1, to=1-2]
	\arrow[from=1-2, to=1-3]
\end{tikzcd}\]
built from the Koszul complexes on the quadrants, half planes and the unbounded infinite plane
\begin{align*}
    C^{0,\bullet}&=y^{M+D}K_{++}^\bullet\oplus x^{L+D}y^{M+D}K_{-+}^\bullet\oplus y^{D-1}K_{+-}^\bullet\oplus x^{L+D}y^{D-1}K_{--}^\bullet,\\
    C^{1,\bullet}&=y^{M+D}K_{\pm+}^\bullet\oplus y^{D-1}K_{\pm-}^\bullet\oplus K_{+\pm}^\bullet\oplus x^{L+D}K_{-\pm}^\bullet\text{ and}\\
    C^{2,\bullet}&=K_{\pm,\pm}^\bullet.
\end{align*}
where the differentials in the vertical direction increasing the first index in $C^{\bullet,\bullet}$ are given by the natural inclusion maps and the horizontal differentials are given by the differentials in the Koszul complexes.
The tile code complex $K^\bullet_{\op{tile}}$ arises as the vertical cohomology of the double complex $C^{\bullet,\bullet}$, namely,
\begin{align*}
    H_{\to}^i(C^{\bullet,\bullet})=\begin{cases}
    K^\bullet_{\op{tile}}&\text{ if }i=1\text{ and }\\
    0&\text{ otherwise.}
    \end{cases}
\end{align*}
Here, by $H_{\to}^i(C^{\bullet,\bullet})$ we denote the complex that arises by taking the $i$-th cohomology along the vertical differential in the double complex $C^{\bullet,\bullet}.$

In this way, the double complex $C^{\bullet,\bullet}$ can be seen as a `resolution' of the tile code complex. In fact, the double complex $C^{\bullet,\bullet}$ naturally arises as the Čech complex computing the higher global section of a Koszul complex on $\mathbb{P}^1\times \mathbb{P}^1$ that we will discuss in the next section.

\end{remark}
\section{Geometry of tile codes}\label{sec:geometryoftilecodes}
We will now explain how tile codes arise naturally when considering Koszul complex on the variety $\mathbb{P}^1 \times \mathbb{P}^1$. We assume some familiarity with algebraic geometry.
\subsection{Line bundles and Koszul complexes on $\mathbb{P}^1 \times \mathbb{P}^1$}
We first introduce some basic facts and notations for line bundles on $\mathbb{P}^1.$ Since we are interested in CSS codes, so chain complexes over $\ftwo$, we always denote by $\mathbb{P}^1=\mathbb{P}^1_{\ftwo}$ the projective line over $\ftwo$.

Recall that (up to isomorphism) all line bundles on $\mathbb{P}^1$ are of the form $\mathcal{O}(n)$ for $n\in \mathbb{Z}$ where $\mathcal{O}(0)=\mathcal{O}$ is the trivial line bundle.
The global sections on the line bundle $\mathcal{O}(n)$ arise as rational functions on $\mathbb{P}^1$ that are defined everywhere but at $\infty$ and are allowed to have a pole of order of up to $n$ at $\infty$, or in other words, polynomials of degree bounded by $n$,
\begin{align}\label{eq:sectionsp1}
    \Gamma(\mathbb{P}^1,\mathcal{O}(n))=\begin{cases}
        \ftwo[x]_{\leq n} & \text{ for }n\geq 0 \text{ and}\\
        0&\text{ otherwise.}
    \end{cases}
\end{align}
While the line bundles $\mathcal{O}(n)$ for $n<0$ have no non-trivial global sections, they have (except for $n=-1$) higher global sections given by
\begin{align}\label{eq:highersectionsp1}
    R^1\Gamma(\mathbb{P}^1,\mathcal{O}(n))=\begin{cases}
        (\ftwo[x]_{\leq -n-2})^* & \text{ for }n<-1 \text{ and}\\
        0&\text{ otherwise.}
    \end{cases}
\end{align}
Moreover, all other higher global sections vanish 
$$R^i\Gamma(\mathbb{P}^1,\mathcal{O}(n))=0 \text{ for all } i>1.$$

We will now explain how this generalizes to $\mathbb{P}^1\times \mathbb{P}^1$. Here, we will use the variables $x$ and $y$ for the first and second copy of $\mathbb{P}^1$, respectively.

Line bundles on $\mathbb{P}^1\times \mathbb{P}^1$ arise from line bundles on each copy of $\mathbb{P}^1$ and are given by $\mathcal{O}(n_1,n_2)=\mathcal{O}(n_1)\boxtimes \mathcal{O}(n_2)$ for two integers $n_1,n_2.$ The (higher) sections of these line bundles can be computed via the Künneth formula
\begin{align}\label{eq:kuennethp1xp1}
    R^i\Gamma(\mathbb{P}^1\times \mathbb{P}^1,\mathcal{O}(n_1,n_2))=\bigoplus_{\substack{0\leq p,q\leq 1\\ p+q=i}}R^p\Gamma(\mathbb{P}^1,\mathcal{O}(n_1))\otimes_{\ftwo} R^q\Gamma(\mathbb{P}^1,\mathcal{O}(n_2)),
\end{align}
where $R^0\Gamma=\Gamma.$
For example, if $D\geq 0,$ then we obtain polynomials in $x,y$ of maximal degree $D$ in both $x$ and $y$ as global sections 
\begin{align*}
    \Gamma(\mathbb{P}^1\times \mathbb{P}^1,\mathcal{O}(D,D))=\ftwo[x]_{\leq D}\otimes_{\ftwo}\ftwo[y]_{\leq D}=\ftwo[x,y]_{\leq D,\leq D}.
\end{align*}
Morphisms between line bundles can be computed via
\begin{align*}
    \operatorname{Hom}(\mathcal{O}(n_1,n_2),\mathcal{O}(m_1,m_2))&=\operatorname{Hom}(\mathcal{O},\mathcal{O}(m_1-n_1,m_2-n_2))\\&=\Gamma(\mathbb{P}^1\times \mathbb{P}^1,\mathcal{O}(m_1-n_1,m_2-n_2)).
\end{align*}
In particular, if we choose $f,g\in \ftwo[x,y]_{\leq D,\leq D},$ we obtain the Koszul complex
% https://q.uiver.app/#q=WzAsMyxbMSwwLCJcXG1hdGhjYWx7T30oLUQsLUQpXjIiXSxbMiwwLCJcXG1hdGhjYWx7T30iXSxbMCwwLCJLXlxcYnVsbGV0KFxcbWF0aGJie1B9XjEsZixnKTpcXG1hdGhjYWx7T30oLTJELC0yRCkiXSxbMCwxLCIoZixnKSJdLFsyLDAsIihnLGYpXnQiXV0=
\begin{equation}\label{eq:koszulcomplexonp1xp1}
    \begin{tikzcd}
	{\mathcal{K}^\bullet(\mathbb{P}^1\times \mathbb{P}^1,f,g):\mathcal{O}(-2D,-2D)} & {\mathcal{O}(-D,-D)^2} & {\mathcal{O}.}
	\arrow["{(g,f)^t}", from=1-1, to=1-2]
	\arrow["{(f,g)}", from=1-2, to=1-3]
\end{tikzcd}
\end{equation}
concentrated in degrees $-2,-1,0.$ 

If we identify the vector bundles with their sheaves of sections, we can consider the cohomology sheaves of the Koszul complex. For example, the $0$-th cohomology sheaf is
\begin{align*}
    H^0(\mathcal{K}^\bullet(\mathbb{P}^1,f,g))=\mathcal{O}/(f,g),
\end{align*}
the sheaf of functions on the joint vanishing set of the polynomials $f,g$ in $\mathbb{P}^1\times \mathbb{P}^1.$
\subsection{Local Sections and codes on infinite lattices}
We know explain how the Koszul complexes in \Cref{eq:allthekoszulcomplexes} considered in \Cref{sec:infinitelatticekoszul} can be recovered from the Koszul complex  of vector bundles on $\mathbb{P}^1\times \mathbb{P}^1$ (\Cref{eq:koszulcomplexonp1xp1}). For this, we consider the following open affine subsets of the projective line  $$\mathbb{A}^1_+=\mathbb{P}^1-\{\infty\},\, \mathbb{A}^1_-=\mathbb{P}^1-\{0\}\text{ and } \mathbb{A}^1_\pm=\mathbb{G}_m=\mathbb{A}^1_+\cap \mathbb{A}^1_-=\mathbb{P}^1-\{0,\infty\}.$$
Here $\mathbb{A}^1_+$ and $\mathbb{A}^1_-$ are copies of the affine line $\mathbb A^1.$

The space of sections into the line bundles $\mathcal{O}(n)$ over these subsets is given by
\begin{align}\label{sec:localsectionsoflinebundlesonp1}
    \Gamma(\mathbb{A}^1_+,\mathcal{O}(n))=\ftwo[x],\,
    \Gamma(\mathbb{A}^1_-,\mathcal{O}(n))=x^n\ftwo[x^{-1}]\text{ and }
    \Gamma(\mathbb{A}^1_\pm,\mathcal{O}(n))=\ftwo[x^\pm]
\end{align}
so that the sections over $\mathbb{P}^1$ arise as intersection
\begin{align*}
    \Gamma(\mathbb{P}^1,\mathcal{O}(n))&=\Gamma(\mathbb{A}^1_+,\mathcal{O}(n))\cap 
    \Gamma(\mathbb{A}^1_-,\mathcal{O}(n))=\ftwo[x]\cap x^n\ftwo[x^{-1}]. 
\end{align*}

By taking products, we obtain open affine subsets of $\mathbb{P}^1\times \mathbb{P}^1$ of the form 
$$\mathbb{A}^2_{\epsilon_1\epsilon_2}=\mathbb{A}^1_{\epsilon_1}\times \mathbb{A}^1_{\epsilon_2}\subset \mathbb{P}^1\times \mathbb{P}^1$$
for $\epsilon_i\in \{+,-,\pm\}.$ The spaces of sections into the line bundles $\mathcal{O}(n_1,n_2)$ can be computed via \Cref{sec:localsectionsoflinebundlesonp1} and
$$\Gamma(\mathbb{A}^2_{\epsilon_1\epsilon_2}, \mathcal{O}(n_1,n_2))=\Gamma(\mathbb{A}^1_{\epsilon_1},\mathcal{O}(n_1))\otimes_{\ftwo}\Gamma(\mathbb{A}^1_{\epsilon_2},\mathcal{O}(n_2))$$
In particular, we obtain the various (Laurent)-polynomial rings defined in \Cref{eq:defoflaurentpolynomials} as functions over these subspaces
$R_{\epsilon_1\epsilon_2}=\Gamma(\mathbb{A}^2_{\epsilon_1\epsilon_2}, \mathcal{O})$
and the Koszul complexes in \Cref{eq:allthekoszulcomplexes} arise as sections of the Koszul complex in \Cref{eq:koszulcomplexonp1xp1}, 
\begin{equation}
    K_{\epsilon_1\epsilon_2}^\bullet=\Gamma(\mathbb{A}^2_{\epsilon_1\epsilon_2},\mathcal{K}^\bullet(\mathbb{P}^1\times \mathbb{P}^1,f,g)).
\end{equation}
Using these facts, one shows that the algebraic topological order condition is equivalent to the following 
\emph{geometric total topologial order} condition that
\begin{enumerate}
    \item $H^i(\mathcal K^\bullet(\mathbb{P}^1,f,g))=0$ for $i\neq 0$ and
    \item the restriction map $\Gamma(\mathbb{P}^1\times \mathbb{P}^1,\mathcal{O}/(f,g))\to \Gamma(\mathbb{A}^2_{\pm\pm},\mathcal{O}/(f,g))=R/(f,g)$ is an isomorphism.
\end{enumerate}
The second condition is equivalent to the condition that the $f$ and $g$ do not vanish simultaneously at either $x=0,\infty$ or $y=0,\infty$, which is ensured by the condition in \Cref{eq:stabilizertilecorners}.
\subsection{Tile codes as higher sections}
We now explain how tile codes arise naturally as higher sections of a shift of the Koszul complex $\mathcal{K}^\bullet(\mathbb{P}^1\times \mathbb{P}^1,f,g)$, see \Cref{eq:koszulcomplexonp1xp1}. For this, for integers $L,M$ defining the size of the tile code, we consider the shifted Koszul complex $$\mathcal{K}_{\op{tile}}:=\mathcal{K}^\bullet(\mathbb{P}^1\times \mathbb{P}^1,f,g)\otimes \mathcal{S}$$
where $\mathcal{S}=\mathcal O(L-1+D,-M-1+D).$
The resulting complex is explicitly given by
% https://q.uiver.app/#q=WzAsMyxbMSwwLCJcXG1hdGhjYWx7T30oTC0xLC1NLTEpXjIiXSxbMiwwLCJcXG1hdGhjYWx7T30oTC0xK0QsLU0tMStEKSJdLFswLDAsIlxcbWF0aGNhbHtPfShMLTEtRCxNLTEtRCkiXSxbMCwxLCIoZixnKSJdLFsyLDAsIihnLGYpXnQiXV0=
\[\begin{tikzcd}
	{\mathcal{O}(L-1-D,-M-D)} & {\mathcal{O}(L-1,-M)^2} & {\mathcal{O}(L-1+D,-M+D) }
	\arrow["{(g,f)^t}", from=1-1, to=1-2]
	\arrow["{(f,g)}", from=1-2, to=1-3]
\end{tikzcd}\]
where we apologize for the unwieldy numbers.

Using the Künneth formula, see \Cref{eq:kuennethp1xp1}, and computation of higher sections of line bundles on $\mathbb{P}^1$, see \Cref{eq:sectionsp1} and \Cref{eq:highersectionsp1}, one can check that the complex $K^\bullet_{\op{tile}}$ arises as the higher global $R^1\Gamma(\mathbb{P}^1\times\mathbb{P}^1,-)$ of the entries of the shifted Koszul complex $\mathcal{K}_{\op{tile}}$, 
% https://q.uiver.app/#q=WzAsNixbMCwwLCJSXjFcXEdhbW1hKFxcbWF0aGJie1B9XjFcXHRpbWVzXFxtYXRoYmJ7UH1eMSxcXG1hdGhjYWx7S31eey17Mn19X3tcXG9we3RpbGV9fSkiXSxbMCwxLCJLXnstMn1fe1xcb3B7dGlsZX19Il0sWzEsMCwiUl4xXFxHYW1tYShcXG1hdGhiYntQfV4xXFx0aW1lc1xcbWF0aGJie1B9XjEsXFxtYXRoY2Fse0t9XnstezF9fV97XFxvcHt0aWxlfX0pIl0sWzIsMCwiUl4xXFxHYW1tYShcXG1hdGhiYntQfV4xXFx0aW1lc1xcbWF0aGJie1B9XjEsXFxtYXRoY2Fse0t9XnswfV97XFxvcHt0aWxlfX0pIl0sWzEsMSwiS157LTF9X3tcXG9we3RpbGV9fSJdLFsyLDEsIkteezB9X3tcXG9we3RpbGV9fSJdLFswLDEsIiIsMCx7ImxldmVsIjoyLCJzdHlsZSI6eyJoZWFkIjp7Im5hbWUiOiJub25lIn19fV0sWzAsMl0sWzIsM10sWzEsNF0sWzQsNV0sWzIsNCwiIiwxLHsibGV2ZWwiOjIsInN0eWxlIjp7ImhlYWQiOnsibmFtZSI6Im5vbmUifX19XSxbMyw1LCIiLDIseyJsZXZlbCI6Miwic3R5bGUiOnsiaGVhZCI6eyJuYW1lIjoibm9uZSJ9fX1dXQ==
\[
\label{eq:tileequalhighersections}
\begin{tikzcd}
	{R^1\Gamma(\mathbb{P}^1\times\mathbb{P}^1,\mathcal{K}^{-{2}}_{\op{tile}})} & {R^1\Gamma(\mathbb{P}^1\times\mathbb{P}^1,\mathcal{K}^{-{1}}_{\op{tile}})} & {R^1\Gamma(\mathbb{P}^1\times\mathbb{P}^1,\mathcal{K}^{0}_{\op{tile}})} \\
	{K^{-2}_{\op{tile}}} & {K^{-1}_{\op{tile}}} & {K^{0}_{\op{tile}}}
	\arrow[from=1-1, to=1-2]
	\arrow[equals, from=1-1, to=2-1]
	\arrow[from=1-2, to=1-3]
	\arrow[equals, from=1-2, to=2-2]
	\arrow[equals, from=1-3, to=2-3]
	\arrow[from=2-1, to=2-2]
	\arrow[from=2-2, to=2-3]
\end{tikzcd}
\]
We observe that the choice of the line bundle $\mathcal{S}$ determines the $L\times M$-shape of the tile code as well as the positioning of the `smooth' and `rough' boundaries according to the positive and negative sign in front of $L$ and $M.$

With this geometric approach, one can easily recover the computations of logicals of tile codes from \Cref{cor:logicalsoftilecode}. The geometric total topological order condition implies that the complex $\mathcal{K}_{\op{tile}}$ is quasi-isomorphic to the complex $\mathcal{O}/(f,g)\otimes \mathcal{S}$ concentrated in degree $0.$

By the geometric total topological order condition, we obtain that
\begin{align*}
    R^n\Gamma(\mathbb{P}^1\times \mathbb{P}^1,\mathcal{O}/(f,g)\otimes\mathcal{S})
    &=
    R^n\Gamma(\mathbb{A}^2_{\pm\pm},\mathcal{O}/(f,g) \otimes\mathcal{S})\\
&=R^n\Gamma(\mathbb{A}^2_{\pm\pm},\mathcal{O}/(f,g))
=\begin{cases}
    R/(f,g)&\text{ for } n=0\text{ and}\\
    0&\text{ otherwise.}
\end{cases}
\end{align*}
Here the second equality uses that all line bundles on $\mathbb{A}^2_{\pm\pm}$ are trivial and the higher sections vanish since $\mathbb{A}^2_{\pm\pm}$ is affine.

We can now consider the spectral sequence
$$E_1^{p,q}=R^q\Gamma(\mathbb{P}^1\times\mathbb{P}^1,\mathcal{K}^{p}_{\op{tile}})\Rightarrow R^{p+q}(\mathbb{P}^1\times\mathbb{P}^1,\mathcal{O}/(f,g)\otimes\mathcal{S})$$
associated to the stupid filtration of the complex $\mathcal{K}_{\op{tile}}$ which compares the higher sections of the terms of the complex $\mathcal{K}_{\op{tile}}$ to the higher sections of the complex itself, see \cite{stacks-project-tag-015X}.

Now, using the computations of higher global sections in \Cref{eq:sectionsp1}, \Cref{eq:highersectionsp1} and \Cref{eq:kuennethp1xp1}, one obtains that $E_1^{p,q}=0$ for $p\neq 0$ and $q\neq -2,-1,0$. In fact, the non-trivial terms on the first page $E_1$ together with the differential are exactly equal to the complex $K^\bullet_{\op{tile}}$, see \Cref{eq:tileequalhighersections}.

This shows that the spectral sequence degenerates at page $2$, which contains the cohomology groups of the tile code complex. We hence obtain in total
$$H^{q}(K^\bullet_{\op{tile}})=E_2^{1,q}=R^{1+q}\Gamma(\mathbb{P}^1\times\mathbb{P}^1,\mathcal{O}/(f,g)\otimes\mathcal{S})=\begin{cases}
    R/(f,g)&\text{ for } q=-1\text{ and}\\
    0&\text{ otherwise.}
\end{cases}$$
\begin{remark}
    In \Cref{thm:logicaldimensiontilecode}, we show that the logical dimension of tile codes is $2D^2.$ By this, we essentially reproved Bezout's theorem which states that the number of intersection points of curves described by equations $f=0$ and $g=0$ in $\mathbb{P}^1\times \mathbb{P}^1$ can be computed as
    \begin{align*}
    \dim H^0(\mathbb{P}^1\times \mathbb{P}^1,\mathcal{O}/(f,g))= \deg_{\op X}f\deg_yg+\deg_yf\deg_{\op X}g.
    \end{align*}
    Here, $\deg_{\op X}$ and $\deg_y$ denotes the degree of $f$ and $g$ in the variables $x$ and $y,$ respectively. By our assumption, all these degrees are equal to $D$ and we obtain a logical dimension of $2D^2.$
\end{remark}
\subsection{Increasing the dimension}

The geometric interpretation of tile codes as higher global section in a shifted Koszul complex on $\mathbb{P}^1\times \mathbb{P}^1$ opens the way to a systematic construction of open boundary versions of  translation invariant systems in any dimension, say $n\geq 0$.

For example, we can construct three-dimensional tile codes using $X=\mathbb{P}^1\times \mathbb{P}^1\times \mathbb{P}^1.$ 
The shape of the stabilizers will be described by three polynomials $$f,g,h\in \ftwo[x,y,z]_{\leq D,\leq D,\leq D}=\Gamma(X,\mathcal{O}(D,D,D))$$
To this, we can associate a three-term Koszul complex in degrees $-3,\dots, 0$ given by
\[
\begin{tikzcd}[ampersand replacement=\&]
    {\mathcal{K}(X,f,g,h): \mathcal{L}^{\otimes 3}} \& {(\mathcal{L}^{\otimes 2})^3} \& {\mathcal{L}^3} \& {\mathcal{O}}
    \arrow["{\begin{psmallmatrix} g \\ f \\ h \end{psmallmatrix}}", from=1-1, to=1-2]
    \arrow["{\begin{psmallmatrix} h & 0 & g \\ 0 & h & f \\ f & g & 0 \end{psmallmatrix}}", from=1-2, to=1-3]
    \arrow["{\begin{psmallmatrix} f & g & h \end{psmallmatrix}}", from=1-3, to=1-4]
\end{tikzcd}
\]
where we denote the line bundle $\mathcal{L}=\mathcal{O}(-D,-D,-D).$ 
We choose to put $\op X$-stabilizers, qubits and $\op Z$-stabilizers in degrees $-2,-1$ and $0$, respectively. The shape of the layout of the resulting code can be determined by the choice of an additional line bundle $\mathcal{S}$, so that we consider the complex of vector bundles
$$\mathcal{K}_{3D\op{-tile}}=\mathcal{K}(X,f,g,h)\otimes \mathcal{S}$$ and define the three dimensional tile code via the space of higher sections

\[\begin{tikzcd}
	{K^{\bullet}_{3D\op{-tile}}:R^1\Gamma(X,\mathcal{K}^{-{2}}_{3D\op{-tile}})} & {R^1\Gamma(X,\mathcal{K}^{-{1}}_{3D\op{-tile}})} & {R^1\Gamma(X,\mathcal{K}^{0}_{3D\op{-tile}}).}
	\arrow[from=1-1, to=1-2]
	\arrow[from=1-2, to=1-3]
\end{tikzcd}\]
We choose the line bundle $\mathcal{S}$ such that $R^i\Gamma(X,\mathcal{S})$ is only non-zero for $i=1.$
For example, for
$$\mathcal{S}=\mathcal{O}(L-1+D, M-1+D, -N+D)$$
we will obtain a code with qubits supported on the $3LMN$ edges of a cube with $\op Z$-boundaries on the boundaries in the $x,y$-directions and $\op X$-boundaries in the $z$-direction, corresponding to the choice of signs. When we assume the appropriate generalization of geometric total topological order, the space of logicals of the code can be computed as
\[
    H^{-1}(K^{\bullet}_{3D\op{-tile}})\cong \ftwo[x^{\pm1},y^{\pm1},z^{\pm1}]/(f,g,h)
\]
and one may show that the dimension of the space of logicals is $6D^3.$

Similarly, we can construct four-dimensional tile codes using $X=\mathbb{P}^1\times \mathbb{P}^1\times
  \mathbb{P}^1\times \mathbb{P}^1.$
  The shape of the stabilizers will be described by four polynomials $$f,g,h,i\in \ftwo[w,x,y,z]_{\leq 
  D,\leq D,\leq D,\leq D}=\Gamma(X,\mathcal{O}(D,D,D,D)).$$
  
  To this, we can associate a Koszul complex $\mathcal{K}(X,f,g,h,i)$. The most natural choice is to put $\op X$-stabilizers, qubits and $\op Z$-stabilizers in the middle degrees degrees $-3,-2$ and $-1$, respectively, where the Koszul complex has the form
  \[
  \begin{tikzcd}[ampersand replacement=\&, column sep=60pt]
      {(\mathcal{L}^{\otimes 3})^4} \& {(\mathcal{L}^{\otimes 2})^6}
   \& {(\mathcal{L})^4}
      \arrow["{\begin{psmallmatrix} h & g & 0 & f & 0 & 0 \\ i & 0 & g & 0 & f & 0 \\ 0 & i & h & 0 & 0 &
  f \\ 0 & 0 & 0 & i & h & g \end{psmallmatrix}^t}", from=1-1, to=1-2]
      \arrow["{\begin{psmallmatrix} g & h & i & 0 & 0 & 0 \\ f & 0 & 0 & h & i & 0 \\ 0 & f & 0 & g & 0 &
  i \\ 0 & 0 & f & 0 & g & h \end{psmallmatrix}}", from=1-2, to=1-3]
  \end{tikzcd}
  \]
  for $\mathcal{L}=\mathcal{O}(-D,-D,-D,-D).$ We choose the line bundle
  $\mathcal{S}=\mathcal{O}(L-1+D, M-1+D, -N+D, -P+D)$ and define shifted Koszul complex $\mathcal{K}_{4D\op{-tile}}=\mathcal{K}(X,f,g,h,i)\otimes \mathcal{S}$. Passing to the second derived higher sections, we obtain the complex
  \[\begin{tikzcd}
	{K^{\bullet}_{4D\op{-tile}}:R^2\Gamma(X,\mathcal{K}^{-{3}}_{4D\op{-tile}})} & {R^2\Gamma(X,\mathcal{K}^{-{2}}_{4D\op{-tile}})} & {R^2\Gamma(X,\mathcal{K}^{-1}_{4D\op{-tile}}).}
	\arrow[from=1-1, to=1-2]
	\arrow[from=1-2, to=1-3]
\end{tikzcd}\]
  The resulting code is supported on a hypercube with $6LMNP$ qubits supported on the faces. When we assume the appropriate generalization of geometric
  total topological order, the space of logicals of the code can be computed as
      $$H^{-1}(K^{\bullet}_{4D\op{-tile}})\cong \ftwo[w^{\pm1},x^{\pm1},y^{\pm1},z^{\pm1}]/(f,g,h,i)$$
  and one may show that the dimension of the space of logicals is $24D^4.$ For example, choosing $D = 1$, $L =M = N = P = 3$ as well as polynomials
  \begin{align*}
  f = 1 + xy + wyz + wxz + x, \quad
  g = xz + w + wxy + wxyz + z, \\
  h = yz + x + wz + wy + wxy, \quad
  i = z + y + xyz + wx + wxz,
  \end{align*}
    we obtain a code with parameters $[[486,24, 10\leq d\leq 15]].$
    
\section{Derived Automorphisms}\label{sec:derivedautomorphisms}

We introduce the notion of derived automorphisms of CSS codes and discuss it in the example of tile codes. This gives the theoretical backbone to the derived automorphisms construct ad-hoc in \Cref{subsec: pedestrianstylederivedautomorphisms}.
\subsection{Definition}
In this section, we introduce a generalized notion of isomorphism of CSS codes, that naturally encapsulates the derived automorphisms considered in \Cref{subsec: pedestrianstylederivedautomorphisms}, but might also be of independent interest.

For this, we first discuss a complication in the definition of a morphism of CSS codes. It might be tempting to define a morphism  between two CSS codes as chain maps between the associated chain complexes, say $C^\bullet$ and $C'^\bullet$.
However, that this introduces a very unnatural asymmetry, since the $\op X$- and $\op Z$-parity checks of the CSS codes are treated in an asymmetric way in the associated chain complexes.
For this reason, it seems to be reasonable to rather consider certain \emph{correspondences} between the chain complexes, that is, diagrams of the form 

\[\begin{tikzcd}[row sep=tiny]
	& {D^\bullet} \\
	{C^\bullet} && {C'^\bullet}
	\arrow[from=1-2, to=2-1]
	\arrow[from=1-2, to=2-3]
\end{tikzcd}\]
where $D^\bullet$ is a third chain complex and the two arrows are chain maps. In particular, correspondences can be transposed by simply switching the role of $C^\bullet$ and $C'^\bullet$. This effectively removes the asymmetry. Now, based on the application, one can put various restriction on the chain maps in the correspondence, for example regarding the preservation of distance or bases.

Secondly, it can be very valuable to weaken the notion of isomorphism. Recall that a chain map $C^\bullet\to D^\bullet$ is called a \emph{quasi-isomorphism} if it induces an isomorphism on all cohomology groups $H^i(C^\bullet)\stackrel{\sim}{\to}H^i(D^\bullet).$ For example, adding a single qubit and a check that is supported on this qubit yields a quasi-isomorphic code. Another example is the inclusion of the tile code in a code supported on a horizontal strip, see \Cref{lem:logicalsoftilecodes}.

Combining these two ideas, we define a \emph{derived isomorphism} between to CSS codes given by chain complexes $C^\bullet$ and $C'^\bullet$ as a diagram of the form
% https://q.uiver.app/#q=WzAsNyxbMCwxLCJDXlxcYnVsbGV0Il0sWzEsMCwiRF8xXlxcYnVsbGV0Il0sWzIsMSwiQ18xXlxcYnVsbGV0Il0sWzMsMCwiRF8yXlxcYnVsbGV0Il0sWzQsMSwiXFxkb3RzIl0sWzUsMCwiRF9uIl0sWzYsMSwiQydeXFxidWxsZXQiXSxbMSwwLCJxLmkuIiwyXSxbMSwyLCJxLmkuIl0sWzMsMiwicS5pLiIsMl0sWzMsNCwicS5pIl0sWzUsNCwicS5pLiIsMl0sWzUsNiwicS5pIl1d
\[\begin{tikzcd}[row sep=tiny]
	& {D_1^\bullet} && {D_2^\bullet} && {D^\bullet_n} \\
	{C^\bullet} && {C_1^\bullet} && \dots && {C'^\bullet}
	\arrow["{q.i.}"', from=1-2, to=2-1]
	\arrow["{q.i.}", from=1-2, to=2-3]
	\arrow["{q.i.}"', from=1-4, to=2-3]
	\arrow["{q.i}", from=1-4, to=2-5]
	\arrow["{q.i.}"', from=1-6, to=2-5]
	\arrow["{q.i}", from=1-6, to=2-7]
\end{tikzcd}\]
where all arrows are quasi-isomorphisms. Derived isomorphism can be composed by concatenation of their diagrams.
While in the diagram defining a derived automorphism, the maps go out of the objects $D_i^\bullet$, one may turn this around by the following construction:
% https://q.uiver.app/#q=WzAsNSxbMSwwLCJDXlxcYnVsbGV0Il0sWzMsMCwiQydeXFxidWxsZXQiXSxbMCwxLCJDXlxcYnVsbGV0Il0sWzIsMSwiRF5cXGJ1bGxldCJdLFs0LDEsIkReXFxidWxsZXQiXSxbMCwzLCJ7cS5pLn0iXSxbMSwzLCJ7cS5pLn0iLDJdLFswLDIsIiIsMix7ImxldmVsIjoyLCJzdHlsZSI6eyJoZWFkIjp7Im5hbWUiOiJub25lIn19fV0sWzEsNCwiIiwwLHsibGV2ZWwiOjIsInN0eWxlIjp7ImhlYWQiOnsibmFtZSI6Im5vbmUifX19XV0=
\[\begin{tikzcd}[row sep=tiny]
	& {C^\bullet} && {C'^\bullet} \\
	{C^\bullet} && {D^\bullet} && {D^\bullet}.
	\arrow[equals, from=1-2, to=2-1]
	\arrow["{{q.i.}}", from=1-2, to=2-3]
	\arrow["{{q.i.}}"', from=1-4, to=2-3]
	\arrow[equals, from=1-4, to=2-5]
\end{tikzcd}\]

A \emph{derived automorphism} is a derived isomorphism between the same chain complex $C^\bullet=C'^\bullet.$ We note that if $C^\bullet$ is quasi-isomorphic to $C'^\bullet$, then each automorphism of $C'^\bullet$ yields a derived automorphism of $C^\bullet$.

We will mostly be interested in derived isomorphisms that fulfill two additional conditions. First, we assume that all chain complexes in the diagram are equipped with bases and all chain maps send each basis vector either to a unique basis vector or to zero. Second, we assume that the distance of $C^\bullet_i$ and $C^\bullet_i$ is at least the distance of $C^\bullet$ and $C'^\bullet.$
With these assumptions, one may construct a fault-tolerant circuit that implements a derived isomorphism.
\begin{remark}
    As suggested by the name, derived isomorphism are closely related to isomorphisms in the derived category of $\ftwo$-vector spaces. However, in this category each complex is isomorphic to a complex with trivial differential and hence a trivial code. So, it is important to impose restrictions on the allowed chain maps.
\end{remark}
\subsection{Derived automorphism of tile codes}
As in \Cref{subsec: Resolving tile codes by Koszul complexes} we denote by $K_{\op{tile}}^\bullet$ denote the chain complex associated to a tile code defined by polynomials $f,g$. We assume that the polynomials fulfill algebraic topological order, see \Cref{sec:infinitelatticekoszul}. In particular, this implies that the space of logical operators is given by
\[
    H^{-1}(K_{\op{tile}}^\bullet)=R/(f,g)=\ftwo[x^\pm, y^\pm]/(f,g).
\]
While, in general, a tile code has no automorphism that sends basis vectors to basis vectors due to the open boundary condition, we will now realize multiplication by $x$ and $y$ as derived automorphisms, say $T_x$ and $T_y$. Moreover, we show that their action on the space of logicals is given by multiplication on $R/(f,g).$

A quick way to see this that however involves infinite dimensional codes is the following. Multiplication by $x$ yields an automorphism of $K^\bullet_{\pm, \op{strip}}$, say $T_x$, which acts by multiplication by $x$ on $H^{-1}(K^\bullet_{\pm, \op{strip}})=R/(f,g).$ The natural inclusion yields a quasi-isomorphism $K_{\op{tile}}^\bullet\to K^\bullet_{\pm, \op{strip}}$, and we obtain the desired derived automorphism $T_x$ of $K_{\op{tile}}^\bullet$. The derived automorphism $T_y$ is obtained similarly by using the quasi-isomorphism $K^\bullet_{\op{strip},\pm}\to K_{\op{tile}}^\bullet$ and the action of $y$ on $K^\bullet_{\op{strip},\pm}.$

These derived automorphism can also be realized using only finite dimensional codes. One can obtain the derived automorphism $T_x$ using the diagram
\[\begin{tikzcd}[row sep=tiny]
	{K^\bullet_{\op{tile}}} && {K^\bullet_{\op{tile}}} \\
	& {K'^\bullet_{\op{tile}}}
	\arrow[from=1-1, to=2-2]
	\arrow[from=1-3, to=2-2]
\end{tikzcd}\]
where $K'^\bullet_{\op{tile}}$ is a tile code of size $(L+1)\times M$. The first map embeds the copy of the tile code $K^\bullet_{\op{tile}}$ of size $L\times M$ on the left side of $K'^\bullet_{\op{tile}}$. The second map embeds $K^\bullet_{\op{tile}}$ on the right side of $K'^\bullet_{\op{tile}}$. 

Similarly, the derived automorphism $T_y$ also arises from the diagram

\[\begin{tikzcd}[row sep=small]
	& {K''^\bullet_{\op{tile}}} \\
	{K^\bullet_{\op{tile}}} && {K^\bullet_{\op{tile}}}
	\arrow[from=1-2, to=2-1]
	\arrow[from=1-2, to=2-3]
\end{tikzcd}\]
where $K''^\bullet_{\op{tile}}$ is a tile code of size $L\times (M+1)$ and the two maps remove the top (respectively bottom) row of qubits and checks in $K''^\bullet_{\op{tile}}$. 

This gives exactly the derived automorphisms that we constructed in \Cref{subsec: pedestrianstylederivedautomorphisms} by extending the lattice and measuring out qubits. 

The group generated by the derived automorphism $T_x$ and $T_y$ now depends on the polynomials $f,g$. We close the section with explaining the structure observed in \Cref{example: pedestrianderivedauto} from an algebraic perspective.
\begin{example}\label{example: algebraofderivedauto}
We can now understand \Cref{example: pedestrianderivedauto} a bit better.  
The primary decomposition of the ideal $(f,g)$ shows that
    \begin{align}
        R/(f,g)\cong \ftwo[x]/(x^3 + x^2 + 1)\oplus \ftwo[x]/(x^5 + x^3 + 1)\cong \mathbb{F}_{8}\oplus \mathbb{F}_{32}.
    \end{align}
    Multiplication by $x$ has order $7$ and $31$ on each component, respectively, and order $7\cdot31=217$ on the ring $R/(f,g)$. Moreover, $y=x^{149}$ in $R/(f,g)$, so that multiplication by $x$ and $y$ generates the same group.

    This shows that the derived automorphism $T_x$ (or equivalently $T_y$) generate a cyclic group of order $217$ acting by derived automorphism on $K^\bullet_{\op{tile}}$.
\end{example}

\textit{Acknowledgements ---}  JNE was supported by Deutsche Forschungsgemeinschaft (DFG), project number 45744154, Equivariant K-motives and Koszul duality. We thank Yu-An Chen and Zijian Liang for helpful discussions.

\bibliographystyle{plain}
\bibliography{slate_codes}

@misc{stacks-project-tag-015X,
  author       = {The {Stacks Project Authors}},
  title        = {The Stacks Project},
  howpublished = {\url{https://stacks.math.columbia.edu}},
  year         = {2023},
  note         = {Tag 015X, \url{https://stacks.math.columbia.edu/tag/015X}}
}

@article{freedman2001projective,
  title={Projective plane and planar quantum codes},
  author={Freedman, Michael H and Meyer, David A},
  journal={Foundations of Computational Mathematics},
  volume={1},
  pages={325--332},
  year={2001},
  publisher={Springer}
}

@misc{eberhardt2024pruningqldpccodesbivariate,
      title={Pruning {qLDPC} codes: Towards bivariate bicycle codes with open boundary conditions}, 
      author={Jens Niklas Eberhardt and Francisco Revson F. Pereira and Vincent Steffan},
      year={2024},
      eprint={2412.04181},
      archivePrefix={arXiv},
      primaryClass={quant-ph},
      url={https://arxiv.org/abs/2412.04181}, 
}

@misc{bravyi1998quantum,
      title={Quantum codes on a lattice with boundary}, 
      author={S. B. Bravyi and A. Yu. Kitaev},
      
      eprint={quant-ph/9811052},
      archivePrefix={arXiv},
      primaryClass={quant-ph},
year={1998}
}

@misc{eberhardt2024logicaloperatorsfoldtransversalgates,
      title={Logical Operators and Fold-Transversal Gates of Bivariate Bicycle Codes}, 
      author={Jens Niklas Eberhardt and Vincent Steffan},
      year={2024},
      eprint={2407.03973},
      archivePrefix={arXiv},
      primaryClass={quant-ph},
      url={https://arxiv.org/abs/2407.03973}, 
}

@article{breuckmannFoldTransversalCliffordGates2024,
  doi = {10.22331/q-2024-06-13-1372},
  url = {https://doi.org/10.22331/q-2024-06-13-1372},
  title = {Fold-{T}ransversal {C}lifford {G}ates for {Q}uantum {C}odes},
  author = {Breuckmann, Nikolas P. and Burton, Simon},
  journal = {{Quantum}},
  issn = {2521-327X},
  publisher = {{Verein zur F{\"{o}}rderung des Open Access Publizierens in den Quantenwissenschaften}},
  volume = {8},
  pages = {1372},
  month = jun,
  year = {2024}
}

@misc{mathews2025placingroutingquantumldpc,
      title={Placing and routing quantum LDPC codes in multilayer superconducting hardware}, 
      author={Melvin Mathews and Lukas Pahl and David Pahl and Vaishnavi L. Addala and Catherine Tang and William D. Oliver and Jeffrey A. Grover},
      year={2025},
      eprint={2507.23011},
      archivePrefix={arXiv},
      primaryClass={quant-ph},
      url={https://arxiv.org/abs/2507.23011}, 
}

@article{linQuantumTwoblockGroup2023,
  title = {Quantum two-block group algebra codes},
  author = {Lin, Hsiang-Ku and Pryadko, Leonid P.},
  journal = {Phys. Rev. A},
  volume = {109},
  issue = {2},
  pages = {022407},
  numpages = {17},
  year = {2024},
  month = {Feb},
  publisher = {American Physical Society},
  doi = {10.1103/PhysRevA.109.022407},
  url = {https://link.aps.org/doi/10.1103/PhysRevA.109.022407}
}

@article{breuckmannQuantumLowDensityParityCheck2021a,
  title = {Quantum {{Low-Density Parity-Check Codes}}},
  author = {Breuckmann, Nikolas P. and Eberhardt, Jens Niklas},
  year = {2021},
  month = oct,
  journal = {PRX Quantum},
  volume = {2},
  number = {4},
  eprint = {2103.06309},
  primaryclass = {quant-ph},
  pages = {040101},
  issn = {2691-3399},
  doi = {10.1103/PRXQuantum.2.040101},
  urldate = {2024-02-01},
  archiveprefix = {arxiv},
  langid = {english}
}

@article{bravyi2023highthreshold,
   title={High-threshold and low-overhead fault-tolerant quantum memory},
   volume={627},
   ISSN={1476-4687},
   url={http://dx.doi.org/10.1038/s41586-024-07107-7},
   DOI={10.1038/s41586-024-07107-7},
   number={8005},
   journal={Nature},
   publisher={Springer Science and Business Media LLC},
   author={Bravyi, Sergey and Cross, Andrew W. and Gambetta, Jay M. and Maslov, Dmitri and Rall, Patrick and Yoder, Theodore J.},
   year={2024},
   month=mar, pages={778–782} }

@article{chen2025anyontheorytopologicalfrustration,
  title = {Anyon Theory and Topological Frustration of High-Efficiency Quantum Low-Density Parity-Check Codes},
  author = {Chen, Keyang and Liu, Yuanting and Zhang, Yiming and Liang, Zijian and Chen, Yu-An and Liu, Ke and Song, Hao},
  journal = {Phys. Rev. Lett.},
  volume = {135},
  issue = {7},
  pages = {076603},
  numpages = {8},
  year = {2025},
  month = {Aug},
  publisher = {American Physical Society},
  doi = {10.1103/86j7-cmsw},
  url = {https://link.aps.org/doi/10.1103/86j7-cmsw}
}

@article{liang2025generalizedtoriccodestwisted,
  title = {Generalized Toric Codes on Twisted Tori for Quantum Error Correction},
  author = {Liang, Zijian and Liu, Ke and Song, Hao and Chen, Yu-An},
  journal = {PRX Quantum},
  volume = {6},
  issue = {2},
  pages = {020357},
  numpages = {22},
  year = {2025},
  month = {Jun},
  publisher = {American Physical Society},
  doi = {10.1103/rmy6-9n89},
  url = {https://link.aps.org/doi/10.1103/rmy6-9n89}
}

@article{vigneau2025quantumerrordetectionqubitresonator,
  title = {Quantum error detection in qubit-resonator star architecture},
  author={Florian Vigneau and Sourav Majumder and Aniket Rath and Pedro Parrado-Rodr{\'\i}guez and Francisco Revson Fernandes Pereira and Hsiang-Sheng Ku and Fedor Šimkovic IV and Stefan Pogorzalek and Tyler Jones and Nicola Wurz and Michael Renger and Jeroen Verjauw and Ping Yang and William Kindel and Frank Deppe and Johannes Heinsoo},
  journal = {PRX Quantum},
  pages = {--},
  year = {2025},
  month = {Nov},
  publisher = {American Physical Society},
  doi = {10.1103/8m33-wn4g},
  url = {https://link.aps.org/doi/10.1103/8m33-wn4g}
}

@misc{renger2025superconductingqubitresonatorquantumprocessor,
      title={A Superconducting Qubit-Resonator Quantum Processor with Effective All-to-All Connectivity}, 
      author={Michael Renger and Jeroen Verjauw and Nicola Wurz and Amin Hosseinkhani and Caspar Ockeloen-Korppi and Wei Liu and Aniket Rath and Manish J. Thapa and Florian Vigneau and Elisabeth Wybo and Ville Bergholm and Chun Fai Chan and Bálint Csatári and Saga Dahl and Rakhim Davletkaliyev and Rakshyakar Giri and Daria Gusenkova and Hermanni Heimonen and Tuukka Hiltunen and Hao Hsu and Eric Hyyppä and Joni Ikonen and Tyler Jones and Shabeeb Khalid and Seung-Goo Kim and Miikka Koistinen and Anton Komlev and Janne Kotilahti and Vladimir Kukushkin and Julia Lamprich and Alessandro Landra and Lan-Hsuan Lee and Tianyi Li and Per Liebermann and Sourav Majumder and Janne Mäntylä and Fabian Marxer and Arianne Meijer - van de Griend and Vladimir Milchakov and Jakub Mrożek and Jayshankar Nath and Tuure Orell and Miha Papič and Matti Partanen and Alexander Plyushch and Stefan Pogorzalek and Jussi Ritvas and Pedro Figuero Romero and Ville Sampo and Marko Seppälä and Ville Selinmaa and Linus Sundström and Ivan Takmakov and Brian Tarasinski and Jani Tuorila and Olli Tyrkkö and Alpo Välimaa and Jaap Wesdorp and Ping Yang and Liuqi Yu and Johannes Heinsoo and Antti Vepsäläinen and William Kindel and Hsiang-Sheng Ku and Frank Deppe},
      year={2025},
      eprint={2503.10903},
      archivePrefix={arXiv},
      primaryClass={quant-ph},
      url={https://arxiv.org/abs/2503.10903}, 
}

@Article{Wallraff2004,
author={Wallraff, A.
and Schuster, D. I.
and Blais, A.
and Frunzio, L.
and Huang, R.-. S.
and Majer, J.
and Kumar, S.
and Girvin, S. M.
and Schoelkopf, R. J.},
title={Strong coupling of a single photon to a superconducting qubit using circuit quantum electrodynamics},
journal={Nature},
year={2004},
month={Sep},
day={01},
volume={431},
number={7005},
pages={162-167},
issn={1476-4687},
doi={10.1038/nature02851},
url={https://doi.org/10.1038/nature02851}
}

@article{blais2004,
  title = {Cavity Quantum Electrodynamics for Superconducting Electrical Circuits: {{An}} Architecture for Quantum Computation},
  shorttitle = {Cavity Quantum Electrodynamics for Superconducting Electrical Circuits},
  author = {Blais, Alexandre and Huang, Ren-Shou and Wallraff, Andreas and Girvin, S. M. and Schoelkopf, R. J.},
  year = {2004},
  month = jun,
  volume = {69},
  pages = {062320},
  doi = {10.1103/PhysRevA.69.062320},
  journal = {Physical Review A},
  number = {6}
}

@article{Song2017,
   title={10-Qubit Entanglement and Parallel Logic Operations with a Superconducting Circuit},
   volume={119},
   ISSN={1079-7114},
   url={http://dx.doi.org/10.1103/PhysRevLett.119.180511},
   number={18},
   journal={Physical Review Letters},
   publisher={American Physical Society (APS)},
   author={Song, Chao and Xu, Kai and Liu, Wuxin and Yang, Chui-ping and Zheng, Shi-Biao and Deng, Hui and Xie, Qiwei and Huang, Keqiang and Guo, Qiujiang and Zhang, Libo and others},
   year={2017},
   month=nov }

@article{Song2019,
   title={Generation of multicomponent atomic Schrödinger cat states of up to 20 qubits},
   volume={365},
   ISSN={1095-9203},
   url={http://dx.doi.org/10.1126/science.aay0600},
   DOI={10.1126/science.aay0600},
   number={6453},
   journal={Science},
   publisher={American Association for the Advancement of Science (AAAS)},
   author={Song, Chao and Xu, Kai and Li, Hekang and Zhang, Yu-Ran and Zhang, Xu and Liu, Wuxin and Guo, Qiujiang and Wang, Zhen and Ren, Wenhui and Hao, Jie and others},
   year={2019},
   month=aug, pages={574–577} }

@article{liang2025planarquantumlowdensityparitycheck,
  title = {Planar Quantum Low-Density Parity-Check Codes with Open Boundaries},
  author = {Liang, Zijian and Eberhardt, Jens Niklas and Chen, Yu-An},
  journal = {PRX Quantum},
  volume = {6},
  issue = {4},
  pages = {040330},
  numpages = {35},
  year = {2025},
  month = {Nov},
  publisher = {American Physical Society},
  doi = {10.1103/qv65-vmzr},
  url = {https://link.aps.org/doi/10.1103/qv65-vmzr}
}

@Article{Haah2013,
author={Haah, Jeongwan},
title={Commuting Pauli Hamiltonians as Maps between Free Modules},
journal={Communications in Mathematical Physics},
year={2013},
month={Dec},
day={01},
volume={324},
number={2},
pages={351-399},
issn={1432-0916},
doi={10.1007/s00220-013-1810-2},
url={https://doi.org/10.1007/s00220-013-1810-2}
}

@article{steffan2025tilecodeshighefficiencyquantum,
  title = {Tile Codes: High-Efficiency Quantum Codes on a Lattice with Boundary},
  author = {Steffan, Vincent and Choe, Shin Ho and Breuckmann, Nikolas P. and Pereira, Francisco Revson Fernandes and Eberhardt, Jens Niklas},
  journal = {Phys. Rev. Lett.},
  volume = {135},
  issue = {17},
  pages = {170601},
  numpages = {7},
  year = {2025},
  month = {Oct},
  publisher = {American Physical Society},
  doi = {10.1103/l4mx-l3xx},
  url = {https://link.aps.org/doi/10.1103/l4mx-l3xx}
}

@article{Litinski_2019,
   title={A Game of Surface Codes: Large-Scale Quantum Computing with Lattice Surgery},
   volume={3},
   ISSN={2521-327X},
   url={http://dx.doi.org/10.22331/q-2019-03-05-128},
   DOI={10.22331/q-2019-03-05-128},
   journal={Quantum},
   publisher={Verein zur Forderung des Open Access Publizierens in den Quantenwissenschaften},
   author={Litinski, Daniel},
   year={2019},
   month=mar, pages={128} }

@misc{yang2025planarfaulttolerantquantumcomputation,
      title={Planar Fault-Tolerant Quantum Computation with Low Overhead}, 
      author={Yingli Yang and Guo Zhang and Ying Li},
      year={2025},
      eprint={2506.18061},
      archivePrefix={arXiv},
      primaryClass={quant-ph},
      url={https://arxiv.org/abs/2506.18061}, 
}

@article{Fowler_2012,
   title={Surface codes: Towards practical large-scale quantum computation},
   volume={86},
   ISSN={1094-1622},
   url={http://dx.doi.org/10.1103/PhysRevA.86.032324},
   DOI={10.1103/physreva.86.032324},
   number={3},
   journal={Physical Review A},
   publisher={American Physical Society (APS)},
   author={Fowler, Austin G. and Mariantoni, Matteo and Martinis, John M. and Cleland, Andrew N.},
   year={2012},
   month=sep }

@misc{yoder2025tourgrossmodularquantum,
      title={Tour de gross: A modular quantum computer based on bivariate bicycle codes}, 
      author={Theodore J. Yoder and Eddie Schoute and Patrick Rall and Emily Pritchett and Jay M. Gambetta and Andrew W. Cross and Malcolm Carroll and Michael E. Beverland},
      year={2025},
      eprint={2506.03094},
      archivePrefix={arXiv},
      primaryClass={quant-ph},
      url={https://arxiv.org/abs/2506.03094}, 
}

@article{Cowtan_2024,
   title={CSS code surgery as a universal construction},
   volume={8},
   ISSN={2521-327X},
   url={http://dx.doi.org/10.22331/q-2024-05-14-1344},
   DOI={10.22331/q-2024-05-14-1344},
   journal={Quantum},
   publisher={Verein zur Forderung des Open Access Publizierens in den Quantenwissenschaften},
   author={Cowtan, Alexander and Burton, Simon},
   year={2024},
   month=may, pages={1344} }

@article{Quintavalle_2023,
   title={Partitioning qubits in hypergraph product codes to implement logical gates},
   volume={7},
   ISSN={2521-327X},
   url={http://dx.doi.org/10.22331/q-2023-10-24-1153},
   DOI={10.22331/q-2023-10-24-1153},
   journal={Quantum},
   publisher={Verein zur Forderung des Open Access Publizierens in den Quantenwissenschaften},
   author={Quintavalle, Armanda O. and Webster, Paul and Vasmer, Michael},
   year={2023},
   month=oct, pages={1153} }

@article{Tillich_2014,
   title={Quantum LDPC Codes With Positive Rate and Minimum Distance Proportional to the Square Root of the Blocklength},
   volume={60},
   ISSN={1557-9654},
   url={http://dx.doi.org/10.1109/TIT.2013.2292061},
   DOI={10.1109/tit.2013.2292061},
   number={2},
   journal={IEEE Transactions on Information Theory},
   publisher={Institute of Electrical and Electronics Engineers (IEEE)},
   author={Tillich, Jean-Pierre and Zemor, Gilles},
   year={2014},
   month=feb, pages={1193–1202} }

@article{Pecorari_2025,
   title={High-rate quantum LDPC codes for long-range-connected neutral atom registers},
   volume={16},
   ISSN={2041-1723},
   url={http://dx.doi.org/10.1038/s41467-025-56255-5},
   DOI={10.1038/s41467-025-56255-5},
   number={1},
   journal={Nature Communications},
   publisher={Springer Science and Business Media LLC},
   author={Pecorari, Laura and Jandura, Sven and Brennen, Gavin K. and Pupillo, Guido},
   year={2025},
   month=jan }

@article{PhysRevLett.85.2208,
  title = {Fast Quantum Gates for Neutral Atoms},
  author = {Jaksch, D. and Cirac, J. I. and Zoller, P. and Rolston, S. L. and C\^ot\'e, R. and Lukin, M. D.},
  journal = {Phys. Rev. Lett.},
  volume = {85},
  issue = {10},
  pages = {2208--2211},
  numpages = {0},
  year = {2000},
  month = {Sep},
  publisher = {American Physical Society},
  doi = {10.1103/PhysRevLett.85.2208},
  url = {https://link.aps.org/doi/10.1103/PhysRevLett.85.2208}
}

@inproceedings{Grassl_2013,
   title={Leveraging automorphisms of quantum codes for fault-tolerant quantum computation},
   url={http://dx.doi.org/10.1109/ISIT.2013.6620283},
   DOI={10.1109/isit.2013.6620283},
   booktitle={2013 IEEE International Symposium on Information Theory},
   publisher={IEEE},
   author={Grassl, Markus and Roetteler, Martin},
   year={2013},
   month=jul, pages={534–538} }

@article{RevModPhys.82.2313,
  title = {Quantum information with Rydberg atoms},
  author = {Saffman, M. and Walker, T. G. and M\o{}lmer, K.},
  journal = {Rev. Mod. Phys.},
  volume = {82},
  issue = {3},
  pages = {2313--2363},
  numpages = {0},
  year = {2010},
  month = {Aug},
  publisher = {American Physical Society},
  doi = {10.1103/RevModPhys.82.2313},
  url = {https://link.aps.org/doi/10.1103/RevModPhys.82.2313}
}

@article{PhysRevLett.107.263001,
  title = {Trapping Rydberg Atoms in an Optical Lattice},
  author = {Anderson, S. E. and Younge, K. C. and Raithel, G.},
  journal = {Phys. Rev. Lett.},
  volume = {107},
  issue = {26},
  pages = {263001},
  numpages = {5},
  year = {2011},
  month = {Dec},
  publisher = {American Physical Society},
  doi = {10.1103/PhysRevLett.107.263001},
  url = {https://link.aps.org/doi/10.1103/PhysRevLett.107.263001}
}

@misc{liang2025operatoralgebraalgorithmicconstruction,
      title={Operator algebra and algorithmic construction of boundaries and defects in (2+1)D topological Pauli stabilizer codes}, 
      author={Zijian Liang and Bowen Yang and Joseph T. Iosue and Yu-An Chen},
      year={2025},
      eprint={2410.11942},
      archivePrefix={arXiv},
      primaryClass={quant-ph},
      url={https://arxiv.org/abs/2410.11942}, 
}

@article{cellular_automaton,
	date = {1984/06/01},
author = {Olivier Martin and Andrew M. Odlyzko and Stephen Wolfram },
	doi = {10.1007/BF01223745},
	id = {Martin1984},
	isbn = {1432-0916},
	journal = {Communications in Mathematical Physics},
	number = {2},
	pages = {219--258},
	title = {Algebraic properties of cellular automata},
	url = {https://doi.org/10.1007/BF01223745},
	volume = {93},
	year = {1984}}

\end{document}